\documentclass[preprint,aps,12pt,showpacs,nofootinbib,tightenlines,amsmath,amssymb]{revtex4}
\usepackage{amsmath}
\usepackage{graphicx}
\usepackage{amssymb}
\usepackage{color}

\newcommand{\be}{\begin{eqnarray}}
\newcommand{\ee}{\end{eqnarray}}
\newcommand{\wt}{\widetilde}
\newcommand{\wh}{\widehat}
\newcommand{\p}{\partial}

\newcommand{\bvnabla}{\breve{\nabla}}
 
\newcommand{\bvE}{\breve{E}}
\newcommand{\bvB}{\breve{B}}
\newcommand{\bvfE}{\breve{\fE}}
\newcommand{\bvfB}{\breve{\fB}}
\newcommand{\bvA}{\breve{A}}
\newcommand{\bvfA}{\breve{\fA}}

\newcommand{\ck}{\check}

\newcommand{\nn}{\nonumber}

\newcommand{\Tr}{\mathop{\rm Tr}\nolimits}
\newcommand{\diag}{\mathop{\rm diag}}
\newcommand{\cL}{{\mathcal L}}
\newcommand{\fkL}{\mathfrak{L}}

\newcommand{\cA}{{\mathcal A}}
\newcommand{\hcA}{ \hat{{\mathcal A}}}
\newcommand{\ckcA}{ \check{{\mathcal A}}}

\newcommand{\hcF}{ \hat{{\mathcal F}}}

\newcommand{\bcF}{ \bar{{\mathcal F}}}
\newcommand{\ckcF}{ \check{{\mathcal F}}}

\newcommand{\mF}{\mathsf F}

\newcommand{\bs}{\boldsymbol}
\newcommand{\bv}{\breve}

\newcommand{\whcF}{ \boldsymbol{{\mathcal F}}}
\newcommand{\whmF}{ \boldsymbol{{\mF}}}

\newcommand{\whF}{\boldsymbol{F} }
\newcommand{\whcA}{\boldsymbol{{\mathcal A}}}
\newcommand{\whA}{\boldsymbol{A}}
\newcommand{\whD}{\boldsymbol{D}}
\newcommand{\whcD}{ \boldsymbol{{\mathcal D}}}
\newcommand{\bsckcD}{ \boldsymbol{{\ckcD}}}

\newcommand{\cB}{\mathcal B }
\newcommand{\cF}{{\mathcal F}}
\newcommand{\cE}{{\mathcal E}}

\newcommand{\fkA}{\mathfrak{A}}

\newcommand{\ckcD}{\check{\cD}}

\newcommand{\fF}{{\mathbf F}}

\newcommand{\bfF}{\bar{{\mathbf F}}}

\newcommand{\fV}{{\mathbf V}}
\newcommand{\mV}{{\mathsf V}}

\newcommand{\cD}{{\mathcal D}}

\newcommand{\hcD}{\hat{{\mathcal D}}}

\newcommand{\fG}{{\mathbf G}}

\newcommand{\cJ}{{\mathcal J}}
\newcommand{\mJ}{{\mathsf J}}
\newcommand{\fJ}{{\mathbf J}}

\newcommand{\cW}{{\mathcal W}}
\newcommand{\cS}{{\mathcal S}}

\newcommand{\bchi}{\bar{\chi} }

\newcommand{\1}{\mspace{1mu}}

\newcommand{\hga}{\hat{\gamma}}

\newcommand{\heth}{\hat{\eth}}

\newcommand{\fGa}{\mathbf \Gamma}

\newcommand{\vGa}{\varGamma}

\newcommand{\hfGa}{\hat{\fGa}}

\newcommand{\vka}{\varkappa}

\newcommand{\vSi}{\varSigma}

\newcommand{\fSi}{\mathbf \Sigma}
\newcommand{\hfSi}{\hat{\fSi}}

\newcommand{\fH}{\mathbf H }
\newcommand{\mH}{\mathsf H }
\newcommand{\hmH}{\hat{\mH}}
\newcommand{\tmH}{\tilde{\mH}}
\newcommand{\bmH}{\bar{\mH}}

\newcommand{\cR}{\mathcal R}
\newcommand{\fR}{\mathbf R}

\newcommand{\mR}{\mathsf R}

\newcommand{\mS}{\mathsf S}

\newcommand{\adjoin}{\,\widetilde{\Join}\, }

\newcommand{\mA}{\mathsf A}

\newcommand{\mD}{\mathsf D}

\newcommand{\rF}{\mathrm F}

\newcommand{\fM}{\mathbf M}

\newcommand{\tL}{\tilde{L}}

\newcommand{\fA}{\mathbf A}
\newcommand{\fB}{\mathbf B}

\newcommand{\fD}{\mathbf D}
\newcommand{\fE}{\mathbf E}

\newcommand{\hfA}{\hat{\mathbf A}}

\newcommand{\ckcC}{\check{\cC}}

\newcommand{\mOm}{\mathsf \Omega }
\newcommand{\fOm}{\mathbf \Omega }

\newcommand{\tLam}{\tilde{\Lambda}}

\newcommand{\chih}{\hat{\chi}}

\newcommand{\tchi}{\tilde{\chi}}

\newcommand{\Mka}{M_{\kappa}}

\newcommand{\ha}{\hat{a}}
\newcommand{\hb}{\hat{b}}

\newcommand{\cC}{\mathcal{C}}

\newcommand{\vPsi}{\varPsi}

\begin{document}
\def\intdk{\int\frac{d^4k}{(2\pi)^4}}
\def\sla{\hspace{-0.22cm}\slash}
\hfill


\title{Gravidynamics, spinodynamics and electrodynamics within the framework of gravitational quantum field theory}

\author{Yue-Liang Wu}\email{ylwu@itp.ac.cn, ylwu@ucas.ac.cn}
\affiliation{$^1$ International Centre for Theoretical Physics Asia-Pacific, University of Chinese Academy of Sciences, Beijing 100190, China \\
$^2$ Institute of Theoretical Physics, Chinese Academy of Sciences, Beijing 100190, China\\
$^3$ Taiji Laboratory for Gravitational Wave Universe (Beijing/Hangzhou), University of Chinese Academy of Sciences, Beijing 100049, China \\
$^4$ School of Fundamental Physics and Mathematical Sciences, Hangzhou Institute for Advanced Study, University of Chinese Academy of Sciences, Hangzhou 310024, China }


\begin{abstract}
By noticing the fact that the charged leptons and quarks in the standard model are chirality-based Dirac spinors since their weak interaction violates maximally parity symmetry though they behave as Dirac fermions in electromagnetic interaction, we show that such a chirality-based Dirac spinor possesses not only electric charge gauge symmetry U(1) but also inhomogeneous spin gauge symmetry WS(1,3) = SP(1,3)$\rtimes$W$^{1,3}$, which reveals the nature of gravity and spacetime. The gravitational force and spin gauge force are governed by the gauge symmetries $W^{1,3}$ and SP(1,3), respectively, and a biframe spacetime with globally flat Minkowski spacetime as base spacetime and locally flat gravigauge spacetime as a fiber is described by the gravigauge field through emergent non-commutative geometry. The gauge-geometry duality and renormalizability in gravitational quantum field theory (GQFT) are carefully discussed.  A detailed analysis and systematic investigation on gravidynamics and spinodynamics as well as electrodynamics are carried out within the framework of GQFT. A full discussion on the generalized Dirac equation and Maxwell equation as well as Einstein equation and spin gauge equation is made in biframe spacetime. New effects of gravidynamics as extension of general relativity are particularly analyzed. All dynamic equations of basic fields are demonstrated to preserve the spin gauge covariance and general coordinate covariance due to the spin gauge symmetry and emergent general linear group symmetry GL(1,3,R), so they hold naturally in any spinning reference frame and motional reference frame. 
\end{abstract}

\pacs{04.50.+h,11.15.-q,12.20.-m \\
{\bf Keywords}: Inhomogeneous spin gauge symmetry, Locally flat gravigauge spacetime, Gravitational relativistic quantum mechanics with generalized Dirac equation, Gravidynamics with generalized Einstein equation, Generalized Maxwell equation in any motional and spinning reference frame}

\maketitle

\begin{widetext}
\tableofcontents
\end{widetext}

\section{Introduction}

The theory of electromagnetism was built by Maxwell to unify the electricity and magnetism, which is regarded as the first classical unified field theory. The constancy of the speed of light in Maxwell’s theory inspired Einstein to build up the special theory of relativity (SR) \cite{SR} which unifies space and time into a four dimensional spacetime. Such a spacetime is referred to as four-dimensional Minkowski spacetime which is characterized by the Lorentz symmetry SO(1,3). Later on, Einstein extended the SR to the theory of general relativity (GR) \cite{GR,FGR} based on the principle of general coordinate covariance under arbitrary transformations of coordinates, which is characterized by the general linear group symmetry GL(1,3, R). In GR, the gravity is described by the dynamics of Riemannian geometry of curved spacetime instead of globally flat dimensional Minkowski spacetime in SR. The GR indicates that space and time cannot be well defined in such a way that the differences of the spatial coordinates or time coordinates can directly be measured by the standard ways in SR. On the other hand, the relativistic quantum mechanics built by Dirac has successfully combined the quantum mechanics and special relativity, which led to the establishment of quantum field theory (QFT) in globally flat four-dimensional Minkowski spacetime. The framework of QFT was initiated to extend the classical electromagnetism into quantum electrodynamics (QED)\cite{QED1,QED2a,QED2b,QED3a,QED3b,QED3c,QED4a,QED4b}, which is governed by the Abelian gauge symmetry U(1). Later on, Yang and Mills extended such an Abelian gauge symmetry to a non-Abelian gauge symmetry SU(2) in describing the isotopic spin gauge invariance\cite{YM}. 

The success of the standard model (SM) shows that the basic forces of electroweak and strong interactions are governed by the gauge symmetries U(1)$\times$SU$_L$(2)$\times$SU(3) \cite{EW1,EW2,EW3,QCD1,QCD2,TW}. Namely, the three basic forces in SM can well be described by the dynamics of gauge fields within the framework of quantum field theory formulated in globally flat Minkowski spacetime, which is in contrast to the gravity described by Einstein's GR built based on the dynamics of Riemannian geometry of curved spacetime. It is such a contradiction that causes an obstacle to make a unified description on the gravity and other three basic forces. Recently, the quantum field theory of gravity was constructed based on the spin and scaling gauge symmetries when treating the gravitational interaction on the same footing as the electroweak and strong interactions, such a gravitational gauge field theory provides a framework of gravitational quantum field theory (GQFT) \cite{GQFT,GQFTTK}. The concept of biframe spacetime was shown to play an essential role in GQFT, where a bicovariant vector field defined in biframe spacetime is introduced as a basic gravitational field instead of a metric field in GR to describe the gravitational interaction. In GQFT, the action was proposed to possess a joint symmetry, namely, the Poincar\'e group symmetry PO(1,3) in Minkowski spacetime of coordinates and a proposed internal Poincar\'e-type gauge symmetry PG(1,3) in noncoordinate spacetime spanned by the bicovariant vector field. Such a bicovariant vector field was postulated to be a gauge-type field in the coset PG(1,3)/SP(1,3) of internal Poincar\'e-type gauge symmetry group PG(1,3) with spin gauge group SP(1,3) as a subgroup. 

It is clear that the key point is to realize explicitly the group structure of internal Poincar\'e-type gauge symmetry PG(1,3), which comes to the main purpose in our present paper. Our starting point in this paper is to examine carefully the chiral structure of Dirac fermion. This is because the electron and charged leptons and quarks as basic constituents of matter in SM are actually Weyl fermions although they behave as Dirac fermions in the electromagnetic interaction. The reason is simple by noticing the well-known fact that the weak interactions of leptons and quarks in SM violate maximally the law of parity conservation\cite{PV1,PV2,VA1,VA2}. To make a systematic analysis and detailed investigation, our paper is organized as follows: after a brief introduction in Sect. 1, we will show in Sect. 2 how an inhomogeneous spin symmetry WS(1,3) in Hilbert space is resulted to provide an internal Poincar\'e-type symmetry. Specifically, by viewing the Dirac fermion in four-dimensional Hilbert space as the superposition of left-handed and right-handed Weyl fermions and representing it into a chirality-based Dirac spinor in a chiral spinor representation of eight dimensional Hilbert space, we are able to demonstrate that such a chirality-based Dirac spinor gets an enlarged inhomogeneous spin symmetry group WS(1,3) = SP(1,3)$\rtimes$W$^{1,3}$, which is a semi-direct product group with SP(1,3) being the spin group and W$^{1,3}$ referring to as $\cW_e$-spin group. Such a $\cW_e$-spin group W$^{1,3}$ is shown to be a translation-like chiral-type spin group in eight dimensional Hilbert space. Meanwhile, an internal Abelian symmetry group U(1) remains as electric charge symmetry group. A free motional chirality-based Dirac spinor is verified to possess a maximal associated symmetry when combining the inhomogeneous spin symmetry WS(1,3) in eight dimensional Hilbert space with the Poincar\'e group symmetry PO(1,3) in four-dimensional Minkowski spacetime of coordinates. In Sect. 3, the inhomogeneous spin symmetry WS(1,3) and electric charge symmetry U(1) are gauged to be local symmetries based on the gauge invariance principle. It is found that the $\cW_e$-spin invariant-gauge field brings on the genesis of gravigauge field, which reveals the nature of gravity and brings on the gravitational origin of spin gauge symmetry. The group structure of internal Poincar\'e-type gauge symmetry PG(1,3) is explicitly realized to be as the inhomogeneous spin gauge symmetry, i.e., PG(1,3)$\equiv$WS(1,3) = SP(1,3)$\rtimes$W$^{1,3}$. We further present in Sect. 4 a general demonstration that the non-coordinate spacetime spanned by the gravigauge field basis generates a locally flat gravigauge spacetime characterized by emergent non-commutative geometry. A fiber bundle structure is analyzed in light of biframe spacetime with globally flat Minkowski spacetime as base spacetime and locally flat gravigauge spacetime as a fiber. In Sect. 5, we will show the scaling property of basic fields and make a general definition on gauge fields and field strengths as vector and tensor fields in locally flat gravigauge spacetime. We are going to formulate in Sect. 6 the various formalisms of gauge and scaling invariant actions for the chirality-based Dirac spinor based on the action principle of path integral formulation, which enables us to prove the gauge-gravity and gravity-geometry correspondences and corroborate a gauge-geometry duality relation within the framework of gravitational quantum field theory. A generalized Dirac equation is derived in Sect. 7 to describe the gravitational relativistic quantum theory. Meanwhile, a gravigauge Dirac equation is obtained in locally flat gravigauge spacetime. In Sect. 8, we will present a detailed investigation on the gravidynamics by deriving the gauge-type gravitational equations of gravigauge field in locally flat gravigauge spacetime and biframe spacetime. A geometric gravidynamics is discussed as the extension of general relativity and also analyzed in locally flat gravigauge spacetime. We provide a particular investigation on the new effects of gravidynamics beyond the Einstein theory of general relativity. The spinodynamics is studied in Sect. 9 by deriving the equation of motion for the spin gauge field. In Sect. 10, we provide a systematic analysis on the electrodynamics in the presence of gravitational interaction. Various generalized Maxwell equations are derived to characterize the so-called gravigeometry-medium electrodynamics in the presence of gravimetric field and the gravigauge-mediated electrodynamics obtained in locally flat gravigauge spacetime. Meanwhile, the electromagnetic-like gravimetric-gauge field is introduced in generalized Maxwell equations to describe the gravimetric-gauge-mediated electrodynamics formulated in dynamic Riemannian spacetime. We will further show in Sect. 11 that the dynamics of all basic fields obeys the general coordinate covariance in curved Riemannian spacetime and the spin gauge covariance in locally flat gravigauge spacetime. it is demonstrated that the gravigeometry-medium electrodynamics holds in any motional reference frame and the gravigauge-mediated electrodynamics is applicable in any spinning reference frame. In particular, the gravigeometry-medium Maxwell equations in special background medium are explicitly demonstrated to maintain the general coordinate covariance in motional reference frame, and the gravigauge Maxwell equations in special background medium are examined to keep the spin gauge covariance. Our conclusions and discussions are presented in the final section.

\section{Inhomogeneous spin symmetry WS(1,3) and maximal associated symmetry for chirality-based Dirac spinor }

The basic constituents of nature are known to be the leptons and quarks as indicated in SM. The observed matter consists of the electrons and nucleons which are made of the up and down quarks, they all appear as Dirac fermions in the electromagnetic interaction. Their free motion is described by the following relativistic quantum Dirac equation\cite{DE}:
\be
& & \left(\mathsf{E} - c \boldsymbol{\alpha}\cdot \mathsf{\bf{p}} - \beta mc^2 \right) \psi = 0 , \mbox{or} \nn \\ 
& &  \left( \mathsf{E} + c \boldsymbol{\alpha}\cdot \mathsf{\bf{p}} + \beta mc^2 \right) \psi =0 , \nn
\ee
with $\boldsymbol{\alpha} =(\alpha_1, \alpha_2, \alpha_3 )$ and $\beta$ the $4\times 4$ Hermitian matrices satisfying the conditions $\alpha _{i}\beta =-\beta \alpha _{i}$, $\alpha _{i}\alpha _{j}=-\alpha _{j}\alpha _{i}$ ( $i\neq j $) and $\alpha _{i}^{2}=\beta ^{2}= 1 $. $\psi$ is the Dirac fermion with a complex four-component entity $\psi^T =(\psi_1\, , \psi_2\, ,\psi_3\, , \psi_4)$ in the spinor representation of four-dimensional Hilbert space. A general covariant formulation of the Dirac equation can be rewritten as follows:
\be \label{DiracEQ}
\left( \gamma^{\mu} i \partial_{\mu}  - m \right) \psi = 0 , \nn 
\ee
with $\mu = 0,1,2,3$, $\gamma^0 = \beta$ and $\gamma^i = \beta\alpha^i$.  Where $\partial_{\mu} = \partial/\partial x^{\mu}$ is the 4-dimensional coordinate derivative and $\gamma_{\mu}$ the 4$\times$4 $\gamma$-matrices which satisfy the anticommutation relations, 
\be
\{ \gamma_{\mu}\, \,  \gamma_{\nu} \} = \eta_{\mu\nu} , \qquad \eta_{\mu\nu} = \diag.(1, -1, -1, -1). \nn
\ee

The hermitian action in obtaining the above Dirac equation can be constructed as follows:
\be \label{action1}
S_{D} = \int d^4x\,  \frac{1}{2} \left( \bar{\psi}(x) \gamma^{\mu} i \partial_{\mu} \psi(x) + H.c. \right)  - m\, \bar{\psi}(x) \psi(x) ,
\ee
with $\bar{\psi}(x) = \psi^{\dagger}(x) \gamma^0$ and $m$ being the mass of Dirac fermion. 

In the SM of electroweak and strong interactions, the charged leptons and quarks are actually Weyl fermions due to their weak interactions which bring on the maximal parity violation. Such a Dirac fermion in the electromagnetic interaction should be regarded as the superposition of left-handed and right-handed Weyl fermions. By representing the Dirac fermion in a chiral spinor representation of eight-dimensional Hilbert space, we can express the action of free motion massive Dirac fermion given in Eq.(\ref{action1}) into the following equivalent form:
\begin{eqnarray}
\label{action2}
S & = &  \int d^{4}x\, \frac{1}{2}\{ \bar{\Psi}_{-}(x)  \Gamma^{a}\Gamma_{-}  \delta_{a}^{\;\;\mu} i D_{\mu} \Psi_{-}(x) - \bar{\Psi}_{-}(x) (m_5\Gamma^5\Gamma_{-}  + m_6 \Gamma^{6}\Gamma_{-} ) \Psi_{-}(x) +  H.c.  \} ,
\end{eqnarray}
where we have introduced the following definitions:
\be
& &  \Psi_{-}(x) = \Gamma_{-}  \Psi(x) \equiv \binom{\psi_L(x)}{\psi_R(x)}, \nn \\
& &  \psi_{L,R}(x) = \gamma_{\mp}  \psi(x)  , \nn \\
& & \Gamma_{\mp} = \frac{1}{2} ( 1 \mp \hga_7 ), \quad \gamma_{\mp} = \frac{1}{2} ( 1 \mp \gamma_5 ),   \nn \\
& & \Gamma^{a} = \sigma_0 \otimes \gamma^{a} , \quad \gamma^a= \delta^{a}_{\;\mu} \gamma^{\mu} , \nn \\
& &   \Gamma^5 =  i\sigma_1\otimes \gamma_5 , \quad  \Gamma^6 = i\sigma_2\otimes \gamma_5, \nn \\
& & \Gamma^{\ha} \equiv  2(\Sigma_{-}^{\ha} +  \Sigma_{+}^{\ha}), \quad \Sigma_{\mp}^{\ha} \equiv \frac{1}{2} \Gamma^{\ha} \Gamma_{\mp} , \nn \\
& & \gamma_5 = -i \gamma^0 \gamma^1 \gamma^2 \gamma^3 = \sigma_3 \otimes \sigma_0, \nn \\
& & \hga_7 = - \Gamma^0 \Gamma^1 \Gamma^2 \Gamma^3 \Gamma^5 \Gamma^6 = \sigma_3\otimes \gamma_5,  
\ee
with $\ha \equiv (a,6,7)$. Where $\gamma^{a} $ and $\gamma_5$ are the usual $\gamma$ matrices and $\delta_{a}^{\;\;\mu} $ is the Kronecker symbol. Note that $\gamma_{\mp}$ and $\Gamma_{-}$ provide the chiral project operators via the chiral $\gamma$-matrices $\gamma_5$ and $\hga_7$ defined in four-dimensional and eight-dimensional Hilbert spaces, respectively. 

It is noticed that $\Psi_{-}(x)$ defines a chirality-based Dirac spinor in the chiral spinor representation of eight-dimensional Hilbert space with $\psi_{L,R}(x)$ denoting the ordinary left-handed and right-handed Weyl fermions in four-dimensional Hilbert space. Note that we have introduced the Greek alphabet ($\mu,\nu = 0,1,2,3$) and Latin alphabet ($a,b,=0,1,2,3$) to distinguish four-vector indices in coordinate spacetime and noncoordinate spacetime, respectively. Both the Greek and Latin indices are raised and lowered by the constant metric matrices, i.e., $\eta^{\mu\nu} $ and $\eta^{ab}$ with $\eta^{\mu\nu}$($\eta_{\mu\nu}$)=$\diag.$(1,-1,-1,-1) and $\eta^{ab}$($\eta_{ab}$)=$\diag.$(1,-1,-1,-1). All the scalar product of vectors and tensors is obtained via the contraction with the constant metric matrices $\eta^{\mu\nu}$ and $\eta^{ab}$, i.e., $V^{\mu} V_{\mu} = \eta^{\mu\nu} V_{\mu} V_{\nu}=  \eta_{\mu\nu} V^{\mu} V^{\nu}$ and $A^{a} A_{a} = \eta^{ab} A_{a} A_{b}=\eta_{ab} A^{a} A^{b}$. The system of units is chosen such that $c = \hbar = 1$. 

The mass term of Dirac fermion is produced via the following relations:
\be 
& & \bar{\Psi}_{-}(x) (m_5\Gamma^5\Gamma_{-} + m_6 \Gamma^{6}\Gamma_{-}) \Psi_{-}(x) =  m \bar{\Psi}_{-}(x)\Gamma^{6}\Gamma_{-} e^{i 2\alpha_p \gamma_7} \Psi_{-}(x) \nn \\
& & \equiv  m \bar{\Psi}_{-}(x)\Gamma^{6}\Gamma_{-} e^{-i 2\alpha_p \gamma_5} \Psi_{-}(x)  \to   m \bar{\Psi}_{-}(x)\Gamma^{6}\Gamma_{-} \Psi_{-}(x) =   m \bar{\psi} \psi ,
\ee
where we have made a replacement in obtaining the last form:
\be
& & \Psi_{-}(x) \to e^{- i \alpha_p \gamma_7} \Psi_{-}(x) \equiv e^{i \alpha_p \gamma_5} \Psi_{-}(x), 
\ee
with 
\be
& & \gamma_7 = \sigma_3\otimes \sigma_0\otimes \sigma_0 , \quad \tan 2\alpha_p = m_5/m_6 , \nn \\
& & m = \sqrt{m_5^2 + m_6^2}  .
\ee
Note that such a replacement for the chirality-based Dirac spinor preserves the kinetic term of the action in Eq.(\ref{action2}) due to its invariance under such a transformation.

The $\Gamma$-matrices $\Gamma^{\ha}$ with $\ha = (a,5,6)$ and $\hga_7$ defined in eight-dimensional Hilbert space satisfy the anticommuting relations of Clifford algebra:
\be
\{ \Gamma^{\ha}, \Gamma^{\hb} \} = \eta^{\ha\hb} , \quad \{ \Gamma^{\ha}, \hga_7 \} =0 , 
\ee
which indicates that the mass of Dirac fermion may have a geometric origin via the extra dimension of spacetime as discussed in ref.\cite{GQFT6D}. 

It can be verified that the action given in Eq.(\ref{action2}) possesses the following associated symmetry:
\be \label{MAS}
G_S & = &  P^{1,3}\ltimes SO(1,3)\adjoin SP(1,3)\rtimes W^{1,3} \times U(1) \nn \\
& = & PO(1, 3) \adjoin WS(1,3) \times U(1),  
\ee
where the symbol ``$\adjoin$" is adopted to notate the associated symmetry in which the transformation of spin symmetry group SP(1,3) in Hilbert space must be coincidental to that of the isomorphic Lorentz symmetry group SO(1,3) in Minkowski spacetime. 

PO(1,3) denotes the inhomogeneous Lorentz symmetry group or Poincar\'e symmetry group in four-dimensional Minkowski spacetime,
\be
 PO(1,3) = P^{1,3}\ltimes SO(1,3) , \nn
\ee
where the symbol `$\ltimes$' is used to indicate that PO(1,3) is a semi-direct product group with P$^{1,3}$ representing the translational symmetry group of coordinates in Minkowski spacetime. 

WS(1,3) is an enlarged spin symmetry group, which is also a semi-direct product group expressed as follows:
\be
WS(1, 3) \equiv  SP(1,3)\rtimes W^{1,3} .
\ee
It can be checked that SP(1,3) group generators $\Sigma^{a b}$ and W$^{1,3}$ group generators $\Sigma_{-}^{a}$ have the following explicit forms:
\be
& & \Sigma^{a b} = \frac{i}{4}  [\Gamma^{a}, \Gamma^{b} ] ,  \nn \\
& &  \Sigma_{-}^{a} = \frac{1}{2}\Gamma^{a}\Gamma_{-}, \quad \Gamma_{-} = \frac{1}{2} ( 1- \hat{\gamma}_7 ) ,
\ee
which satisfy the group algebra:
\be
& & [\Sigma^{ab}, \Sigma^{cd}] =  i (\Sigma^{ad}\eta^{bc} -\Sigma^{bd}  \eta^{ac} - \Sigma^{ac} \eta^{bd} + \Sigma^{bc} \eta^{ad}) , \nn \\
& &  [\Sigma^{ab}, \Sigma_{-}^{c}] = i( \Sigma_{-}^{a}\eta^{bc} -\Sigma_{-}^{b}  \eta^{ac}  ) , \nn \\
& &  [\Sigma_{-}^{a}, \Sigma_{-}^{b}] = 0 ,
\ee
where $W^{1,3}$ appears as a translation-like Abelian-type symmetry group in Hilbert space. 

Let us now discuss explicitly the translation-like Abelian-type symmetry group $W^{1,3}$. The chirality-based Dirac spinor $\Psi_{-}(x)$ transforms under $W^{1,3}$ group operation as follows:
\be
& &  \Psi_{-}(x) \to  \Psi'_{-}(x) =  S(\varpi)  \Psi_{-}(x) , \nn \\
& & S(\varpi) = e^{i \varpi_{a}\Sigma_{-}^{a}/2 } = 1 + i \varpi_{a} \Sigma_{-}^{a}/2, 
\ee
which can be rewritten into the following forms:
\be \label{WES}
& & \Psi'_{-}(x) \equiv \Psi_{-}^{(w)}(x)  + \tilde{\Psi}_{-}^{(e)}(x)  , \nn \\
& &  \Psi_{-}^{(w)}(x) \equiv  \Psi_{-}(x), \quad \tilde{\Psi}_{-}^{(e)}(x) \equiv i \frac{1}{2}\varpi_{a} \Sigma_{-}^{a} \Psi_{-}(x) ,
\ee
where $\Psi_{-}^{(w)}(x)$  and $\tilde{\Psi}_{-}^{(e)}(x)$ have opposite chiral property under the operation of chiral operator $\hga_7$ in eight-dimensional Hilbert space, i.e.: 
\be
\hga_7 \Psi_{-}^{(w)}(x) = - \Psi_{-}^{(w)}(x) , \quad \hga_7  \tilde{\Psi}_{-}^{(e)}(x) = + \tilde{\Psi}_{-}^{(e)}(x) .
\ee
For that, $\Psi_{-}^{(w)}(x)$  and $\tilde{\Psi}_{-}^{(e)}(x)$ are referred to as westward and eastward chirality-based Dirac spinors indicated by the upper indices $``w"$ and $``e"$, respectively, in order to distinguish with the left-handed and right-handed Weyl fermions $\psi_L(x)$ and $\psi_R(x)$ defined in four-dimensional Hilbert space. 

As such an Abelian-type spin symmetry group W$^{1,3}$ is associated with the sign flip in chirality of westward and eastward spinor fields for the initial and shifted new parts, we may refer to the translation-like chiral-type symmetry W$^{1,3}$ as {\it $\cW_e$-spin symmetry} for short. So we may refer to the enlarged spin symmetry group WS(1,3) = SP(1,3)$\rtimes$ W$^{1,3}$ as {\it inhomogeneous spin symmetry group} acting on the chirality-based Dirac spinor $\Psi_{-}(x)$, which is a semi-direct product group as indicated by the symbol `$\rtimes$'.  It is intriguing to notice that such an inhomogeneous spin symmetry group is in correspondence to the inhomogeneous Lorentz symmetry group (or Poincar\'e symmetry group) PO(1,3) = P$^{1,3}\ltimes$ SO(1,3). 

It is essential to distinguish the spin symmetry SP(1,3) with the Lorentz symmetry SO(1,3) as they operate on different spaces. The former acts on the chirality-based Dirac spinor field $\Psi_{-}(x)$ in Hilbert space and the latter operates on the coordinates in Minkowski spacetime. They become the associated symmetry just for the free motion Dirac fermion in globally flat Minkowski spacetime. 

It is easy to check that the action in Eq.(\ref{action2}) possesses an Abelian group symmetry U(1) under the following transformation:
\be
& &  \Psi_{-}(x) \to  \Psi'_{-}(x) =  e^{i \varpi }\Psi_{-}(x) ,
\ee
which describes the electric charge symmetry.

It is clear that the symmetry shown in Eq.(\ref{MAS}) presents the maximal associated symmetry for a free motion Dirac fermion expressed in the chiral spinor representation of eight-dimensional Hilbert space.  


\section{ Inhomogeneous spin gauge symmetry with the genesis of gravigauge field and the gravitational origin of spin gauge symmetry }

It is the essential point to distinguish the group symmetries in Hilbert space and Minkowski spacetime when proposing the gauge invariance principle to study the fundamental interaction. The gauge invariance principle states that {\it the laws of nature should be independent of the choice of local field configurations}, from which the inhomogeneous spin symmetry WS(1,3) and electric charge symmetry U(1) of chirality-based Dirac spinor $\Psi_{-}(x)$ are postulated to be local gauge symmetries. Meanwhile, the inhomogeneous Lorentz group symmetry (Poincar\'e group symmetry) PO(1,3) of coordinates is considered to remain a global symmetry. In such a consideration, the inhomogeneous spin gauge symmetry WS(1, 3) in Hilbert space of spinor field does become distinguishable from the inhomogeneous Lorentz group symmetry PO(1,3) in Minkowski spacetime of coordinates. It becomes natural that the inhomogeneous spin gauge symmetry group WS(1,3) provides the so-called internal Poincar\'e-type gauge symmetry group PG(1,3) proposed in\cite{GQFT}, i.e., PG(1,3)$\equiv$WS(1,3) = SP(1,3)$\rtimes$W$^{1,3}$.

\subsection{Inhomogeneous spin gauge symmetry with the genesis of gravigauge field }

Based on the gauge invariance principle, the gauge fields $\hcA_{\mu}(x)$ and $A_{\mu}(x)$ in correspondence to the inhomogeneous spin gauge symmetry WS(1, 3) and electromagnetic gauge symmetry U(1) are introduced to ensure the theory be gauge invariance. To realize such a gauge invariant theory, it is practically carried out by replacing the usual derivative operator of coordinates in Minkowski spacetime into the following covariant derivative operator: 
\be \label{CDO}
& &  i\p_{\mu} \to i\hcD_{\mu} \equiv  i\p_{\mu} + \hcA_{\mu}(x) +  A_{\mu}(x) , 
\ee
with
\be
& & \hcA_{\mu}(x) \equiv  \cA_{\mu}(x) +  \ckcA_{\mu}(x) , \nn \\
& & \cA_{\mu}(x) \equiv \cA_{\mu}^{\; ab}(x)\, \frac{1}{2}\varSigma_{ab}  , \quad 
\ckcA_{\mu}(x) \equiv \cA_{\mu}^{a}(x)\frac{1}{2}\varSigma_{- a} , 
\ee
where $\hcA_{\mu}(x)$ is referred to as inhomogeneous spin gauge field with $\cA_{\mu}^{\; ab}(x)$ representing the spin gauge field relevant to the spin gauge symmetry SP(1, 3) and $\cA_{\mu}^{a}(x)$ denoting the $\cW_e$-spin gauge field relating to the translation-like $\cW_e$-spin Abelian gauge symmetry $W^{1,3}$. $A_{\mu}(x)$ is the electromagnetic gauge field corresponding to the U(1) gauge symmetry. 

From the general covariant derivative $\hcD_{\fM}$, the field strength of gauge field $\hcA_{\fM}(x)$ is obtained as follows: 
\be \label{GFS}
& & \hcF_{\mu\nu} \equiv i [\hcD_{\mu}, \hcD_{\nu} ] = \cF_{\mu\nu} + \ckcF_{\mu\nu} + F_{\mu\nu}, \nn \\
& & \cF_{\mu\nu} \equiv  \cF_{\mu\nu}^{ab}(x)\frac{1}{2}\vSi_{ab} = \p_{\mu} \cA_{\nu}(x) - \p_{\nu} \cA_{\mu}(x) - i [ \cA_{\mu}(x) ,  \cA_{\nu}(x) ], \nn \\
& & \ckcF_{\mu\nu} \equiv \cF_{\mu\nu}^{a}(x)\frac{1}{2}\vSi_{- a}= \cD_{\mu} \ckcA_{\nu}(x) - \cD_{\nu} \ckcA_{\mu}(x) \nn \\
& & \qquad \equiv \p_{\mu} \ckcA_{\nu}(x) - \p_{\nu} \ckcA_{\mu}(x) - i (\cA_{\mu}(x)\ckcA_{\nu}(x) - \cA_{\nu}(x)\ckcA_{\mu}(x) ), \nn \\
& & F_{\mu\nu} \equiv  F_{\mu\nu}(x) = \p_{\mu} A_{\nu}(x) - \p_{\nu} A_{\mu}(x) ,
 \ee 

It can be checked that the gauge transformation properties of gauge fields can explicitly be given as follows:
\be
& & \cA_{\mu}(x) \to \cA'_{\mu}(x) =  S(\Lambda) \cA_{\mu}(x) S^{-1}(\Lambda) + S(\Lambda) i\p_{\mu} S^{-1}(\Lambda) , \nn \\
& & \ckcA_{\mu}(x) \to \ckcA'_{\mu}(x) =  S(\Lambda) \ckcA_{\mu}(x) S^{-1}(\Lambda), \quad S(\Lambda) \in SP(1,3) ,
\ee
or 
\be
& & \cA_{\mu}^{' ab}(x) = \Lambda^{a}_{\; c}(x)  \Lambda^{b}_{\; d}(x) \cA_{\mu}^{cd}(x) + \frac{1}{2} ( \Lambda^{a}_{\; c}(x)  \p_{\mu}  \Lambda^{bc}(x) - \Lambda^{b}_{\; c}(x)  \p_{\mu}  \Lambda^{ac}(x) ) , 
\nn \\
& & \cA_{\mu}^{' a}(x) = \Lambda^{a}_{\; c}(x) \cA_{\mu}^{c}(x) , \quad  \Lambda^{a}_{\; c}(x) \in SP(1,3)\cong SO(1,3) , 
\ee 
under the gauge transformation of SP(1,3) gauge symmetry, and 
\be
& & \cA_{\mu}^{a}(x) \to  \cA_{\mu}^{' a}(x) =  \cA_{\mu}^{a}(x) + \cD_{\mu}\varpi^{a}(x), \nn \\
& & \cD_{\mu}\varpi^{a}(x) = \p_{\mu} \varpi^{a}(x) + \cA_{\mu\, b}^{a}(x) \varpi^{b}(x) ,
\ee
under the $\cW_e$-spin gauge transformation of W$^{1,3}$ gauge symmetry.

It is noticed that the presence of gauge symmetry in the action involves actually redundant degrees of freedom. To eliminate the redundant degrees of freedom arising from the gauge symmetry, it needs to make a gauge prescription by imposing a gauge fixing condition. In order to extract the redundant degrees of freedom, we are initiated to decompose the $\cW_e$-spin gauge field $\ckcA_{\mu}^{a}(x)$ into the following two parts: 
\be
\ckcA_{\mu}(x) \equiv \cA_{\mu}^{a}(x) \frac{1}{2}\vSi_{- a} =  \left(\fOm_{\mu}^{a}(x) +  \fA_{\mu}^{a}(x) \right) \frac{1}{2}\vSi_{- a} , 
\ee
where $\fOm_{\mu}^{a}(x)$ is supposed to get an inhomogeneous gauge transformation and $\fA_{\mu}^{a}(x)$ becomes gauge invariant under the $\cW_e$-spin gauge transformation. Meanwhile, they should transform as covariant vector fields under the gauge transformation of spin gauge symmetry SP(1,3). The explicit transformations are presented as follows:
\be
& & \fOm_{\mu}^{a}(x) \to  \fOm_{\mu}^{' a}(x) =  \fOm_{\mu}^{a}(x) + \cD_{\mu}\varpi^{a}(x), \nn \\
& & \fA_{\mu}^{a}(x) \to  \fA_{\mu}^{' a}(x) =  \fA_{\mu}^{a}(x) , 
\ee
under the gauge transformation of $\cW_e$-spin gauge symmetry W$^{1,3}$, and
\be
& &  \fOm_{\mu}^{a}(x) \to  \fOm_{\mu}^{' a}(x) = \Lambda^{a}_{\;\; b}(x) \fOm_{\mu}^{b}(x), \nn \\
& & \fA_{\mu}^{a}(x) \to  \fA_{\mu}^{' a}(x) = \Lambda^{a}_{\;\; b}(x) \fA_{\mu}^{b}(x),  
\ee
under the gauge transformation of spin gauge symmetry SP(1,3). 

We may refer $\fOm_{\mu}^{a}(x)$ to be as a $\cW_e$-spin non-homogeneous gauge field and $\fA_{\mu}^{a}(x)$ as a $\cW_e$-spin invariant-gauge field. From the property of $\cW_e$-spin gauge transformation, it is natural to express $\fOm_{\mu}^{a}(x)$ into the following form:
\be \label{NHGF}
& & \fOm_{\mu}^{a}(x) \equiv  \cD_{\mu}\vka^{a}(x) = \p_{\mu} \vka^{a}(x) + \cA_{\mu b}^{a}(x) \vka^{b}(x), 
\ee
where we have introduced a vector field $\vka^{a}(x)$ in Hilbert space. Correspondingly, the field strength of $\cW_e$-spin gauge field can also be decomposed into two parts:
\be \label{WESFS}
& & \ckcF_{\mu\nu}^{a}(x) \equiv \cR_{\mu\nu}^{a}(x) + \cF_{\mu\nu}^{a}(x), \nn \\
& & \cR_{\mu\nu}^{a}(x) = \cD_{\mu}\fOm_{\nu}^{a}(x) -  \cD_{\nu}\fOm_{\mu}^{a}(x) =  \cF_{\mu\nu}^{ab}(x) \vka_{b}(x) , \nn \\
& & \cF_{\mu\nu}^{a}(x) = \cD_{\mu}\fA_{\nu}^{a}(x) -  \cD_{\nu}\fA_{\mu}^{a}(x) \nn \\
& & \qquad \quad \; \equiv \p_{\mu} \fA_{\nu}^{a}(x) - \p_{\nu} \fA_{\mu}^{a}(x) + \cA_{\mu\, b}^{a}(x) \fA_{\nu}^{b}(x) - \cA_{\nu\, b}^{a}(x) \fA_{\mu}^{b}(x).
\ee

To maintain the $\cW_e$-spin gauge transformation, the vector field $\vka^{a}(x)$ transforms as follows:
\be \label{VF}
\vka^{a}(x) \to \vka^{' a}(x) = \vka^{a}(x) + \varpi^{a}(x) .
\ee
which behaves as a translation of vector field $\vka^{a}(x)$ in non-coordinate field spacetime. As such a transformation reflects the symmetry property of translation-like $\cW_e$-spin Abelian-type gauge group W$^{1,3}$, we may mention $\vka^{a}(x)$ as $\cW_e$-spin vector field. It is seen that under the $\cW_e$-spin gauge transformation, $\cF_{\mu\nu}^{a}(x)$ becomes $\cW_e$-spin invariant-gauge field strength and the field strength $\cR_{\mu\nu}^{a}(x)$ transforms as follows:
\be
\cR_{\mu\nu}^{a}(x) \to \cR_{\mu\nu}^{' a}(x) = \cR_{\mu\nu}^{a}(x) + \cF_{\mu\nu}^{ab}(x) \varpi_{b}(x) .
\ee

It indicates that to remove the degrees of freedom, we are able to take an appropriate $\cW_e$-spin gauge transformation to make a gauge fixing condition, so that the following relation holds, i.e.:
\be
& & \vka^{' a}(x) = \vka^{a}(x) + \varpi^{a}(x) = 0, \quad \mbox{i.e.}, \quad \varpi^{a}(x)  = -  \vka^{a}(x)  ,\nn \\
& & \fOm_{\mu}^{' a}(x) = 0, \quad \cR_{\mu\nu}^{' a}(x)= 0  .
\ee
Therefore, by choosing the gauge fixing condition, we can always take the gauge field and field strength to be as follows: 
\be
& & \ckcA_{\mu}(x) \equiv \cA_{\mu}^{a}(x) \frac{1}{2}\vSi_{- a} = \fA_{\mu}^{a}(x)  \frac{1}{2}\vSi_{- a} ,  \nn \\
& &  \ckcF_{\mu\nu}(x) \equiv \ckcF_{\mu\nu}^{a}(x) \frac{1}{2} \vSi_{- a} \equiv  \cF_{\mu\nu}^{a}(x)\frac{1}{2} \vSi_{- a} .
\ee


\subsection{Gravitational origin of spin gauge symmetry }

In analogous, to clarify the redundant degrees of freedom from the spin gauge symmetry, we are motivated to decompose the spin gauge field $\cA_{\mu}(x)$ into the following two parts:
\be \label{SGFDC}
\cA_{\mu}(x) = \fOm_{\mu}(x) + \fA_{\mu}(x) \equiv \left(\fOm_{\mu}^{ab}(x) + \fA_{\mu}^{ab}(x)\right)\frac{1}{2}\vSi_{ab} , 
\ee 
where $\fOm_{\mu}(x)$ obeys inhomogeneous gauge transformation of the spin gauge symmetry SP(1,3) and $\fA_{\mu}(x)$ transforms homogeneously under the spin gauge transformation, i.e.:
\be \label{SGFGT}
& & \fOm_{\mu}(x) \to \fOm'_{\mu}(x) = S(\Lambda) i\p_{\mu} S^{-1}(\Lambda)  + S(\Lambda) \fOm_{\mu}(x) S^{-1}(\Lambda) , \nn \\
& & \fA_{\mu}(x) \to \fA'_{\mu}(x) = S(\Lambda) \fA_{\mu}(x) S^{-1}(\Lambda), \quad S(\Lambda) \in SP(1,3).
\ee
Unlike the usual internal gauge field for which the inhomogeneous gauge transformation part is thought to be characterized by a pure gauge field with a vanishing field strength, {\it the gauge field $\fOm_{\mu}^{ab}$ under the inhomogeneous gauge transformation is considered to be relevant to the $\cW_e$-spin invariant-gauge field $\fA_{\mu}^{\; a}$ in order to maintain the independent degrees of freedom arising from the inhomogeneous spin gauge symmetry WS(1,3)}. Namely, the total numbers of independent degrees of freedom for the gauge field should be fixed to be the same as the initial ones required from the inhomogeneous spin gauge symmetry WS(1,3) which brings about both the spin gauge field $\cA_{\mu}^{ab}(x)$ and $\cW_e$-spin gauge field $\fA_{\mu}^{\; a}(x)$.

Indeed, it can be verified that $\fOm_{\mu}^{ab}$ is completely determined through the $\cW_e$-spin invariant-gauge field $\fA_{\mu}^{\; a}$ with the following explicit form:
\be \label{SGGF}
& & \fOm_{\mu}^{ab} = \frac{1}{2}\left( \hat{\fA}^{a\nu} \mF_{\mu\nu}^{b} - \hat{\fA}^{b\nu} \mF_{\mu\nu}^{a} -  \hat{\fA}^{a\rho}  \hat{\fA}^{b\sigma}  \mF_{\rho\sigma}^{c} \fA_{\mu c } \right) , \nn \\
& & \mF_{\mu\nu}^{a}(x)=  \p_{\mu}\fA_{\nu}^{a}(x) -  \p_{\nu}\fA_{\mu}^{a}(x) ,
\ee
where we have introduced the dual $\cW_e$-spin invariant-gauge field $\hat{\fA}_{a}^{\;\;\mu}(x)$ defined as follows:
\be
& & \hat{\fA}_{a}^{\;\;\mu}(x) \fA_{\nu}^{a}(x)  = \hat{\fA}_{a}^{\;\;\mu}(x) \fA_{\nu b}(x) \eta^{ab}= \eta_{\nu}^{\;\; \mu} , \nn \\
& &  \hat{\fA}_{a}^{\;\;\mu}(x) \fA_{\mu}^{b}(x)  = \hat{\fA}_{a}^{\;\;\mu}(x) \fA^{\nu b}(x) \eta_{\mu\nu}= \eta_{a}^{\;\; b} ,
\ee
which indicates that when regarding $\fA_{\mu}^{a}(x)$ as a matrix field, $\hat{\fA}_{a}^{\;\;\mu}(x)$ is viewed as the inverse matrix field of $\fA_{\mu}^{a}(x)$. The existence of $\hat{\fA}_{a}^{\;\;\mu}(x)$ requires a non-zero determinant of matrix field $\fA_{\mu}^{a}(x)$,
\be
 \fkA(x) \equiv \det \fA_{\mu}^{a}(x) \neq 0 , \quad \mbox{or} \quad 
 \hat{\fkA}(x) \equiv \det \hfA_{a}^{\;\mu}(x) \neq 0.
\ee
Note that the field strength $\mF_{\mu\nu}^{a}(x)$ is $\cW_e$-spin gauge invariant but no longer spin gauge covariant, which distinguishes, as indicated by a different letter style, from the spin gauge covariant field strength $\cF_{\mu\nu}^{a}(x)$ defined in Eq.(\ref{WESFS}). 

It can be checked that the gauge transformation of $\fA_{\mu}^{\; a}$ in the vector representation of spin gauge symmetry SP(1,3) does bring $\fOm_{\mu}^{ab}$ to a proper gauge transformation in the adjoint representation of spin gauge symmetry SP(1,3), i.e.:
\be \label{SGFGT1}
& & \fA_{\mu}^{'\; a}(x) =  \Lambda^{a}_{\;\; c}(x) \fA_{\mu}^{\; c}(x) , \quad  \hfA_{a}^{' \; \mu}(x)=  \Lambda_{a}^{\;\; c}(x) \hfA_{c}^{\; \mu}(x) ,\quad  \Lambda^{a}_{\; c}(x)  \in \mbox{SP(1,3)} , \nn \\
& & \fOm_{\mu}^{' ab}(x) =  \Lambda^{a}_{\; c}(x)  \Lambda^{b}_{\; d}(x) \fOm_{\mu}^{cd}(x) + \frac{1}{2} ( \Lambda^{a}_{\; c}(x)  \p_{\mu}  \Lambda^{bc}(x) - \Lambda^{b}_{\; c}(x)  \p_{\mu}  \Lambda^{ac}(x) ) .
\ee
In general, there is no way to eliminate the spin gauge field part $\fOm_{\mu}^{ab}(x)$ by making a spin gauge transformation, which differs completely from the usual internal gauge field. This can be verified from the spin gauge field strength which can also be decomposed into two parts:
\be
\cF_{\mu\nu}^{ab}(x) \equiv  \fR_{\mu\nu}^{ab}(x) + \fF_{\mu\nu}^{ab}(x) , \nn \\
\ee
with the following explicit forms:
\be \label{DCFS0}
& & \fR_{\mu\nu}^{ab}(x) = \p_{\mu} \fOm_{\nu}^{ab} - \p_{\nu} \fOm_{\mu}^{ab} 
+ \fOm_{\mu c}^{a} \fOm_{\nu}^{cb} -  \fOm_{\nu c}^{a} \fOm_{\mu}^{cb},  
\nn \\  
& & \fF_{\mu\nu}^{ab}(x) =  \fD_{\mu} \fA_{\nu}^{ab} - \fD_{\nu} \fA_{\mu}^{ab} +  \fA_{\mu c}^{a} \fA_{\nu}^{cb} -  \fA_{\nu c}^{a} \fA_{\mu}^{cb}  , \nn \\
& & \fD_{\mu} \fA_{\nu}^{ab}(x) = \p_{\mu}  \fA_{\nu}^{ab}  +  \fOm_{\mu c}^{a} \fA_{\nu}^{cb} + \fOm_{\mu c}^{b} \fA_{\nu}^{ac}  ,
\ee
where $\fR_{\mu\nu}^{ab}$ is purely the field strength of $\fOm_{\mu}^{ab}(x)$. Since $\fOm_{\mu}^{ab}(x)$ is uniquely determined by the $\cW_e$-spin invariant-gauge field $\fA_{\mu}^{\; a}$, there are no additional independent degrees of freedom involved in the decomposition of spin gauge field $\cA_{\mu}^{ab}(x)$. 

It can be verified that the field strength $\fR_{\mu\nu}^{ab}$ is related to Riemann-type curvature tensor as follows:
\be \label{RMC1}
\fR_{\mu\nu}^{ab}(x) = \fA_{\rho}^{\; a}(x) \hfA^{b \sigma}(x) \fR^{\; \rho}_{\mu\nu\sigma}(x)
\ee
with the Riemann-type curvature tensor $\fR^{\rho}_{\mu\nu\sigma}(x)$ given by
\be \label{RMC2}
& & \fR_{\mu\nu\sigma}^{\;\rho}(x)  = \p_{\mu} \fGa_{\nu\sigma}^{\rho} - \p_{\nu} \fGa_{\mu\sigma}^{\rho}  + \fGa_{\mu\lambda}^{\rho} \fGa_{\nu\sigma}^{\lambda}  - \fGa_{\nu\lambda}^{\rho} \fGa_{\mu\sigma}^{\lambda} .
\ee
Where $\fGa_{\mu\sigma}^{\rho}(x)$ is defined as follows: 
\be \label{SGMF}
\fGa_{\mu\sigma}^{\rho}(x)  & \equiv &  \hfA_{a}^{\;\; \rho} \p_{\mu} \fA_{\sigma}^{\;\; a} +  \hfA_{a}^{\;\; \rho}   \fOm_{\mu\1 b}^{a} \fA_{\sigma}^{\;\;b} , \nn \\
& = & \frac{1}{2}\hmH^{\rho\lambda} (\p_{\mu} \mH_{\lambda\sigma} + \p_{\sigma} \mH_{\lambda\mu} - \p_{\lambda}\mH_{\mu\sigma} ) =\fGa_{\sigma\mu}^{\rho}.
 \ee
In the second equality we have introduced the gauge invariant dual tensors,
\be \label{DTF}
 & &  \mH_{\mu\nu}(x)  \equiv \fA_{\mu}^{\;a}(x)\fA_{\nu}^{\;b}(x) \eta_{ab} \equiv -\Tr \Gamma^{6} \fA_{\mu}^{\;a}\Sigma_{a -} \Gamma^{6} \fA_{\nu}^{\;b}\Sigma_{b -} , \nn \\
 & &   \hmH^{\mu\rho}(x) \mH_{\rho\nu}(x) = \eta^{\mu}_{\; \; \nu} , \quad \hmH^{\mu\nu}(x) \equiv \hfA_{a}^{\;\mu}(x)\hfA_{b}^{\;\nu}(x) \eta^{ab} .
\ee

Geometrically, we will show later on that the tensor $\mH_{\mu\nu}(x)$ ($\hmH^{\mu\nu}(x)$) composed of $\fA_{\mu}^{\; a}(x)$ ($\hfA_{a}^{\;\mu}(x)$) represents a metric-like tensor which characterizes the gravitational interaction in Einstein's general theory of relativity. It is clear that the $\cW_e$-spin invariant-gauge field $\fA_{\mu}^{\; a}(x)$ should be the basic gauge field which brings the gravitational interaction as fundamental gauge interaction. For convenience of mention, $\fA_{\mu}^{\; a}(x)$ is referred to as {\it gravigauge field}.  Correspondingly, $\mH_{\mu\nu}(x)$ is referred to as {\it gravimetric field}. Meanwhile, $\fOm_{\mu}^{ab}(x)$ is referred to as {\it spin gravigauge field} and $\fA_{\mu}^{ab}(x)$ as {\it spin covariant-gauge field}.

It is obvious that the decomposition of spin gauge field $\cA_{\mu}^{ab}(x)$ into the sum of spin gravigauge field $\fOm_{\mu}^{ab}(x)$ and spin covariant-gauge field $\fA_{\mu}^{ab}(x)$ exhibits the gravitational origin of gauge symmetry, which allows us to bring about the corresponding gauge covariant field strengths $\fR_{\mu\nu}^{ab}(x)$ and $\fF_{\mu\nu}^{ab}(x)$. The spin gravigauge field $\fOm_{\mu}^{ab}(x)$ and its gauge covariant field strength $\fR_{\mu\nu}^{ab}(x)$ describe the gravitational interaction along with the genesis of gravigauge field $\fA_{\mu}^{\; a}(x)$, which realizes the gauge-gravity and gravity-geometry correspondences. 

In conclusion, the inhomogeneous spin gauge symmetry WS(1,3) brings on the introduction of spin gauge field $\cA_{\mu}^{ab}(x)$ and $\cW_e$-spin gauge field $\fA_{\mu}^{\; a}(x)$ via appropriate gauge fixing condition. Meanwhile, the gravitational origin of spin gauge symmetry SP(1,3) for chirality-based Dirac fermion also allows us to choose alternatively the spin covariant-gauge field $\fA_{\mu}^{ab}(x)$ as independent degrees of freedom, where the spin gravigauge field $\fOm_{\mu}^{ab}(x)$ that reflects the non-homogeneous spin gauge symmetry is essentially governed by the $\cW_e$-spin gauge field $\fA_{\mu}^{\; a}(x)$. Such a $\cW_e$-spin gauge field $\fA_{\mu}^{\; a}(x)$ as gravigauge field transforms as a bicovariant vector field under the transformations of spin gauge symmetry SP(1,3) and Lorentz group symmetry SO(1,3). Later on, we shall discuss in detail how the general linear group symmetry GL(1,3,R) emerges automatically as the generalization of scaling and gauge invariance principle and brings on the gauge-geometry duality.

\section{Locally flat gravigauge spacetime with emergence of non-commutative geometry and biframe spacetime with fiber bundle structure}

The $\cW_e$-spin invariant-gauge field $\fA_{\mu}^{\; a}(x)$ as gravigauge field reveals the nature of gravity as a gauge interaction. The $\cW_e$-spin vector field $\vka^{\fA}(x)$ introduced to characterize the translation-like $\cW_e$-spin gauge symmetry is regarded as a dimensionless {\it gravivector field} in a non-coordinate field spacetime. In analogous to the ordinary translation operator $\p_{\mu}\equiv \frac{\p}{\p x^{\mu}}$ of coordinates in globally flat Minkowski spacetime, the $\cW_e$-spin gauge symmetry as a local translation of gravivector field may be represented via a field translation operator defined in formal by the following coordinate-like field derivative:
\be \label{HGCD}
\hat{\eth}_{a} \equiv \frac{\delta }{\delta \vka^{a}}, 
\ee
Similarly, we should introduce a coordinate-like field displacement $\delta\vka^{\fA}$ in correspondence to the ordinary coordinate displacement $dx^{\mu}$. 

The $\cW_e$-spin invariant-gauge field $\fA_{\mu}^{\; a}(x)$ as gravigauge field is regarded as a bi-covariant vector field under the transformations of spin gauge symmetry SP(1,3) in Hilbert space and global Lorentz symmetry SO(1,3) in coordinate spacetime. It is natural to propose the displacement correspondence that the coordinate-like field displacement $\delta\vka^{a}$ is directly associated to the ordinary coordinate displacement $dx^{\mu}$ via the bi-covariant vector field $\fA_{\mu}^{\; a}(x)$. To be  explicit, such a displacement correspondence is presented by the following relations:
\be \label{GCDO}
\delta\vka^{a} \equiv \fA_{\mu}^{\;\;a}(x)dx^{\mu} ,\qquad
\hat{\eth}_{a} \equiv \hat{\fA}_{a}^{\;\;\mu}(x)\p_{\mu} ,
\ee 
where we may refer to the displacement $\delta\vka^{a}$ as {\it dimensionless  gravicoordinate displacement} and the derivative $\hat{\eth}_{a}$ as {\it dimensionless gravicoordinate derivative}.

It is useful to introduce the corresponding dimensionful gravicoordinate displacement and derivative. For that, let us express the gravigauge field $\fA_{\mu}^{\; a}(x)$ and its dual field $\hat{\fA}_{a}^{\;\;\mu}(x)$ into the following forms:
\be \label{SGIGGF}
\fA_{\mu}^{\; a}(x) \equiv \phi(x) \chi_{\mu}^{\; a}(x), \quad \hfA_{a}^{\;\;\mu}(x)\equiv \phi^{-1}(x) \chih_{a}^{\;\; \mu}(x) ,
\ee
where $\phi(x)$ is considered as a scalar field with unit canonical dimension, which is referred to as {\it graviscalar field}. $\chi_{\mu}^{\; a}(x)$ provides a dimensionless gravigauge field with $\chih_{a}^{\;\; \mu}(x)$ as a dimensionless dual gravigauge field, they satisfy the following conditions:
\be \label{DTF1}
 & &  \chi_{\mu\nu}(x) \equiv \chi_{\mu}^{\;a}(x)\chi_{\nu}^{\;b}(x) \eta_{ab}, 
 \quad \chih^{\mu\nu}(x) \equiv \chih_{a}^{\;\mu}(x)\chih_{b}^{\;\nu}(x) \eta^{ab} ,  \nn \\
& & \chi_{\mu\rho}(x)  \chih^{\rho\nu}(x) = \eta_{\mu}^{\; \; \nu} , \quad \chih^{\mu\rho}(x) \chi_{\rho\nu}(x) = \eta^{\mu}_{\; \; \nu} ,
\ee
which enables us to define the {\it dimensionful gravicoordinate displacement and derivative} as follows:
\be \label{DGCDO}
 \delta\zeta^{a} & \equiv & \chi_{\mu}^{\;\;a}(x) dx^{\mu} \equiv \phi^{-1}(x) \delta\vka^{a} , \nn \\
\eth_{a} & \equiv & \frac{\delta}{\delta \zeta^{a}} = \chih_{a}^{\;\;\mu}(x) \p_{\mu} = \phi(x) \hat{\eth}_{a} ,
\ee 
where the dimensionless gravicoordinate displacement $\delta\vka^{a}$ and derivative $\hat{\eth}_{a}$ with respect to gravivector field $\vka^{a}$ are related to the corresponding dimensionful gravicoordinate displacement $\delta\zeta^{a}$ and derivative $\eth_{a}$ via the graviscalar field $\phi(x)$. 

Mathematically, it is known that the ordinary derivative vector operator $\partial_{\mu} \equiv \partial/\partial x^{\mu}$ at point $x$ of Minkowski spacetime $M_4$ defines a tangent basis $\{\partial_{\mu}\}\equiv \{\partial/\partial x^{\mu}\} $ for tangent spacetime $T_4$ over $M_4$. A displacement vector $dx^{\mu}$ at point $x$ of $M_4$ forms a dual tangent basis $\{dx^{\mu}\}$ for dual tangent spacetime $T_4^{\ast}$ over $M_4$. These dual bases satisfy the dual condition:
\be
 \langle dx^{\mu},\, \p_{\nu}  \rangle = \frac{\partial x^{\mu}}{\partial x^{\nu}} = \eta_{\nu}^{\; \mu} .
\ee
In analogous, the gravicoordinate displacement $\delta\vka^{a}$ and gravicoordinate derivative $\hat{\eth}_{a}$ form a pair of dual field bases $\{\delta\vka_{a} \}$ and $\{\hat{\eth}_{a}\}$. Such dual field bases meet to the following dual condition:
 \be
 & & \langle \delta\vka^{a},  \hat{\eth}_{b}\rangle  \equiv \fA_{\mu}^{\; a}(x)  \hfA_{b}^{\;\nu} (x)  \langle dx^{\mu} , \partial_{\nu} \rangle = \fA_{\mu}^{\;\; a}(x)  \hfA_{b}^{\;\; \nu} (x)  \eta_{\nu}^{\; \mu} = \eta_{b}^{\;\, a} ,
\ee
which span a pair of dual locally flat spacetimes over globally flat Minkowski spacetime $M_4$. 

For convenience, the dual field bases $ \{\hat{\eth}_{a} \} $ and $ \{\delta\vka^{a}\} $ determined by the gravigauge field $\fA_{\mu}^{\; a}(x)$ as bi-covariant vector field are referred to as a pair of dual gravigauge field bases. Meanwhile, the dual locally flat field spacetimes spanned by the dual gravigauge field bases are referred to as locally flat dual {\it gravigauge spacetimes}, where the gravigauge spacetime spanned from gravigauge field basis $\{\hat{\eth}_{a}\}$ is mentioned as tangent-like gravigauge spacetime $G_4$, and the dual gravigauge spacetime formed from the dual gravigauge field basis $ \{\delta\vka^{\fA}\} $ is referred to as dual tangent-like gravigauge spacetime $G_4^{\ast}$. Therefore, the gravigauge field $\fA_{\mu}^{\;\; a}(x)$ as bi-covariant vector field is thought to be sided on the dual tangent Minkowski spacetime $T_4^{\ast}$ and valued on the dual gravigauge spacetime $G_4^{\ast}$, while the dual gravigauge field $\hfA_{a}^{\;\;\mu}(x)$ is defined on the gravigauge spacetime $G_4$ and valued on the tangent Minkowski spacetime $T_4$. Thus $\hfA_{a}^{\;\;\mu}(x)$ transforms as a bi-covariant vector field under the transformations of both spin gauge symmetry SP(1,3) and global Lorentz group symmetry SO(1,3). 

It is noticed that the gravigauge field basis $\{\hat{\eth}_{a}\}$ is no longer commutative, its non-commutation relation is explicitly given by: 
\be \label{NCG}
 & & [\heth_{c}\, , \heth_{d} ] =  \fOm_{[cd]}^{a} \1 \heth_{a} , \nn \\
 & & \fOm_{[cd]}^{a} = -\hfA_{c}^{\; \mu}(x)\hfA_{d}^{\; \nu}(x) \mF_{\mu\nu}^{a}(x) \equiv \fOm_{c\1 d}^{a} - \fOm_{d\1 c}^{a},\nn \\& &  \mF_{\mu\nu}^{a}(x) =  \p_{\mu}\fA_{\nu}^{a}(x) -  \p_{\nu}\fA_{\mu}^{a}(x) ,
\ee
which brings on a non-Abelian Lie algebra of gravicoordinate derivative operator $\heth_{c}$. Where the structure factor $\fOm_{[cd]}^{a} $ is no longer a constant and determined by the gravigauge field strenghth $\mF_{\mu\nu}^{a}$. Such a non-Abelian Lie algebra leads to the emergence of non-commutative geometry in locally flat gravigauge spacetime $\fG_4$. To explore the property of non-commutative geometry in locally flat gravigauge spacetime $\fG_4$, it is necessary to study the dynamics of spin gravigauge field $\fOm_{[c d]}^{a}$. 

In general, the gravigauge spacetime $G_4$ and tangent Minkowski spacetime $T_4$ bring on a {\it biframe spacetime} $T_4\times G_4$ which is associated with a dual {\it biframe spacetime} $T_4^{\ast}\times G_4^{\ast}$ over coordinate spacetime $M_4$. mathematically, the globally and locally flat vector spacetimes allow for a canonical identification of vectors in tangent Minkowski spacetime $T_4$ at points with vectors in Minkowski spacetime itself $M_4$, and also for a canonical identification of vectors at points with its dual vectors at the same points. In physics, either tangent or dual tangent Minkowski spacetime over globally flat Minkowski spacetime is viewed as a free-motion spacetime $\fM_4$. The locally flat gravigauge spacetime characterized by the gravigauge field is regarded as an {\it emergent spacetime} $\fG_4$.  

The canonical identification for the vector spacetimes simplifies the structure of biframe spacetime to be,
\be
& & \fB_4 =  \fM_4 \times \bf{G}_4 , \nn \\
 & & \fM_4 \equiv T_4 \cong T_4^{\ast} \cong  M_4 , \quad \fG_4 \equiv G_4 \cong G_4^{\ast} .
\ee
Mathematically, such a biframe spacetime structure brings about a {\it gravigauge fiber bundle} $\fE_4$, where the globally flat free-motion Minkowksi spacetime is considered as a base spacetime $\fM_4$ and the locally flat emergent gravigauge spacetime is viewed as a fiber $\fG_4$. For a trivial case, we have
\be
& & \fE_4 \sim  \fB_4 = \fM_4 \times \fG_4 .
\ee
In general, the gravigauge fiber bundle $\fE_4$ is related to the product biframe spacetime $\fM_4\times \fG_4$ through a continuous surjective map $\Pi_{\chi}$ which projects the bundle $\fE_4$ to the base spacetime $\fM_4$. Thus the gravigauge fiber bundle $\fE_4$ with the surjective map $\Pi_{\chi}$ is in general expressed as follows: 
\be
& & \Pi_{\chi}: \; \; \fE_4 \to \fM_4 .
\ee
Geometrically, the gravigauge fiber bundle structure of biframe spacetime is represented by  $(\fE_4, \fM_4, \Pi_{\chi}, \fG_4) $. 


\section{Scaling property of basic fields and the gauge fields and field strengths in gravigauge spacetime}

It is useful to analyze the scaling property of basic fields in order to construct a scaling invariant action. Meanwhile, the introduction of biframe spacetime enables us to redefine the gauge fields and field strengths in locally flat gravigauge spacetime, which is crucial to build a gauge invariant and coordinate independent action.

\subsection{Scaling property of basic fields}

From the covariant derivative operator presented in Eq.(\ref{CDO}), it is clear that the gauge field has the same scaling property as coordinate derivative. Under the scaling transformation of coordinates, the gauge fields transform as follows:
\be
& & x^{\mu} \to x^{' \mu} = \lambda^{-1}x^{\mu}, \quad \p_{\mu}\to \p'_{\mu} = \lambda \p_{\mu} ,\nn \\
& & \left(\cA_{\mu}(x), \ckcA_{\mu}(x), A_{\mu}(x)\right) \to \left(\cA'_{\mu}(x'), \ckcA'_{\mu}(x'), A'_{\mu}(x') \right) = \lambda \left(\cA_{\mu}(x), \ckcA_{\mu}(x), A_{\mu}(x) \right), 
\ee
with $\lambda$ the constant scaling factor. Where the power of $\lambda$ defines the canonical dimension of coordinates and gauge fields. 

The dimensionless gravivector $\vka^{\mu}(x)$ in gravigauge spacetime should be a scaling invariant vector field indicated from the definitions of gravicoodinator displacement $\delta \vka^a$ and derivative $\hat{\eth}_{a}$ operators given in Eq.(\ref{GCDO}). A graviscalar field $\phi(x)$ is rescaled from the gravigauge field in Eq.(\ref{SGIGGF}) to provide the dimensionful gravicoodinator displacement $\delta \zeta^a$ and derivative $\eth_{a}$ operators in Eq.(\ref{DGCDO}). The presence of graviscalar field enables us to analyze the local and global scaling properties of basic fields:
\be
& & \phi(x) \to \phi'(x) = \xi(x) \phi(x), \nn \\
& & \chi_{\mu}^{\;\; a}(x) \to  \chi_{\mu}^{' \;a}(x) = \xi^{-1}(x) \chi_{\mu}^{\;\; a}(x), \quad \chih_{\mu}^{\;\; a}(x) \to  \chih_{\mu}^{' \;a}(x) = \xi(x) \chih_{\mu}^{\;\; a}(x) , \nn \\
& & \chi_{\mu\nu}(x) \equiv \chi_{\mu}^{\;\; a}(x) \chi_{\mu}^{\;\; b}(x)\eta_{ab} \to \chi'_{\mu\nu}(x)  = \xi^{-2}(x)\chi_{\mu\nu}(x) , \nn \\
& & \chih^{\mu\nu}(x) \equiv \chih_{a}^{\;\; \mu}(x) \chih_{b}^{\;\; \nu}(x)\eta^{ab} \to \chih^{' \mu\nu}(x)  = \xi^{2}(x)\chih^{\mu\nu}(x) ,
\ee
under the scaling gauge transformation, and 
\be
& & x^{\mu} \to x^{'\mu} = \lambda^{-1} x^{\mu}, \quad \phi(x) \to \phi'(x') =  \phi(x), \nn \\
& & \chi_{\mu}^{\;\; a}(x) \to  \chi_{\mu}^{' \;a}(x') = \lambda\chi_{\mu}^{\;\; a}(x), \quad \chih_{\mu}^{\;\; a}(x) \to  \chih_{\mu}^{' \;a}(x') = \lambda^{-1} \chih_{\mu}^{\;\; a}(x) , \nn \\
& & \chi_{\mu\nu}(x)  \to \chi'_{\mu\nu}(x)  = \lambda^2 \chi_{\mu\nu}(x) , \quad \chih^{\mu\nu}(x) \to \chih^{' \mu\nu}(x')  = \lambda^{-2}\chih^{\mu\nu}(x) ,
\ee
under the global scaling transformation. 

Let us assign the powers of $\lambda$ and $\xi(x)$ under the global and local scaling transformations of fields to be as the global and local scaling charges $\cC_s$ and $\ckcC_s$ of fields, respectively. The canonical dimension $\cC_d$ of field is defined as the sum of the global and local scaling charges, i.e.:
\be
\cC_d = \cC_s + \ckcC_s .
\ee
It is obvious that the gauge fields $\cA_{\mu}^{ab}(x)$ and $\fA_{\mu}^{\;\; a}(x)$ as well as $A_{\mu}(x)$ all have global and local scaling charges $\cC_s=1$ and $\ckcC_s=0$ with the canonical dimension $\cC_d = 1$, while the gravigauge field $\chi_{\mu}^{\;\; a}(x)$($\chih_{a}^{\;\; \mu}(x)$) has global and local scaling charges $\cC_s=1$($\cC_s=-1$) and $\ckcC_s=-1$($\ckcC_s=1$) with the canonical dimension $\cC_d = 0$. A field with zero canonical dimension $\cC_d = 0$ is referred to as {\it dimensionless field} and a field with zero local scaling charge $\ckcC_s = 0$ is called as {\it scaling gauge invariant field}. In general, a field with both zero global and local scaling charges $\cC_s=0$ and $\ckcC_s=0$ is referred to as {\it scaling gauge invariant dimensionless field}.

In general, taking the dimensionless gravigauge field $\chi_{\mu}^{\;\; a}(x)$ and graviscalar field $\phi(x)$, we can equivalently decompose, instead of using the explicit scaling gauge invariant ones $\fOm_{\mu}^{ab}(x)$ and $\fA_{\mu}^{ab}(x)$, the spin gauge field into the following two alternative parts:
\be \label{SGFDC1}
\cA_{\mu}^{ab}(x)  \equiv \mOm_{\mu}^{ab}(x) + \mA_{\mu}^{ab}(x) \equiv \fOm_{\mu}^{ab}(x) + \fA_{\mu}^{ab}(x), 
\ee 
where the alternative spin gravigauge field $\mOm_{\mu}^{ab}(x)$ and spin covariant gauge field $\mA_{\mu}^{ab}(x)$ are defined as follows:
\be \label{SGGF1}
& & \mOm_{\mu}^{ab}(x) = \frac{1}{2}\left( \chih^{a\nu} F_{\mu\nu}^{b} - \chih^{b\nu} F_{\mu\nu}^{a} -  \chih^{a\rho}  \chih^{b\sigma}  F_{\rho\sigma}^{c} \chi_{\mu c } \right) , \nn \\
& & F_{\mu\nu}^{a}(x) =  \p_{\mu}\chi_{\nu}^{\; a}(x) -  \p_{\nu}\chi_{\mu}^{\; a}(x) , 
\ee
with the relations:
\be \label{Relation}
& &  \fOm_{\mu}^{ab}(x) \equiv \mOm_{\mu}^{ab}(x) + \mS_{\mu}^{ab}(x) , \quad \fA_{\mu}^{ab}(x) \equiv \mA_{\mu}^{ab}(x) -  \mS_{\mu}^{ab}(x) , \nn \\
& & \mS_{\mu}^{ab}(x) \equiv \left(\chi_{\mu}^{\; a}(x) \chih^{b\nu}(x) - \chi_{\mu}^{\; b}(x) \chih^{a\nu}(x) \right) S_{\nu} , 
\ee
and the scaling invariant gravigauge field strength,
\be
& & \rF_{\mu\nu}^a(x) \equiv d_{\mu}\chi_{\nu}^{\; a}(x) -  d_{\nu}\chi_{\mu}^{\; a}(x)  \equiv F_{\mu\nu}^{a} + S_{\mu\nu}^a , \nn \\
& & S_{\mu\nu}^a \equiv S_{\mu}\chi_{\nu}^{\; a}(x) -  S_{\nu}\chi_{\mu}^{\; a}(x) ,\nn \\
& &  d_{\nu} \equiv \p_{\nu} + S_{\nu}, \quad S_{\nu} \equiv \p_{\nu} \ln \phi(x) , 
\ee
which indicates that $\mOm_{\mu}^{ab}(x)$ and $\mA_{\mu}^{ab}(x)$ become no longer scaling gauge invariance. Under the scaling gauge transformation, they transform explicitly as follows :
\be
& & \mOm_{\mu}^{ab}(x) \to \mOm_{\mu}^{' ab}(x) = \mOm_{\mu}^{ab}(x) + \left(\chi_{\mu}^{\; a}(x) \chih^{b\nu}(x) - \chi_{\mu}^{\; b}(x) \chih^{a\nu}(x) \right) \p_{\nu}\ln \xi(x) , \nn \\
& & \mA_{\mu}^{ab}(x) \to \mA_{\mu}^{' ab}(x) = \mA_{\mu}^{ab}(x) - \left(\chi_{\mu}^{\; a}(x) \chih^{b\nu}(x) - \chi_{\mu}^{\; b}(x) \chih^{a\nu}(x) \right) \p_{\nu}\ln \xi(x)
\ee

Correspondingly, the spin gauge field strength is decomposed into the following two alternative parts:
\be \label{DCFS2}
\cF_{\mu\nu}^{ab}(x) \equiv  R_{\mu\nu}^{ab}(x) + \mF_{\mu\nu}^{ab}(x) \equiv  \fR_{\mu\nu}^{ab}(x) + \fF_{\mu\nu}^{ab}(x), 
\ee
with the definitions:
\be \label{DCFS3}
& & R_{\mu\nu}^{ab}(x) = \p_{\mu} \mOm_{\nu}^{ab} - \p_{\nu} \mOm_{\mu}^{ab} 
+ \mOm_{\mu c}^{a} \mOm_{\nu}^{cb} -  \mOm_{\nu c}^{a} \mOm_{\mu}^{cb},  
\nn \\  
& & \mF_{\mu\nu}^{ab}(x) =  \mD_{\mu} \mA_{\nu}^{ab} - \mD_{\nu} \mA_{\mu}^{ab} +  \mA_{\mu c}^{a} \mA_{\nu}^{cb} -  \mA_{\nu c}^{a} \mA_{\mu}^{cb}  , \nn \\
& & \mD_{\mu} \mA_{\nu}^{ab}(x) = \p_{\mu}  \mA_{\nu}^{ab}  +  \mOm_{\mu c}^{a} \mA_{\nu}^{cb} + \mOm_{\mu c}^{b} \mA_{\nu}^{ac}  ,
\ee
and the relations:
\be \label{DCFS4}
& &  \fR_{\mu\nu}^{ab}(x) \equiv  R_{\mu\nu}^{ab}(x) +  \mS_{\mu\nu}^{ab}(x) , \quad  \fF_{\mu\nu}^{ab}(x) \equiv  \mF_{\mu\nu}^{ab}(x) -  \mS_{\mu\nu}^{ab}(x) , \nn \\
& & \mS_{\mu\nu}^{ab}(x) =  \mD_{\mu} \mS_{\nu}^{ab} - \mD_{\nu} \mS_{\mu}^{ab} +  \mS_{\mu c}^{a} \mS_{\nu}^{cb} -  \mS_{\nu c}^{a} \mS_{\mu}^{cb}  , 
\ee

\subsection{Gauge fields and field strengths in locally flat gravigauge spacetime}

The gravitational origin of spin gauge symmetry indicates that the scaling gauge invariant gravigauge field $\fA_{\mu}^{\;\; a}$ ($\hfA_{a}^{\;\, \mu}$) as bi-covariant vector field in biframe spacetime can be regarded as Goldstone-like boson, which enables us to project the spin gauge field $\cA_{\mu}^{ab}(x)$ and electromagnetic gauge field $A_{\mu}(x)$ as vector fields in free-motion Minkowski spacetime $\fM_4$ into the vector fields in locally flat gravigauge spacetime $\fG_4$,
\be \label{SGFGG}
 & & \boldsymbol{\cA}_{c}^{ab} \equiv \hfA_{c}^{\;\, \mu}(x) \cA_{\mu}^{ab}(x)  \equiv \phi^{-1} \cA_{c}^{ab}, \nn \\
 & & \whA_{c} \equiv \hfA_{c}^{\;\, \mu}(x) A_{\mu}(x) \equiv \phi^{-1} A_c ,  \nn \\
 & & \cA_{c}^{ab} \equiv \chih_{c}^{\;\, \mu}(x) \cA_{\mu}^{ab}(x)  , \quad A_{c} \equiv \chih_{c}^{\;\, \mu}(x) A_{\mu}(x) ,
\ee
where $\whcA_{c}^{ab}$ and $\whA_{c}$ appear as scaling gauge invariant dimensionless fields, and $\cA_{c}^{ab}$ and $A_{c}$ become local scaling charged gauge fields. The scaling gauge invariant dimensionless covariant derivative operator in locally flat gravigauge spacetime is defined as: 
\be
& & \whcD_{c} \equiv \heth_{c} -i \whcA_c \equiv \heth_{c} - i \whcA_{c}^{ab} \frac{1}{2}\Sigma_{ab} \equiv \phi^{-1} \cD_{c} , \nn \\
& & \cD_{c} \equiv \eth_{c} -i \cA_c \equiv \eth_{c} - i \cA_{c}^{ab} \frac{1}{2}\Sigma_{ab} ,
\ee
for the spin gauge field, and 
\be
& & \whD_{c} \equiv \heth_{c} -i \whA_c \equiv \phi^{-1} D_{c}, \nn \\
& & D_{c} \equiv \eth_{c} -i A_c ,
 \ee
for the electromagnetic gauge field. Where $\heth_{c} \equiv \hfA_{c}^{\;\,\mu} \p_{\mu}$ is the  scaling gauge invariant dimensionless gravicoordinate derivative operator given in Eq.(\ref{GCDO}) and $\eth_{c} \equiv \chih_{c}^{\;\,\mu} \p_{\mu}$ is the local scaling charged one. The commutator of the covariant derivative operator is given by 
\be \label{NCR}
& & [\whcD_{c}, \whcD_{d}] = \fOm_{[cd]}^{a} \whcD_{a} -i \whcF_{cd} , \quad [ \whD_{c}, \whD_{d}] = \fOm_{[cd]}^{a} \whD_{a} -i \whF_{cd} , \nn \\
& & [\cD_{c}, \cD_{d}] = \mOm_{[cd]}^{a} \cD_{a} -i \cF_{cd} , \quad [ D_{c}, D_{d}] = \mOm_{[cd]}^{a} D_{a} -i F_{cd} , 
\ee
which define the scaling gauge invariant dimensionless gauge field strengths in locally flat gravigauge spacetime:
\be \label{SIGFS}
& & \fOm_{[cd]}^{a} = - \hfA_{c}^{\; \mu}(x)\hfA_{d}^{\; \nu}(x) \mF_{\mu\nu}^{a}(x) \equiv -\whmF_{cd}^{a} , \nn \\
& & \whcF_{cd} = \heth_{c} \whcA_d -  \heth_{d} \whcA_c + [\whcA_{c}, \whcA_{d}] - \fOm_{[cd]}^{a} \whcA_a  \equiv \whcF_{cd}^{\; ab} \frac{1}{2}\Sigma_{ab} ,  \nn \\
& & \whF_{cd} = \heth_{c} \whA_d -  \heth_{d} \whA_c - \fOm_{[cd]}^{a} \whA_a ,
\ee
and the corresponding local scaling charged gauge field strengths: 
\be \label{SCGFS}
& & \mOm_{[cd]}^a  \equiv- F_{cd}^a \equiv- \chih_{c}^{\; \mu}(x)\chih_{d}^{\; \nu}(x) F_{\mu\nu}^{a}(x)  , \nn \\
& & \cF_{cd}  \equiv \cF_{cd}^{\; ab} \frac{1}{2}\Sigma_{ab} = \eth_{c} \cA_d -  \eth_{d} \cA_c + [\cA_{c}, \cA_{d}] - \mOm_{[cd]}^{a} \cA_a  \equiv \phi^{2} \whcF_{cd},  \nn \\
& & F_{cd} = \eth_{c} A_d -  \eth_{d} A_c - \mOm_{[cd]}^{a} A_a \equiv \phi^{2} \whF_{cd} , \nn \\
& & \rF_{cd}^a \equiv \chih_{c}^{\; \mu}(x)\chih_{d}^{\; \nu}(x) \rF_{\mu\nu}^{a}(x) \equiv F_{cd}^a + S_{cd}^{a} \equiv \phi\, \whmF_{cd}^{a} , \nn \\ 
& & S_{cd}^a \equiv \chih_{c}^{\; \mu}(x)\chih_{d}^{\; \nu}(x) S_{\mu\nu}^{a}(x) = (S_c \, \eta_{d}^{\; a} -  S_d \, \eta_{c}^{\; a}) ,\quad S_c \equiv \eth_c\ln \phi .
\ee
The spin gauge and scaling gauge invariant gravigauge field strength is defined as
\be \label{GIGGFS}
& & \whcF_{cd}^{a} \equiv \hfA_{c}^{\; \mu}(x)\hfA_{d}^{\; \nu}(x) \cF_{\mu\nu}^{a}(x) \equiv \phi^{-1} \cF_{cd}^{a} \nn \\
& & \cF_{cd}^{a} \equiv \fF_{cd}^{a} + S_{cd}^a , \quad \fF_{cd}^{a} \equiv \chih_{c}^{\; \mu} \chih_{d}^{\; \nu} \fF_{\mu\nu}^{a} ,\nn \\
& &  \fF_{\mu\nu}^{a} \equiv \p_{\mu} \chi_{\nu}^{a}(x) - \p_{\nu} \chi_{\mu}^{a}(x) + \cA_{\mu\, b}^{a}(x) \chi_{\nu}^{b}(x) - \cA_{\nu\, b}^{a}(x) \chi_{\mu}^{b}(x) ,
\ee
with $ \cF_{\mu\nu}^{a}(x)$ being the spin gauge and scaling gauge invariant gravigauge field strength defined in Eq.(\ref{WESFS}).

The field strengths $\cF_{cd}^{\; ab}$ and $F_{cd}$ possess the local scaling charge $\ckcC_s=2$, while $\cF_{cd}^{a}$ and $\mOm_{[cd]}^a  \equiv F_{cd}^a$ have the local scaling charge $\ckcC_s=1$ as indicated from their relations to the scaling gauge invariant dimensionless gauge field strengths. Where $\fOm_{[cd]}^{a} \equiv \whmF_{cd}^{\;a}$ ($\mOm_{[cd]}^{a} \equiv F_{cd}^{\;a}$) brings on the gravigauge field strength arising from the non-commutation relation of the gravicoordinate derivative operator. $\whcF_{cd}^{\;ab}$($\cF_{cd}^{\;ab}$) defines the field strength of spin gauge field and $\whF_{cd}$($F_{cd}$) represents the field strength of electromagnetic gauge field.

The gravitational origin of spin gauge symmetry brings the scaling gauge invariant dimensionless spin gauge field $\whcA_{c}^{ab}$ into the two parts:
\be \label{SGFDC2}
\whcA_{c}^{ab} =  \fOm_{c}^{ab} + \fA_{c}^{ab} \equiv \phi^{-1} \cA_{c}^{ab} \equiv \phi^{-1}  (\mOm_{c}^{ab} + \mA_{c}^{ab} ) ,
\ee
where $\fOm_{c}^{ab}$ and $\fA_{c}^{ab}$ correspond to the scaling gauge invariant dimensionless spin gravigauge field and spin covariant-gauge field in locally flat gravigauge spacetime. $\fOm_{c}^{ab}$ characterizes the gravitational origin of gauge symmetry determined by the dual gravigauge fields $\fA_{\mu}^{\;\, a}$ and $\hfA_{a}^{\;\, \mu}$ with the following explicit form: 
\be \label{SGGFHC}
\fOm_{c}^{ab}  & = &  \frac{1}{2} [\, \hfA^{a\mu} \heth_{c}\fA_{\mu}^{\;\, b} - \hfA^{b\mu} \heth_{c}\fA_{\mu}^{\;\, a} - \hfA_{c}^{\;\,\mu} ( \heth^{a}\fA_{\mu}^{\;\, b} - \heth^{b}\fA_{\mu}^{\;\, a} ) \nn \\
& + & ( \heth^{a}\hfA^{b\mu} -  \heth^{b}\hfA^{a\mu} ) \fA_{\mu c}] \equiv \eta^{ad}\eta^{bd'}  \fOm_{c d d'} .
\ee
 
The scaling gauge invariant dimensionless field strength of spin gauge field is correspondingly expressed into the following two parts:
\be
\whcF_{cd}^{\;ab} = \fR_{cd}^{\; ab} + \fF_{cd}^{\; ab} \equiv \phi^{-2} \left( \mR_{cd}^{\; ab} +  \mF_{cd}^{\; ab} \right) , 
\ee
with the explicit forms: 
\be \label{SGFS}
& & \whcF_{cd}^{\;ab} = \bsckcD_{c} \whcA_{d}^{ab} - \bsckcD_{d} \whcA_{c}^{\;ab} + ( \whcA_{c b'}^{a} \whcA_{d}^{b' b} -  \whcA_{d b'}^{a} \whcA_{c}^{b' b} ) \equiv \phi^{-2} \cF_{cd}^{\;ab} , \nn \\ 
& & \fR_{cd}^{\; ab} = \bsckcD_{c} \fOm_{d}^{ab} - \bsckcD_{d} \fOm_{c}^{ab} + \fOm_{c b'}^{a} \fOm_{d}^{b' b} -  \fOm_{d b'}^{a} \fOm_{c}^{b' b} \equiv \phi^{-2} \left( \mR_{cd}^{\; ab} +  \mS_{cd}^{\; ab} \right) , \nn \\
 & & \fF_{cd}^{\;ab} = \bsckcD_{c} \fA_{d}^{ab} - \bsckcD_{d} \fA_{c}^{ab}  +  \fA_{c b'}^{a} \fA_{d}^{b' b} -  \fA_{d b'}^{a} \fA_{c}^{b' b} \nn \\
 & & \qquad \quad + \fOm_{c b'}^{a} \fA_{d}^{b' b} + \fOm_{c b'}^{b} \fA_{d}^{a b'}  -  \fOm_{d b'}^{a} \fA_{c}^{b' b} - \fOm_{d b'}^{b} \fA_{c}^{a b'} \equiv \phi^{-2} \left( \mF_{cd}^{\; ab} -\mS_{cd}^{\; ab} \right), \nn \\
& &  \bsckcD_{c} \whcA_{d}^{ab}  =  \heth_{c} \whcA_{d}^{ab}  - \fOm_{cd}^{c'} \whcA_{c'}^{ab} ,
\quad \bsckcD_{c} \fA_{d}^{ab}  =  \heth_{c} \fA_{d}^{ab}  - \fOm_{cd}^{c'} \fA_{c'}^{ab} ,\nn\\
& & \bsckcD_{c} \fOm_{d}^{ab}  =  \heth_{c} \fOm_{d}^{ab}  - \fOm_{cd}^{c'} \fOm_{c'}^{ab} ,
\ee
where $\fR_{cd}^{\;ab}$ and $\fF_{cd}^{\; ab}$ correspond to the scaling gauge invariant dimensionless spin gravigauge field strength and spin covariant-gauge field strength. $\bsckcD_{c}$ is regarded as the scaling gauge invariant dimensionless covariant derivative of gravicoordinate derivative $\heth_{c}$ under the presence of spin gravigauge field in locally flat gravigauge spacetime. It can be checked that $\mR_{cd}^{\;ab}$ and $\mF_{cd}^{\; ab}$ with corresponding derivatives $\ckcD_{c}$ and $\eth_{c}$ get the similar definitions in light of the scaling charged quantities.

\section{Gauge and scaling invariant action and gauge-gravity-geometry correspondence in the framework of gravitational quantum field theory}

With the above discussions and analyses, we are able to construct the gauge and scaling invariant action in locally flat gravigauge spacetime and biframe spacetime. Meanwhile, we are going to present a general demonstration on the gauge-gravity and gravity-geometry correspondences within the framework of gravitational quantum field theory. We shall begin with discussing the scaling gauge fixing and the Einstein basis.

\subsection{Scaling gauge fixing and Einstein basis}

The gravigauge field expressed in Eq.(\ref{SGIGGF}) indicates a hidden scaling gauge invariance, which allows us to take an essential scaling gauge transformation $\xi_e(x)$ to set a special gauge fixing condition, so that the graviscalar field $\phi(x)$ can be transformed into a constant scale $\Mka$:
\be
\label{SGFEB1}
\phi(x) \to  \phi^{(e)}(x) = \xi_e(x) \phi(x) \equiv \Mka .
\ee
The dimensionless gravigauge field $\chi_{\mu}^{\;\; a}(x)$ gets a corresponding scaling gauge transformation in order to preserve the scaling gauge invariance of gravigauge field $\fA_{\mu}^{\;\; a}(x)$, i.e.: 
\be \label{SGFEB2}
& & \chi_{\mu}^{\;\; a}(x) \to \chi_{\mu}^{(e)\1 a}(x) \equiv \xi_e^{-1}(x) \chi_{\mu}^{\;\; a}(x) .
\ee
In such a special scaling gauge transformation, the scaling gauge invariant gravigauge field and gravimetric field can be written into the following forms:
\be \label{SGFEB3}
\fA_{\mu}^{\;\; a}(x) & \equiv & \phi(x)\chi_{\mu}^{\;\; a}(x)  \to \fA_{\mu}^{(e)\1 a}(x) = \Mka \chi_{\mu}^{(e)\1 a}(x) , \nn \\
\mH_{\mu\nu}(x) & \equiv & \phi^2(x) \chi_{\mu\nu}(x) \to \mH_{\mu\nu}^{(e)}(x) = \Mka^2 \chi^{(e)}_{\mu\nu}(x) ,
\ee
which fixes the scaling gauge symmetry into the so-called Einstein basis. Where the constant mass $\Mka$ plays a role as a {\it fundamental mass scale}. $\chi_{\mu}^{(e)a}(x)$ and $\chi_{\mu\nu}^{(e)}(x)$ are mentioned to be the dimensionless gravigauge field and gravimetric field in Einstein basis.

\subsection{Gauge and scaling invariant action in locally flat gravigauge spacetime}

We are now in the position to construct the gauge and scaling invariant action for the chirality-based Dirac spinor field appearing in the SM. It is believed that the SM is an effective theory as it involves 18 unknown parameters, so the action that we are going to build should also be an effective action. For our present purpose, we just provide an action with keeping the lowest order terms by following along the gauge and scaling invariance principle. The explicit form for such an action in locally flat gravigauge spacetime is found to have the following simple form:
\be
\label{HSGaction}
\cS_D & \equiv & \int [\delta^4\vka]\, \fkL \equiv  \int [\delta^4\vka]\; \{ (\bar{\Psi}_{-} \Sigma_{-}^{a} i \whcD_{a} \Psi_{-}  - \frac{m}{\Mka} \bar{\Psi}_{-}\Sigma_{-}^{6}\Psi_{-}  + H.c. ) \nn \\
 & - & \frac{1}{4}  g_E^{-2}\eta^{cc'}\eta^{dd'} \whF_{cd} \whF_{c'd'} - \frac{1}{4} g_G^{-2}  \eta^{cc'}\eta^{dd'}  \whcF_{cd}^{\; ab} \whcF_{c'd' a b}  \nn \\
 & + & \frac{1}{4} \bar{\eta}^{c d c' d'}_{a a'}  \frac{m_G^2}{\Mka^2} \whcF_{cd}^{a}\whcF_{c'd'}^{a'} +    \frac{1}{4} g_G^{-2}\tilde{\eta}^{cd\1 c'd'}_{a a' } \whmF_{cd}^{a}\whmF_{c'd'}^{a'}  \} ,
\ee
with the total covariant derivative,
\be
& & i\whcD_{a} \equiv i\heth_{a}+  \whcA_a + \whA_a \equiv i \heth_{a} + \whcA_{a}^{bc} \frac{1}{2}\Sigma_{bc} + \whA_a .
\ee
It is noted that the gravigauge field $\fA_{\mu}^{\;\; a}(x)$ does not couple to the spinor field via the usual covariant derivative due to the chirality property of the group generators $\Sigma_{-}^{a}\Sigma_{-}^{b} =0$, it couples alternatively to the spinor field through the gravicoordinate derivative $\Sigma_{-}^{a}\hat{\eth}_{a} \equiv \Sigma_{-}^{a}\hfA_{a}^{\;\;\mu} \p_{\mu} \equiv \hmH^{\mu\nu} \fA_{\mu a}\Sigma_{-}^{a}\p_{\nu}$ in locally flat gravigauge spacetime. In fact, the kinematics of all basic fields acts as the source of gravitational interaction along with the gravicoordinate derivative characterized by the dual gravigauge field $\hat{\fA}_{a}^{\;\;\mu}(x)$, which actually brings on the comprehension why all motional fields get universal gravitational interactions.

It is noticed that all basic fields in the action appear as scaling gauge invariant dimensionless fields. Namely, all basic fields and operators in the above action have zero global and local scaling charges $\cC_s=0$ and $\ckcC_s=0$. The mass terms are also rescaled into dimensionless quantities in light of the fundamental mass scale, i.e., $m/\Mka$ and $m_G/\Mka$. 

The tensor $\bar{\eta}^{c d\1 c' d'}_{a a'}$ is given by the following general structure:
\be \label{CTensor1}
& & \bar{\eta}^{c d\1 c' d'}_{a a'}   \equiv \alpha_G \eta^{c c'} \eta^{d d'} \eta_{a a'}  - \frac{1}{2}\alpha_W (\eta^{c c'} \eta_{a'}^{d} \eta_{a}^{d'} + \eta^{d d'} \eta_{a'}^{c} \eta_{a}^{c'} ) 
\ee
with $\alpha_G$ and $\alpha_W$ being two coupling constants, and the tensor structure $\tilde{\eta}^{cd c'd'}_{a a'}$ is defined as follows:
\be \label{CTensor3}
\tilde{\eta}^{cd c'd'}_{a a'}  & \equiv &    \eta^{c c'} \eta^{d d'} \eta_{a a'}  
+  \eta^{c c'} ( \eta_{a'}^{d} \eta_{a}^{d'}  -  2\eta_{a}^{d} \eta_{a'}^{d'}  ) +  \eta^{d d'} ( \eta_{a'}^{c} \eta_{a}^{c'} -2 \eta_{a}^{c} \eta_{a'}^{c'} )  , 
\ee
which preserves the spin gauge invariance for gravitational interaction of gravigauge field. It can be verified explicitly from the following relation and identity:
\be \label{Relation2}
& & \frac{1}{4} \tilde{\eta}^{cd\1 c'd'}_{a a' } \fkA \whmF_{cd}^{a}\whmF_{c'd'}^{a'}  =   \eta_{a}^{\; d} \eta_{b}^{\; c} \fkA \fR_{cd}^{\;ab} + 2 \p_{\mu} (\fkA \hfA^{a\mu} \whmF_{ca}^{c}) \nn \\
& & \qquad \equiv  \frac{1}{4} \tilde{\eta}^{cd\1 c'd'}_{a a' } \chi \phi^{2} \rF_{cd}^{a}\rF_{c'd'}^{a'} =  \frac{1}{4} \tilde{\eta}^{cd\1 c'd'}_{a a' } \chi \phi^{2} ( F_{cd}^{a} +  S_{cd}^{a}) (F_{c'd'}^{a'} + S_{cd}^{a}) \nn \\
& & \qquad =  \eta_{a}^{\; d} \eta_{b}^{\; c}  \chi \phi^{2}  ( R_{cd}^{\;ab} + \mS_{cd}^{\;ab}) + 2 \p_{\mu} (\chi \phi^2\chih^{a\mu} \rF_{ca}^{c} )  ,
\ee
where the term $\p_{\mu} (\fkA \hfA^{a\mu} \whmF_{ca}^{c} )$(or $\p_{\mu} (\chi \phi^2\chih^{a\mu} \rF_{ca}^{c} ) $) on the right-hand side of equality appears as a total derivative in the action. 

From the above identity and relation, the above spin gauge invariant action via a hidden scaling gauge formalism (Eq.(\ref{HSGaction})) can be rewritten into the following spin gauge and scaling gauge invariant action:
 \be
\label{SGIaction}
\cS_D & \equiv &   \int [\delta^4\zeta]\, \cL \equiv \int [\delta^4\zeta]\; \{( \bar{\vPsi}_{-} \Sigma_{-}^{a} i \cD_{a} \vPsi_{-} - \frac{m}{\Mka}\phi\, \bar{\vPsi}_{-}\Sigma_{-}^{6}\vPsi_{-}  + H.c. )\nn \\
& - & \frac{1}{4} g_E^{-2}  \eta^{cc'}\eta^{dd'} F_{cd} F_{c'd'}  - \frac{1}{4} g_G^{-2}  \eta^{cc'}\eta^{dd'} \cF_{cd}^{\; ab} \cF_{c'd' a b} \nn \\
& + & \frac{1}{4} \bar{\eta}^{c d c' d'}_{a a'}  \frac{m_G^2}{\Mka^2}\phi^2 \cF_{cd}^{a} \cF_{c'd'}^{a'}  +    \frac{1}{4} g_G^{-2}  \tilde{\eta}^{cd\1 c'd'}_{a a' } \phi^2 \rF_{cd}^{a}\rF_{c'd'}^{a'} 
\} , \nn \\
 & \equiv & \int [\delta^4\zeta]\1 \{ ( \bar{\vPsi}_{-} \Sigma_{-}^{a} i \cD_{a} \vPsi_{-} - \frac{m}{\Mka}\phi\, \bar{\vPsi}_{-}\Sigma_{-}^{6}\vPsi_{-} +  H.c. ) \nn \\
& - & \frac{1}{4}  \eta^{cc'}\eta^{dd'} F_{cd} F_{c'd'}  - \frac{1}{4} \eta^{cc'}\eta^{dd'} \cF_{cd}^{\; ab} \cF_{c'd' a b } \nn \\
& + &  \frac{1}{4} \bar{\eta}^{c d c' d'}_{a a'}  \frac{m_G^2}{\Mka^2}\phi^2 (\fF_{cd}^{a} + S_{cd}^a)(\fF_{c'd'}^{a'} + S_{c'd'}^{a'}) \nn \\
& + & g_G^{-2}  ( \frac{1}{4}\tilde{\eta}^{cd\1 c'd'}_{a a' } \phi^2 F_{cd}^{a} F_{c'd'}^{a'} - 6 \eta^{cd}\eth_c \phi \eth_d \phi ) \} ,
\ee
where we have introduced the following definition:
\be
& & \Psi_{-} \to \vPsi_{-} \equiv \phi^{3/2} \Psi_{-} ,  
\ee
and made the replacement in the second formalism of the action:
\be
& & \mA_{a}^{bc} \to g_G \mA_{a}^{bc}, \quad A_a \to g_E A_a, \nn \\
& &  i\cD_{a} \equiv i\eth_{a} + \cA_a + A_a \to  i\cD_{a}  \equiv i \eth_{a} + g_G \cA_{a}^{bc} \frac{1}{2}\Sigma_{bc}  + g_E A_a , 
\ee
Such a redefined chirality-based Dirac spinor $\vPsi_{-}$ has a local scaling charge $\ckcC_s = 3/2$ and canonical dimension $\cC_d = 3/2$. Note that the covariant derivative $\cD_{c}$ and the integral measure $[\delta^4\zeta]$ have local scaling charges $\ckcC_s = 1 $ and $\ckcC_s = -4$, respectively. 

When making special scaling gauge transformation and taking the scaling gauge fixing condition to be Einstein basis, i.e., $\phi \to \Mka$ and $\chi_{\mu}^{\;\; a} \to \chi_{\mu}^{(e)a}$ as shown in Eqs.(\ref{SGFEB1})-(\ref{SGFEB3}), we are able to express the gauge invariant action in Eq.(\ref{SGIaction}) into the following simple form:
 \be
\label{EBaction}
\cS_D & \equiv &  \int [\delta^4\zeta^{(e)}]\, \cL^{(e)} \nn \\
& \equiv & \int [\delta^4\zeta^{(e)}]\1 \{  
(\bar{\vPsi}_{-} \Sigma_{-}^{a} i \cD_{a} \vPsi_{-}  - m \1 \bar{\vPsi}_{-}\Sigma_{-}^{6}\vPsi_{-}  + H.c. ) \nn \\
& - &  \frac{1}{4} \eta^{cc'}\eta^{dd'} F_{cd} F_{c'd'}  - \frac{1}{4}  \eta^{cc'}\eta^{dd'} \cF_{cd}^{\; ab} \cF_{c'd' a b}  \nn \\
& + & \frac{1}{4} \bar{\eta}^{c d c' d'}_{a a'}  m_G^2 \fF_{cd}^{a}\fF_{c'd'}^{a'}  +  g_G^{-2} \Mka^2 \frac{1}{4} \tilde{\eta}^{cd\1 c'd'}_{a a' } F_{cd}^{a} F_{c'd'}^{a'} \} .
\ee


\subsection{ Gauge and scaling invariant action with gauge-gravity-geometry correspondence in gravitational quantum field theory}

As the gravigauge field $\fA_{\mu}^{\;\; a}$ ($\hfA_{a}^{\;\, \mu}$) or $\chi_{\mu}^{\;\; a}$ ($\chih_{a}^{\;\, \mu}$) plays a role as Goldstone-like boson and appears as a bi-covariant vector field in biframe spacetime, it enables us to project the above actions built in locally flat gravigauge spacetime into the ones represented in biframe spacetime within the framework of gravitational quantum field theory. 

In light of the scaling gauge invariant fields, we can rewrite the spin gauge invariant action into the following form:
\be
\label{HSGaction2}
\cS_D & \equiv & \int [d^4x] \fkA(x)\, \fkL \nn \\
& \equiv & \int [d^4x] \fkA(x)\1 \{ (\hmH^{\mu\nu}\bar{\Psi}_{-} \Sigma_{-}^{a} \fA_{\mu a} i\cD_{\nu} \Psi_{-}  - \frac{m}{\Mka} \bar{\Psi}_{-}\Sigma_{-}^{6}\Psi_{-}  + H.c. )\nn \\
& - &  \frac{1}{4}  \hmH^{\mu\mu'}\hmH^{\nu\nu'} F_{\mu\nu} F_{\mu'\nu'}  - \frac{1}{4}  \hmH^{\mu\mu'}\hmH^{\nu\nu'} \cF_{\mu\nu}^{\; ab} \cF_{\mu'\nu' a b} \nn \\
& + & \frac{1}{4} \frac{m_G^2}{\Mka^2}  \bmH_{a a'}^{\mu\nu \mu'\nu'} \cF_{\mu\nu}^{\; a} \cF_{\mu'\nu'}^{a'} + \frac{1}{4} g_G^{-2} \tmH_{aa'}^{\mu\nu \mu'\nu'} \mF_{\mu\nu}^{a}\mF_{\mu'\nu'}^{a'} \} , 
\ee
where we have used the definitions:
\be \label{CDTensors}
& & \mA_{\mu}^{ab} \to g_G \mA_{\mu}^{ab}, \quad A_{\mu} \to g_E A_{\mu}, \nn \\
& & i\cD_{\nu} \equiv i\p_{\nu}+  \cA_{\nu} + A_{\nu} \equiv i \p_{\nu} + g_G \cA_{\nu}^{bc} \frac{1}{2}\Sigma_{bc} +g_E A_{\nu} , \nn \\
& & \bmH_{aa'}^{\mu\nu \mu'\nu'} \equiv \hfA_{c}^{\;\, \mu}\hfA_{d}^{\;\, \nu} \hfA_{c'}^{\;\, \mu'} \hfA_{d'}^{\;\, \nu'}  \bar{\eta}^{c d c' d'}_{a a'} , \nn \\
& & \tmH_{aa'}^{\mu\nu \mu'\nu'} \equiv \hfA_{c}^{\;\, \mu}\hfA_{d}^{\;\, \nu} \hfA_{c'}^{\;\, \mu'} \hfA_{d'}^{\;\, \nu'}  \tilde{\eta}^{c d c' d'}_{a a'} .
\ee

In terms of the local scaling charged fields, we arrive at the following spin gauge and scaling gauge invariant action:
\be
\label{SGIaction2}
\cS_D & \equiv &  \int [d^4x] \chi(x)\, \cL  \nn \\
& \equiv & \int [d^4x] \chi(x)\1 \{( \chih^{\mu\nu}\bar{\vPsi}_{-} \Sigma_{-}^{a} \chi_{\mu a} i\cD_{\nu} \vPsi_{-}  - \frac{m}{\Mka} \phi \bar{\vPsi}_{-}\Sigma_{-}^{6}\vPsi_{-} + H.c. )
\nn \\
& - &  \frac{1}{4} \chih^{\mu\mu'}\chih^{\nu\nu'} F_{\mu\nu} F_{\mu'\nu'}- \frac{1}{4} \chih^{\mu\mu'}\chih^{\nu\nu'} \cF_{\mu\nu}^{\; ab} \cF_{\mu'\nu' a b}  \nn \\
& + & \frac{1}{4} \frac{m_G^2}{\Mka^2}\phi^2 \bchi_{a a'}^{\mu\nu \mu'\nu'} ( \fF_{\mu\nu}^{\; a} +  S_{\mu\nu}^{\; a} )( \fF_{\mu'\nu'}^{a'} +  S_{\mu'\nu'}^{a'} )  \nn \\
& + &  g_G^{-2} ( \frac{1}{4} \phi^2 \tchi_{aa'}^{\mu\nu \mu'\nu'} F_{\mu\nu}^{a}F_{\mu'\nu'}^{a'} - 6 \chih^{\mu\nu} \p_{\mu}\phi \p_{\nu}\phi  \} ,
\ee
with the definitions:
\be \label{CDTensors2}
& & i\cD_{\nu} \equiv i\p_{\nu}+  \cA_{\nu} + A_{\nu} \equiv i \p_{\nu} + g_G \cA_{\nu}^{bc} \frac{1}{2}\Sigma_{bc} + g_E A_{\nu} , \nn \\
& &  \bchi_{aa'}^{\mu\nu \mu'\nu'} \equiv \chih_{c}^{\;\, \mu}\chih_{d}^{\;\, \nu} \chih_{c'}^{\;\, \mu'} \chih_{d'}^{\;\, \nu'}  \bar{\eta}^{c d c' d'}_{a a'}  =  \phi^4 \bmH_{a a'}^{\mu\nu \mu'\nu'} , \nn \\
& & \tchi_{aa'}^{\mu\nu \mu'\nu'} \equiv \chih_{c}^{\;\, \mu}\chih_{d}^{\;\, \nu} \chih_{c'}^{\;\, \mu'} \chih_{d'}^{\;\, \nu'}  \tilde{\eta}^{c d c' d'}_{a a'} =  \phi^4 \tmH_{a a'}^{\mu\nu \mu'\nu'} , \nn \\
& & \chi(x) \equiv \det \chi_{\mu}^{\; a}(x) = \phi^{-4} \fkA(x) , \quad \chih \equiv \det \chih_{a}^{\; \mu} .
\ee

When making the scaling gauge fixing condition to be in Einstein basis, we are able to simplify the above action into the following form:
\be
\label{EBaction1}
\cS_D & \equiv &  \int [d^4x] \chi(x) \1 \cL^{(e)} \nn \\
& \equiv & \int [d^4x] \chi(x) \1 \{ (\chih^{\mu\nu}\bar{\vPsi}_{-} \Sigma_{-}^{a} \chi_{\mu a} i\cD_{\nu} \vPsi_{-}  - m \bar{\vPsi}_{-}\Gamma^{6}\vPsi_{-} + H.c. ) \nn \\
& - &  \frac{1}{4} \chih^{\mu\mu'}\chih^{\nu\nu'} F_{\mu\nu} F_{\mu'\nu'}- \frac{1}{4} \chih^{\mu\mu'}\chih^{\nu\nu'} \cF_{\mu\nu}^{\; ab} \cF_{\mu'\nu' a b}  \nn \\
& + & \frac{1}{4} m_G^2\bchi_{a a'}^{\mu\nu \mu'\nu'} \fF_{\mu\nu}^{\; a} \fF_{\mu'\nu'}^{a'}  + \frac{1}{4}  g_G^{-2}  \Mka^2 \tchi_{aa'}^{\mu\nu \mu'\nu'} F_{\mu\nu}^{a}F_{\mu'\nu'}^{a'}  \} ,
 \ee
where all scaling charged fields are understood to be in Einstein basis and the index `($e$)'  appearing in the gravigauge field $\chi_{\mu}^{(e)a}$ ($\chih_{a}^{(e) \mu}$) have been ignored for convenience. 

It can be verified from the relation in Eq.(\ref{Relation2}) that the gravitational gauge interaction characterized by the gravigauge field is related to Einstein-Hilbert action up to a total derivative as follows:
\be \label{GGGR}
& & \frac{1}{4} \chi\, \tchi_{aa'}^{\mu\nu \mu'\nu'} F_{\mu\nu}^{a} F_{\mu'\nu'}^{a'} = \chi\, R
+ 2 \p_{\mu} (\chi \chih^{\mu\rho} \chih_{a}^{\;\sigma} F_{\sigma\rho}^{a} ) , 
\ee
where we have used the identities:
\be \label{GGGI}
& & R \equiv \chih_{b}^{\; \mu} \chih_{a}^{\; \nu} R_{\mu\nu}^{ab} \equiv \eta^{cb}R_{cb} \equiv \eta_{b}^{\; c} \eta_{a}^{\; d} R_{cd}^{\; ab}  \equiv \chih^{\mu\sigma} R_{\mu\sigma} \equiv \chih^{\mu\sigma} \chih^{\nu\rho} R_{\mu\nu\rho\sigma} .
\ee
Such relation and identities reveal the gauge-gravity-geometry correspondence.

Geometrically, $R_{\mu\nu\rho\sigma}$ is the so-called Riemann curvature tensor, $R_{\mu\sigma}$ and $R$ correspond to the Ricci curvature tensor and Ricci curvature scalar. $R_{\mu\nu\sigma}^{\rho}$ is explicitly given by,  
\be \label{RMC3}
& & R_{\mu\nu\sigma}^{\;\rho}(x)  = \p_{\mu} \Gamma_{\nu\sigma}^{\rho} - \p_{\nu} \Gamma_{\mu\sigma}^{\rho}  + \Gamma_{\mu\lambda}^{\rho} \Gamma_{\nu\sigma}^{\lambda}  - \Gamma_{\nu\lambda}^{\rho} \Gamma_{\mu\sigma}^{\lambda} ,
\ee
with $\Gamma_{\mu\sigma}^{\rho}(x)$ defined as follows: 
\be \label{SGMF}
\Gamma_{\mu\sigma}^{\rho}(x)  & \equiv &  \chih_{a}^{\;\; \rho} \p_{\mu} \chi_{\sigma}^{\;\; a} +  \chih_{a}^{\;\; \rho}   \mOm_{\mu\1 b}^{a} \chi_{\sigma}^{\;\;b} , \nn \\
& = & \frac{1}{2}\chih^{\rho\lambda} (\p_{\mu} \chi_{\lambda\sigma} + \p_{\sigma} \chi_{\lambda\mu} - \p_{\lambda}\chi_{\mu\sigma} ) =\Gamma_{\sigma\mu}^{\rho},
 \ee
which is the so-called affine connection or Christoffel symbol in geometry. 

When applying for the gauge-gravity-geometry correspondence and adopting the vector-like spinor representation of Dirac fermion in four-dimensional Hilbert space, the action in Eq.(\ref{EBaction1}) can be further simplified into the following form:
\be
\label{EBaction2}
\cS_D & \equiv & \int [d^4x] \chi(x) \1 \cL^{(e)} \nn \\
& \equiv &  \int [d^4x] \chi(x) \; \{\frac{1}{2} ( \chih^{\mu\nu}\bar{\psi} \gamma^{a} \chi_{\mu a} i\cD_{\nu} \psi + H.c.) - m \bar{\psi}\psi \nn \\
& - &  \frac{1}{4} \chih^{\mu\mu'}\chih^{\nu\nu'} F_{\mu\nu} F_{\mu'\nu'}- \frac{1}{4} \chih^{\mu\mu'}\chih^{\nu\nu'} \cF_{\mu\nu}^{\; ab} \cF_{\mu'\nu' a b}  \nn \\
& + & \frac{1}{4} m_G^2\bchi_{a a'}^{\mu\nu \mu'\nu'} \fF_{\mu\nu}^{\; a} \fF_{\mu'\nu'}^{a'} +  \frac{1}{16\pi G_N}\, R \, \}  - 2 \p_{\mu} (\chi \chih^{\mu\rho} \chih_{a}^{\;\sigma} F_{\rho\sigma}^{a} ) ,
\ee
with $G_N$ the gravitational constant,
\be
\frac{1}{16\pi G_N} \equiv \frac{\Mka^2}{g_G^2} \equiv \frac{M_P^2}{16\pi} .
\ee

So far, we have provided various formalisms for the gauge and scaling invariant action of Dirac spinor field within the framework of gravitational quantum field theory following along the action principle of path integral formulation, which indicate the gauge-gravity and gravity-geometry correspondences and lead to the gauge-geometry duality relation.

\subsection{Hidden gauge formalism of the action with flowing unitary gauge and the gauge-geometry duality with emergent general linear group symmetry GL(1,3,R)}

The above formalisms of the action are built based on the scaling and gauge invariance principle through the inhomogeneous spin gauge symmetry in locally flat gravigauge spacetime. The coordinate spacetime remains to maintain the Poincar\'e group symmetry in globally flat Minkowski spacetime. The gauge fixing $\cW_e$-spin gauge field as a bicovariant vector field spanned in biframe spacetime is shown to play an essential role as a projection operator. In general, such a projection operator should allow us to define a spacetime gauge field from the spin gauge field and provide an equivalent action in hidden gauge formalism, so that we are able to investigate more profound correlation between the gravitational interaction and Riemann geometry of spacetime. 

Let us demonstrate explicitly how the general linear group symmetry GL(1,3,R) emerges automatically as a consequence of gauge invariance principle, which corroborates the gauge-geometry duality in curved Riemannian spacetime. The $\cW_e$-spin gauge field with imposing the gauge fixing condition turns out to be characterized by the gravigauge field $\fA_{\mu}^{\;\; a}$, which is associated in general with the spin gauge field $\cA_{\mu}^{a b}$ through the following covariant relation of spin gauge symmetry:
\be
 \p_{\mu} \fA_{\nu}^{\;\; a} +  \cA_{\mu\1 b}^{a} \fA_{\nu}^{\;\; b} - \cA_{\mu\nu}^{\rho} \fA_{\rho}^{\;\; a} = 0 . 
\ee
Such a relation enables us to define a hidden spin gauge invariant field $\cA_{\mu\nu}^{\rho}$ from the spin gauge field as follows:
\be \label{STGF}
 & & \cA_{\mu\nu}^{\rho}   \equiv  \hfA_{a}^{\;\; \rho} \p_{\mu} \fA_{\nu}^{\;\; a} + \hfA_{a}^{\;\; \rho}   \cA_{\mu\1 b}^{a} \fA_{\nu}^{\;\; b} \equiv  \hfA_{a}^{\;\; \rho} \cD_{\mu} \fA_{\nu}^{\;\; a}  , 
\ee
which brings on a gauge field represented in coordinate spacetime. For short, we may refer to such a gauge field $\cA_{\mu\nu}^{\rho}$ as a {\it spacetime gauge field}.

As shown in Eq.(\ref{SGFDC}), the spin gauge symmetry gets a gravitational origin by decomposing the spin gauge field $ \cA_{\mu}^{ab}$ into the sum of spin gravigauge field $\fOm_{\mu}^{ab}$ and covariant-gauge field $\fA_{\mu}^{ab}$. In light of such a decomposition, the spacetime gauge field $\cA_{\mu\nu}^{\rho}$ can be devided into the corresponding two parts:
\be \label{STGF}
 \cA_{\mu\nu}^{\rho}   \equiv   \fGa_{\mu\nu}^{\rho}  +  \fA_{\mu\nu}^{\rho} ,
\ee
with $\fGa_{\mu\nu}^{\rho}$ satisfying the following gauge covariant relation: 
\be \label{STGR}
 \p_{\mu} \fA_{\nu}^{\;\; a} +  \fOm_{\mu\1 b}^{a} \fA_{\nu}^{\;\; b} - \fGa_{\mu\nu}^{\rho} \fA_{\rho}^{\;\; a} = 0 , 
\ee
and $\fA_{\mu\nu}^{\rho}$ being defined as follows:
\be 
 \fA_{\mu\nu}^{\rho} & \equiv &  \hfA_{a}^{\;\; \rho} \fA_{\mu b}^{a} \fA_{\nu}^{\;\; b} .
 \ee
From the relation Eq.(\ref{STGR}) and the explicit form of $\fOm_{\mu\1 b}^{a}$ presented in Eq.(\ref{SGGF}), we arrive at the following form for $\fGa_{\mu\nu}^{\rho}$: 
 \be
 \fGa_{\mu\nu}^{\rho}  & \equiv &  \hfA_{a}^{\;\; \rho} \p_{\mu} \fA_{\nu}^{\;\;a} +  \hfA_{a}^{\;\; \rho}   \fOm_{\mu\1 b}^{a} \fA_{\nu}^{\;\;b} , \nn \\
& = & \frac{1}{2}\hmH^{\rho\sigma} (\p_{\mu} \mH_{\sigma\nu} + \p_{\nu} \mH_{\sigma\mu} - \p_{\sigma}\mH_{\mu\nu} ) =\fGa_{\nu\mu}^{\rho} , \nn
\ee
which comes to the identity given in Eq.(\ref{SGMF}). 

The gravigauge field $\fA_{\mu}^{\;\;a}$ may be regarded as {\it Goldstone-like boson field} which transforms as bicovariant vector field under both spin gauge group and Lorentz group transformations. In light of the spacetime gauge field, the spin gauge symmetry SP(1,3) is transmuted into a hidden gauge symmetry. The symmetric tensor fields $\mH_{\mu\nu}$ and $\hmH^{\mu\nu}$ as the product of two gravigauge fields by contracting the vector indices in locally flat gravigauge spacetime are considered as composite fields with both hidden spin gauge symmetry and scaling gauge symmetry, which are referred to as Goldstone-like dual {\it gravimetric fields} geometrically. 

The symmetric gauge field $\fGa_{\mu\nu}^{\rho}=\fGa_{\nu\mu}^{\rho}$ is completely determined by the dual Goldstone-like gravimetric fields $\mH_{\mu\nu}$ and $\hmH^{\mu\nu}$, we may refer to $\fGa_{\mu\nu}^{\rho}$ as scaling gauge invariant {\it spacetime gravimetric-gauge field}. For convenience of mention, the gauge field $\fA_{\mu\nu}^{\rho} $ is called as {\it spacetime covariant-gauge field} . In terms of the spacetime gauge field, we obtain the corresponding field strength: 
\be
 & & \bs{\cF}_{\mu\nu\sigma}^{\rho}  = \nabla_{\mu} \cA_{\nu\sigma}^{\rho} - \nabla_{\nu} \cA_{\mu\sigma}^{\rho} + \cA_{\mu\lambda}^{\rho} \cA_{\nu\sigma}^{\lambda}   - \cA_{\nu\lambda}^{\rho} \cA_{\mu\sigma}^{\lambda} , \nn \\
& & \nabla_{\nu} \cA_{\nu\sigma}^{\rho} = \p_{\mu} \cA_{\nu\sigma}^{\rho} - \fGa_{\mu\sigma}^{\lambda} \cA_{\nu\lambda}^{\rho} + \fGa_{\mu\lambda}^{\rho}  \cA_{\nu\sigma}^{\lambda} ,
\ee
which can be rewritten into the following two parts in correspondence to the decomposition of spacetime gauge field shown in Eq.(\ref{STGF}):
\be \label{SGFS}
& & \bs{\cF}_{\mu\nu\sigma}^{\rho} \equiv \fR_{\mu\nu\sigma}^{\rho} + \fF_{\mu\nu\sigma}^{\rho}  , \nn \\
& & \fR_{\mu\nu\sigma}^{\;\rho} = \p_{\mu} \fGa_{\nu\sigma}^{\rho} - \p_{\nu} \fGa_{\mu\sigma}^{\rho}  + \fGa_{\mu\lambda}^{\rho} \fGa_{\nu\sigma}^{\lambda}  - \fGa_{\nu\lambda}^{\rho} \fGa_{\mu\sigma}^{\lambda}, \nn \\
& & \fF_{\mu\nu\sigma}^{\rho} = \nabla_{\mu} \fA_{\nu\sigma}^{\rho} - \nabla_{\nu} \fA_{\mu\sigma}^{\rho} + \fA_{\mu\lambda}^{\rho} \fA_{\nu\sigma}^{\lambda}   - \fA_{\nu\lambda}^{\rho} \fA_{\mu\sigma}^{\lambda}  ,\nn  \\
& & \nabla_{\mu} \fA_{\nu\sigma}^{\rho} = \p_{\mu} \fA_{\nu\sigma}^{\rho} - \fGa_{\mu\sigma}^{\lambda} \fA_{\nu\lambda}^{\rho} + \fGa_{\mu\lambda}^{\rho}  \fA_{\nu\sigma}^{\lambda} .
\ee
Such defined tensor fields $\fR_{\mu\nu\sigma}^{\;\rho}$ and $\fF_{\mu\nu\sigma}^{\;\rho}$ in the hidden gauge formalism are referred to as scaling gauge invariant {\it spacetime gravimetric-gauge field strength} and {\it spacetime covariant-gauge field strength}, respectively.

In terms of the dimensionless gravigauge field $\chi_{\mu}^{\;\; a}$ ($\chih_{a}^{\;\; \mu}$), we arrive at the following gauge covariant relation:
\be
\p_{\mu} \chi_{\nu}^{\;\; a} +  \cA_{\mu\1 b}^{a} \chi_{\nu}^{\;\; b} - \mA_{\mu\nu}^{\rho} \chi_{\rho}^{\;\; a} = 0 .
\ee
Again decomposing the spin gauge field into two parts:
\be
\cA_{\mu\1 b }^{a} \equiv \mOm_{\mu\1 b}^{a} + \mA_{\mu\1 b}^{a} , \nn
\ee
with the spin gravigauge field $\mOm_{\mu\1 b}^{a}$ being the so-called {\it spin connection}, which is completely determined by the dimensionless gravigauge field $\chi_{\mu}^{\;\; a}$ ($\chih_{a}^{\;\; \mu}$) as presented in Eqs.(\ref{SGFDC1}) and (\ref{SGGF1}). In this case, the spacetime gauge field $\mA_{\mu\nu}^{\rho}$  can also be expressed into the following form: 
\be \label{STGF2}
 \mA_{\mu\nu}^{\rho}   \equiv   \Gamma_{\mu\nu}^{\rho}  +  A_{\mu\nu}^{\rho} ,
\ee
where the spacetime gravimetric-gauge field $\Gamma_{\mu\nu}^{\rho}$ satisfies the well-known identity:
\be \label{STGR2}
 \p_{\mu} \chi_{\nu}^{\;\; a} +  \mOm_{\mu\1 b}^{a} \chi_{\nu}^{\;\; b} - \Gamma_{\mu\nu}^{\rho} \chi_{\rho}^{\;\; a} = 0 , 
\ee
with the explicit form:
\be \label{SGMF2}
\Gamma_{\mu\nu}^{\rho}(x)  = \frac{1}{2}\chih^{\rho\lambda} (\p_{\mu} \chi_{\lambda\nu} + \p_{\nu} \chi_{\lambda\mu} - \p_{\lambda}\chi_{\mu\nu} ) =\Gamma_{\nu\mu}^{\rho}, \nn
 \ee
which is called as {\it affine connection} or {\it Christoffel symbol} represented in Eq.(\ref{SGMF}). The spacetime covariant-gauge field $A_{\mu\nu}^{\rho}$ is defined as follows:
\be
 A_{\mu\nu}^{\rho} \equiv \chih_{a}^{\;\; \rho} \mA_{\mu b}^{a} \chi_{\nu}^{\;\; b} .
\ee

It is noticed that by directly taking the symmetric feature of $\Gamma_{\mu\nu}^{\rho}$, i.e., $\Gamma_{\mu\nu}^{\rho}=\Gamma_{\nu\mu}^{\rho}$, one can obtain from the relation in Eq.(\ref{STGR2}) the explicit form of spin gravigauge field $\mOm_{\mu\1 b}^{a}$ in Eq.(\ref{SGGF1}) by simply solving the following equation:
\be
 \p_{\mu} \chi_{\nu}^{\;\; a}  -  \p_{\nu} \chi_{\mu}^{\;\; a} +  \mOm_{\mu\1 b}^{a} \chi_{\nu}^{\;\; b} -  \mOm_{\nu\1 b}^{a} \chi_{\mu}^{\;\; b} = 0 .
\ee

The corresponding spacetime gauge field strength can be written into the following form: 
\be \label{SGFS}
& & \cF_{\mu\nu\sigma}^{\rho} \equiv R_{\mu\nu\sigma}^{\rho} + F_{\mu\nu\sigma}^{\rho}  , \nn \\
& & R_{\mu\nu\sigma}^{\;\rho} = \p_{\mu} \Gamma_{\nu\sigma}^{\rho} - \p_{\nu} \Gamma_{\mu\sigma}^{\rho}  + \Gamma_{\mu\lambda}^{\rho} \Gamma_{\nu\sigma}^{\lambda}  - \Gamma_{\nu\lambda}^{\rho} \Gamma_{\mu\sigma}^{\lambda}, \nn \\
& & F_{\mu\nu\sigma}^{\rho} = \nabla_{\mu} A_{\nu\sigma}^{\rho} - \nabla_{\nu} A_{\mu\sigma}^{\rho} + A_{\mu\lambda}^{\rho} A_{\nu\sigma}^{\lambda}   - A_{\nu\lambda}^{\rho} A_{\mu\sigma}^{\lambda}  ,\nn  \\
& & \nabla_{\mu} A_{\nu\sigma}^{\rho} = \p_{\mu} A_{\nu\sigma}^{\rho} - \Gamma_{\mu\sigma}^{\lambda} A_{\nu\lambda}^{\rho} + \Gamma_{\mu\lambda}^{\rho}  A_{\nu\sigma}^{\lambda} .
\ee
where $R_{\mu\nu\sigma}^{\rho}$ is called as Riemann curvature tensor presented in Eq.(\ref{RMC3}). In general, we can verify the following relations:
\be \label{GFSR}
& & \cF_{\mu\nu}^{\rho\sigma} = \cF_{\mu\nu\sigma'}^{\rho}\chih^{\sigma'\sigma} = \cF_{\mu\nu\rho'\sigma'}\chih^{\rho'\rho}\chih^{\sigma'\sigma} =  \cF_{\mu\nu}^{ab} \chih_{a}^{\; \rho} \chih_{b}^{\; \sigma} =  - \cF_{\mu\nu}^{\sigma\rho},  \nn \\
& & R_{\mu\nu}^{\rho\sigma} = R_{\mu\nu\sigma'}^{\rho}\chih^{\sigma'\sigma} = R_{\mu\nu\rho'\sigma'}\chih^{\rho'\rho}\chih^{\sigma'\sigma} =  R_{\mu\nu}^{ab} \chih_{a}^{\; \rho} \chih_{b}^{\; \sigma} =  - R_{\mu\nu}^{\sigma\rho},  \nn \\
& & F_{\mu\nu}^{\rho\sigma} = F_{\mu\nu\sigma'}^{\rho}\chih^{\sigma'\sigma} = F_{\mu\nu\rho'\sigma'}\chih^{\rho'\rho}\chih^{\sigma'\sigma} =  F_{\mu\nu}^{ab} \chih_{a}^{\; \rho} \chih_{b}^{\; \sigma} =  - F_{\mu\nu}^{\sigma\rho},  \nn \\
& & \cF_{\mu\nu}^{ab} =  \cF_{\mu\nu}^{\rho\sigma} \chi_{\rho}^{\; a} \chi_{\sigma}^{\; b}= \chih^{a\rho} \chih^{b\sigma}  \cF_{\mu\nu\rho\sigma}  
\ee
by using the following general identities of covariant derivative: 
\be \label{GID}
& & \nabla_{\mu}\chi_{\nu}^{\; a} = \p_{\mu} \chi_{\nu}^{\; a} + \mOm_{\mu b}^{a} \chi_{\nu}^{\; b}  
- \Gamma_{\mu\nu}^{\rho} \chi_{\rho}^{\; a} =0 , \nn \\
& & \nabla_{\mu}\chih_{b}^{\; \rho} = \p_{\mu} \chih_{b}^{\; \nu} - \mOm_{\mu b}^{a} \chih_{a}^{\; \rho}  
+ \Gamma_{\mu\nu}^{\rho} \chih_{a}^{\; \nu} =0 , \nn \\
& & \nabla_{\mu} \chi_{\rho\sigma} = \p_{\mu} \chi_{\rho\sigma}  - \Gamma_{\mu\rho}^{\nu} \chi_{\nu\sigma} - \Gamma_{\mu\sigma}^{\nu} \chi_{\rho\nu} = 0 , \nn \\
& & \nabla_{\mu} \chih^{\rho\sigma} = \p_{\mu} \chih^{\rho\sigma}  + \Gamma_{\mu\nu}^{\rho} \chih^{\nu\sigma} + \Gamma_{\mu\nu}^{\sigma} \chih^{\rho\nu} = 0 , 
\ee

By taking the above identities and relations, we can reformulate the actions in Eqs.(\ref{HSGaction}) and (\ref{HSGaction2}) into the following alternative form in hidden gauge formalism:
\be
\label{HGFaction}
\cS_D & \equiv & \int [d^4x] \fkA(x)\, \fkL \nn \\
& \equiv & \int [d^4x] \fkA(x)\1 \{ (\bar{\Psi}_{-} \hat{\Sigma}_{-}^{\mu} i\cD_{\mu} \Psi_{-}  - \frac{m}{\Mka} \bar{\Psi}_{-}\Sigma_{-}^{6}\Psi_{-}  + H.c. )\nn \\
& - &  \frac{1}{4}  \hmH^{\mu\mu'}\hmH^{\nu\nu'} F_{\mu\nu} F_{\mu'\nu'}  - \frac{1}{4}  \hmH^{\mu\mu'}\hmH^{\nu\nu'} \bs{\cF}_{\mu\nu}^{\; \rho\sigma} \bs{\cF}_{\mu'\nu' \rho\sigma} \nn \\
& + & \frac{1}{4} \frac{m_G^2}{\Mka^2}  \bmH_{\rho\rho' }^{\mu\nu \mu'\nu'} \bs{\cF}_{\mu\nu}^{\; \rho} \bs{\cF}_{\mu'\nu'}^{\rho'} + \frac{1}{4} g_G^{-2} \tmH_{\rho\rho'}^{\mu\nu \mu'\nu'} \bs{\mF}_{\mu\nu}^{\rho}\bs{\mF}_{\mu'\nu'}^{\rho'} \} ,
\ee
where we have used the following definitions in the hidden gauge formalism:
\be \label{CDTensors3}
& & \bs{\cF}_{\mu\nu}^{\; \rho}  \equiv \hfA_{a}^{\; \rho} \cF_{\mu\nu}^{\; a} , \quad  \bs{\mF}_{\mu\nu}^{\; \rho}  \equiv \hfA_{a}^{\; \rho} \mF_{\mu\nu}^{\; a} , \nn \\
& & \bmH_{\rho'\rho}^{\mu\nu \mu'\nu'} \equiv \fA_{\rho}^{\; a} \fA_{\rho'}^{\; a'} \bmH_{aa'}^{\mu\nu \mu'\nu'} , \quad  \tmH_{\rho\rho'}^{\mu\nu \mu'\nu'} \equiv \fA_{\rho}^{\; a} \fA_{\rho'}^{\; a'} \tmH_{aa'}^{\mu\nu \mu'\nu'} , 
\ee
with $\bmH_{aa'}^{\mu\nu \mu'\nu'}$ and $\tmH_{aa'}^{\mu\nu \mu'\nu'}$ defined in Eq.(\ref{CDTensors}), and 
\be
& & i\cD_{\mu} \equiv i\p_{\mu} + (\fOm_{[\mu\rho\sigma]}+ g_G  \cA_{[\mu\rho\sigma]})\frac{1}{2}\hfSi^{\rho\sigma}, \nn \\
& & \cA_{\mu\rho\sigma}  \equiv  \mH_{\rho\rho'} \cA_{\mu\sigma}^{\rho'}, \quad \fOm_{\mu\rho\sigma}  \equiv  \fA_{\rho}^{\;\; a} \fOm_{\mu ab} \fA_{\sigma}^{\;\; b}, \nn \\
& & \cA_{[\mu\rho\sigma]} \equiv \frac{1}{3} (\cA_{\mu\rho\sigma} + \cA_{\rho\sigma\mu}  + \cA_{\sigma\mu\rho} ) , \nn \\
& & \fOm_{[\mu\rho\sigma]} \equiv  \frac{1}{3} (\fOm_{\mu\rho\sigma} + \fOm_{\rho\sigma\mu}  + \fOm_{\sigma\mu\rho} ) \nn \\
& & \hfSi_{-}^{\mu} \equiv \frac{1}{2}\hat{\fGa}^{\mu} \vGa_{-} , \quad \hfGa^{\mu} \equiv \hfA_{a}^{\;\;\mu} \vGa^{a}, \nn \\
& &  \hfSi^{\mu\nu} \equiv \frac{i}{4} [\hfGa^{\mu}, \hfGa^{\nu} ]  = \hfA_{a}^{\;\;\mu} \hfA_{b}^{\;\;\nu} \Sigma^{a b} .  
\ee
It is noted that the {\it spacetime gravigauge field} $\fOm_{[\mu\rho\sigma]}$ actually reflects the gravitational origin of gauge symmetry. The $\Gamma$-matrices $\hfGa^{\mu}$ are regarded as {\it gravigauge-dressed $\Gamma$-matrices} as they are no longer constant matrices. The matrices $\hfSi^{\mu\nu}$ and $\hfSi_{-}^{\mu}$ provide the {\it gravigauge-dressed inhomogeneous spin group generators}. 
  
By choosing the scaling gauge fixing condition to be in Einstein basis, the above action can be expressed as follows:
\be
\label{HGFactionEB}
\cS_D & \equiv & \int [d^4x] \chi(x) \1 \cL^{(e)} \nn \\
& \equiv &  \int [d^4x] \chi(x) \; \{\frac{1}{2} ( \bar{\psi} \hat{\gamma}^{\mu} i\cD_{\mu} \psi + H.c.) - m \bar{\psi}\psi \nn \\
& - &  \frac{1}{4} \chih^{\mu\mu'}\chih^{\nu\nu'} F_{\mu\nu} F_{\mu'\nu'}- \frac{1}{4} \chih^{\mu\mu'}\chih^{\nu\nu'} \cF_{\mu\nu}^{\; \rho\sigma} \cF_{\mu'\nu' \rho\sigma} + \frac{1}{16\pi G_N}\, R  \nn \\
& + &    \frac{1}{4}(\alpha_G-\frac{1}{2}\alpha_W) m_G^2 g_G^2 \chih^{\mu\mu'} \chih^{\rho\rho'} \chih^{\sigma\sigma'}(\mA_{\mu\rho\sigma} - \Gamma_{\mu\rho\sigma}/g_G ) (\mA_{\mu'\rho'\sigma'} -\Gamma_{\mu'\rho'\sigma'}/g_G ) \nn \\
& + & \frac{3}{4}( \alpha_G+\frac{3}{2}\alpha_W) m_G^2 g_G^2 \chih^{\mu\mu'} \chih^{\rho\rho'} \chih^{\sigma\sigma'} \mA_{[\mu\rho\sigma]} \mA_{[\mu'\rho'\sigma']} ,
\ee
with the definitions:
\be
& & i\cD_{\mu} \equiv i\p_{\mu} + (\mOm_{[\mu\rho\sigma]} + g_G \mA_{[\mu\rho\sigma]})\frac{1}{2}\hat{\Sigma}^{\rho\sigma}, \nn \\
& &  \mA_{[\mu\rho\sigma]} \equiv \frac{1}{3} (\mA_{\mu\rho\sigma} + \mA_{\rho\sigma\mu}  + \mA_{\sigma\mu\rho} ) , \nn \\
& & \mOm_{[\mu\rho\sigma]} \equiv  \frac{1}{3} (\mOm_{\mu\rho\sigma} + \mOm_{\rho\sigma\mu}  + \mOm_{\sigma\mu\rho} ) \nn \\
& & \quad \quad \;\;\; \equiv \frac{1}{3} ( \p_{\mu} \chi_{\rho}^{\;\;c}\, \chi_{\sigma c} + \p_{\rho} \chi_{\sigma}^{\;\; c}\, \chi_{\mu c}  + \p_{\sigma} \chi_{\mu}^{\;\; c}\, \chi_{\rho c}) , \nn \\
& &  \hat{\Sigma}^{\mu\nu} \equiv \frac{i}{4} [\hat{\gamma}^{\mu}, \hat{\gamma}^{\nu} ]  = \chih_{a}^{\;\;\mu} \chih_{b}^{\;\;\nu} \Sigma^{a b} , \quad \hat{\gamma}^{\mu} \equiv \chih_{a}^{\; \mu}\gamma^{a} . 
\ee
  
It becomes manifest that the actions presented in Eqs.(\ref{HGFaction}) and (\ref{HGFactionEB}) bring on an emergent general linear group symmetry in coordinate spacetime:
\be
G_S = \mbox{GL(}1,3,\mbox{R)} ,
\ee
which lays the foundation for Einstein theory of GR and governs the gravitational interaction in curved Riemannian spacetime. It indicates that such a group symmetry actually appears as a hidden local symmetry in the action built based on the scaling and gauge invariance principle. In general, the action in Eq.(\ref{HGFaction}) reformulated in the hidden gauge formalism possesses the following maximal joint symmetry:
\be \label{JGLS}
G_S = \mbox{GL(}1,3,\mbox{R)} \Join \mbox{WS(}1, 3),
\ee
which extends naturally the global Poincar\'e group symmetry PO(1,3) in globally flat Minkowski spacetime to be a local group symmetry GL(1,3,R) in curved Riemannian spacetime with emergence of Riemann geometry. 

It is noted that the inhomogeneous spin gauge symmetry WS(1,3) still holds for the chirality-based Dirac spinor field in the action Eq.(\ref{HGFaction}) though it appears as a hidden spin gauge symmetry. This can be seen from the coupling of totally antisymmetric spacetime gravigauge field $\mOm_{[\mu\rho\sigma]}$ which is characterized by the gravigauge field $\chi_{\mu}^{\;\; a}$ ($\chih_{a}^{\;\; \mu}$) rather than gravimetric field $\chi_{\mu\nu}$ ($\chih^{\mu\nu}$). It is interesting to notice that the spacetime gravimetric-gauge field $\Gamma_{\mu\rho}^{\sigma}$ actually decouples from the Dirac spinor field due to its symmetric property $\Gamma_{\mu\rho}^{\sigma}=\Gamma_{\rho\mu}^{\sigma}$ and the hermiticity requirement of the action. Nevertheless, in the purely bosonic interactions of the action, the spacetime gauge interaction is described by the antisymmetric spacetime gauge field $\mA_{\mu}^{\rho\sigma}= -\mA_{\mu}^{\sigma\rho}$ together with the symmetric gravimetric field $\chi_{\mu\nu}$ which characterizes the symmetric spacetime gravimetric-gauge field $\Gamma_{\mu\nu}^{\rho} = \Gamma_{\nu\mu}^{\rho} $ as affine connection/Christoffel symbols. 

It is observed that the symmetric Goldstone-like gravimetric field $\chi_{\mu\nu}$ concerns $10 $ degrees of freedom, while the basic gravitational field characterized by the Goldstone-like gravigauge field $\chi_{\mu}^{\;\; a}$ contains $16$ degrees of freedom, which involves additional six degrees of freedom. Such an extra degrees of freedom reflect exactly the equivalence classes of spin gauge symmetry SP(1,3), which corresponds to the six group parameters of SP(1,3) gauge transformation. Such an observation motivates us to eliminate the redundant degrees of freedom caused from the spin gauge symmetry by simply making a gauge prescription of spin gauge symmetry SP(1,3). 

Unlike the usual internal gauge symmetry, the spin gauge symmetry gets a gravitational origin characterized by the gravigauge field $\chi_{\mu a}(x)$, a simple gauge prescription should be realized by taking an appropriate spin gauge transformation. Let us choose such a spin gauge transformation $\tilde{\Lambda}_{\;\, a}^{b}(x)$ that transmutes the Goldstone-like gravigauge field into the following symmetric one: 
\be
& & \chi_{\mu a }(x) \to  \tchi_{\mu a}(x) = \chi_{\mu b}(x) \tilde{\Lambda}^{b}_{\; \, a}(x)  
=  \tchi_{a\mu}(x) , 
\ee 
which presents a natural gauge fixing condition of spin gauge symmetry SP(1,3). We may refer to such a gauge prescription as {\it flowing unitary gauge}, which holds locally for a coordinate system at point to point in spacetime. In such a flowing unitary gauge, the Goldstone-like symmetric gravigauge field $\tchi_{\mu a}(x)$ gets exactly the same degrees of freedom as the gravimetric field $\chi_{\mu\nu}(x)$ via the hidden gauge symmetry, i.e.:
\be
\chi_{\mu\nu}(x) = \chi_{\mu a}(x) \eta^{ab} \chi_{\nu b}(x) \equiv \tchi_{\mu a}(x) \eta^{ab} \tchi_{\nu b}(x)  \equiv (\tchi_{\mu a}(x) )^2 .
\ee

When fixing gauge to be in the flowing unitary gauge, the independent degrees of freedom for the spacetime gauge interactions are represented by the symmetric Goldstone-like gravigauge field $\tchi_{\mu a}(x)=\tchi_{a \mu}(x)$ (or symmetric Goldstone-like gravimetric field $\chi_{\mu\nu}(x) \equiv (\tchi_{\mu a}(x) )^2$) and the spacetime gauge field $\mA_{\mu}^{\rho\sigma}(x)$. Meanwhile, the total independent degrees of freedom in the theory should remain unchanged since the extra degrees of freedom appearing in the Goldstone-like gravigauge field $\chi_{\mu a}(x)$ are actually absorbed into the antisymmetric spacetime gauge field $\mA_{\mu}^{\rho\sigma}(x)$ which behaves as a massive-like gauge field as shown in Eq.(\ref{HGFactionEB}). It can be checked that in such a flowing unitary gauge, the action can still possess an associated global symmetry:
\be \label{GAS}
G_S = \mbox{SO}(1,3) \adjoin \mbox{SP}(1, 3) ,
\ee
which transforms the gauging fixing symmetric gravigauge field $\tchi_{\mu a}(x)$ to remain symmetric, i.e., $\tchi'_{\mu a}(x')=\tchi'_{a \mu}(x')$.

We now come to the conclusion that {\it the laws of nature should be independent of the choice of coordinate systems}, which arises from the intriguing observation that when the fundamental symmetry of basic fields in Hilbert space is gauged to be local symmetry by following along the scaling and gauge invariance principle, the fundamental symmetry of coordinates in Minkowski spacetime automatically obeys the general covariance principle of coordinate system with the emergence of general linear group symmetry GL(1,3,R) and the genesis of Riemann geometry in curved Riemannian spacetime, which indicates that the scaling and gauge invariance principle plays an essential role not only to the fundamental symmetry of basic fields in Hilbert space but also to the fundamental symmetry in coordinate spacetime. In fact, such a feature reflects intrinsically a {\it gauge-geometry duality} in GQFT.


\subsection{Finiteness and renormalizability of gravitational quantum field theory }

Before studying the gravidynamics and spinodynamics as well as electrodynamics, it is useful to present a general analysis and discussion on the finiteness and renormalizability of GQFT though the investigation on the quantum contributions to the dynamics of basic fields is beyond the scope of present paper. 
Actually, the finiteness and renormalizability have been long-standing issues involved in all QFTs established based on the special relativity and quantum mechanics. In the GQFT, there gets the universal coupling with the dual gravigauge field $\chih_{a}^{\mu}(x)$ or dual gravimetric field $\chih^{\mu\nu}(x)$ as gravitational interactions, which is usually thought to bring on the nonlinear nature of the theory and result in nonrenormalizable high dimensionful interaction terms which leads to the non-renormalizability of the theory in the usual sense of renormalizability, it means that the divergencies of quantum loop effects cannot be absorbed into the basic fields and coupling constants of the theory. 

To get better understanding on the finiteness and renormalizability of GQFT, let us first briefly look through the issues on the finiteness and renormalizability of QFT. QFT has provided a very successful description on the microscopic world, especially it has been well tested in applying to the SM which was built to describe the three basic forces of nature governed by the gauge symmetries $SU(3)_c\times SU(2)_L\times U(1)_Y$. Nevertheless, in the perturbative treatment of QFT, it still meets the ultraviolet (UV) divergence problem due to infinite Feynman integrals involved in the closed loops of virtual particles, which may destroy the basic symmetries of original theory. Mathematically, the divergence is caused from the integral region where all virtual particles will get infinite large energy momentum. Physically, it arises from the very short wavelength/high frequency fluctuations of basic fields in the path integral. To deal with the divergences, the so-called renormalization scheme is developed to absorb all divergences into basic fields and coupling constants in the theory through their redefinitions, which is referred to as the renormalizability of the theory. 

The theoretical consistency of the SM mainly relies on its renormalizability within the framework of QFT. Nevertheless, such a renormalization scheme does not satisfy to all physicists, like Dirac and Feynman, which may be seen from their criticisms\cite{PDirac,HKragh,RFeynman}. 

Dirac has commented that ``Most physicists are very satisfied with the situation. They say: `Quantum electrodynamics (QED) is a good theory and we do not have to worry about it any more.' I must say that I am very dissatisfied with the situation, because this so-called `good theory' does involve neglecting infinities which appear in its equations, neglecting them in an arbitrary way. This is just not sensible mathematics. Sensible mathematics involves neglecting a quantity when it is small, not neglecting it just because it is infinitely great and you do not want it."  

Feynman has also remarked that  ``The shell game that we play $\dots$ is technically called `renormalization'. But no matter how clever the word, it is still what I would call a dippy process! Having to resort to such hocus-pocus has prevented us from proving that the theory of quantum electrodynamics (QED) is mathematically self-consistent. It's surprising that the theory still hasn't been proved self-consistent one way or the other by now; I suspect that renormalization is not mathematically legitimate."

Therefore, QFT becomes well-defined only when it can be regularized properly to avoid the infinities through appropriate regularization method. Namely, such a renormalization scheme should eliminate divergences and get only finite quantities. On the other hand, the development of renormalization group technique has inspired us to figure out a well-defined QFT with the presence of physically meaningful energy scale. Nevertheless, the usual regularization schemes always bear some limitations for realizing a satisfied description on QFT. To achieve an infinity-free and symmetry-preserving regularization scheme, we have developed a novel regularization method which is the so-called loop regularization (LORE) method\cite{LORE1,LORE2}. Such a LORE method has turned out to be a consistent regularization scheme which enables us to introduce intrinsically the physically meaningful energy scale and avoid infinities without spoiling basic symmetries of original theory. For a detailed description on the LORE method, it is referred to the review article\cite{LORE3}.

The key concept in LORE method is the so-called irreducible loop integrals (ILIs) which are evaluated from the Feynman integrals by using Feynman parametrization and ultraviolet divergence-preserving (UVDP) parametrization\cite{LORE1,LORE2}. For illustration on the ILIs, let us represent the one loop calculation of Feynman diagrams, where all Feynman integrals of the one-particle irreducible graphs can be expressed into the following sets of loop integrals by adopting the Feynman parametrization method:
    \be
  & & I_{-2\alpha} = \int \frac{d^4 k}{(2\pi)^4}\ \frac{1}{(k^2 - {\cal M}^2)^{2+\alpha}} , \nn \\
  & & I_{-2\alpha\ \mu\nu} = \int \frac{d^4 k}{(2\pi)^4}\
    \frac{k_{\mu}k_{\nu}}{(k^2 - {\cal M}^2)^{3 + \alpha} } , \nn \\
  & & I_{-2\alpha\ \mu\nu\rho\sigma} = \int \frac{d^4 k}{(2\pi)^4}\
    \frac{k_{\mu}k_{\nu} k_{\rho}k_{\sigma} }{(k^2 - {\cal M}^2)^{4+ \alpha} },
    \ee
with $\alpha =-1, 0, 1, \cdots$. Where the subscript ($-2\alpha$) labels the power counting dimension of energy momentum in the loop integrals. For the cases with $\alpha = -1$ and $\alpha = 0$, we arrive at the scalar and tensor type ILIs corresponding to the quadratic divergent integrals ($I_2$, $I_{2 \mu\nu} \cdots $) and logarithmic divergent integrals ($I_0$, $I_{0 \mu\nu} \cdots$). ${\cal M}^2$ is the mass factor resulted as a function of Feynman parameters and external momenta $p_i$, ${\cal M}^2 = {\cal M}^2 (m_1^2, p_1^2, \cdots)$. For the high loop overlapping Feynman integrals, the ILIs are evaluated to get rid of, in the denominator,  the overlapping momentum factors $(k_i-k_j + p_{ij})^2$ $(i\ne j)$ which appear in the original overlapping Feynman integrals of loop momenta $k_i$ ($i=1,2,\cdots $), and eliminate the scalar momentum factors $k^2$ in the numerator. 

The essential feature demanded for a consistent regularization scheme is that it must preserve the symmetry principles of original theory, such as gauge invariance, Lorentz invariance and translational invariance, and meanwhile avoid the infinities of loop integrals with maintaining the initial divergent behavior and structure of original theory, such as quadratic and logarithmic divergent behaviors and structures. It can generally be proved that to preserve the gauge invariance principle in QFT, the regularized tensor-type and scalar-type ILIs should satisfy the following necessary and sufficient conditions:
\begin{eqnarray}\label{CC}
& & I_{2\mu\nu}^R = \frac{1}{2} g_{\mu\nu}\ I_2^R, \quad
I_{2\mu\nu\rho\sigma }^R = \frac{1}{8} (g_{\mu\nu}g_{\rho\sigma} +
g_{\mu\rho}g_{\nu\sigma} +
g_{\mu\sigma}g_{\rho\nu})\ I_2^R  , \nonumber \\
& & I_{0\mu\nu}^R = \frac{1}{4} g_{\mu\nu} \ I_0^R, \quad
I_{0\mu\nu\rho\sigma }^R = \frac{1}{24} (g_{\mu\nu}g_{\rho\sigma} +
g_{\mu\rho}g_{\nu\sigma} + g_{\mu\sigma}g_{\rho\nu})\ I_0^R ,
\end{eqnarray}
which is referred to as the gauge invariance consistency conditions\cite{LORE1,LORE2,LORE3}. 

To achieve such consistency conditions, a simple regularization prescription in LORE method has consistently be realized as follows: firstly rotating the momentum to the Euclidean space via a Wick rotation and then replacing the loop integrating variable $k^2$ and loop integrating measure $\int{d^4k}$ of the ILIs through the corresponding regularized ones $[k^2]_l$ and $\int[d^4k]_l$, i.e.:
\begin{eqnarray}
 k^2 & \rightarrow & [k^2]_l \equiv k^2+M^2_l\ ,  \\
 \int{d^4k}\  {\cal F}(k^2)  & \rightarrow & \int[d^4k]_l \   {\cal F}(k^2 +M^2_l )  \nn \\
& \equiv & \int{d^4k}  \lim_{N, M_i^2}\sum_{l=0}^{N}c_l^N \  {\cal F}(k^2 +M^2_l )  \nn \\
&  = & \lim_{N, M_i^2}\sum_{l=0}^{N}c_l^N \  \int{d^4k}  \  {\cal F}(k^2 +M^2_l ) .
\end{eqnarray}
Where $M_l^2$ ($ l= 0,1,\ \cdots $) are regarded as regulator masses and ${\cal F}(k^2)$ is considered to be any integration function. The notation $\lim_{N, M_i^2}$ denotes the limiting case $\lim_{N\to \infty, M_i^2\rightarrow \infty}$ ($i=1,2,\cdots, N$). 

The key point in LORE is to choose the appropriate coefficients $c_l^N$, so that any loop divergence with the power counting dimension larger than or equal to the space-time dimension will vanish. From that, we come to the general conditions for the coefficients $c_l^N$ as follows:
\begin{eqnarray} \label{CC}
 \int d^4 k \  \lim_{N, M_i^2}\sum_{l=0}^{N}c_l^N  (k^2 + M_l^2)^n = 0, \quad  (n= 0, 1, \cdots),
\end{eqnarray}
which brings on the following relations for the regulator masses:
\begin{eqnarray} \label{RMC}
\sum_{l=0}^{N}c_l^N  (M_l^2)^n = 0, \quad  (n= 0, 1, \cdots) ,
\end{eqnarray}
where we shall choose $M_0^2 \equiv \mu_s^2 \to 0$ and $c_0^N = 1$ as the initial conditions in order to recover the original integrals in the limits $M_i^2 \to \infty$ ($i=1,2,\cdots,N$ ) and $N\to \infty$.

To completely fix the coefficients $c_l^N$ and make them independent of the regulator masses,  it is simple to take the following string-mode regulators:
\begin{equation}
M_l^2=\mu_s^2+lM_R^2, \qquad l=0,1,2,\cdots ,
\end{equation}
which enables us to determine uniquely the coefficients $c_l^N$ from the general relations of regulator masses given in Eq.(\ref{RMC}) and obtain the following explicit solution:  
\begin{equation} \label{CLC}
 c_l^N=(-1)^l\frac{N!}{(N-l)!l!}
\end{equation}
which is known as the sign-associated Combinations. The factor $(-1)^l c_l^N$ may be regarded as the number of combinations of $N$ regulators taken $l$ at a time.

To be more explicit and concrete, when applying such a regularization prescription in the LORE method with the simple solution of the regulators to the quadratically and logarithmically divergent ILIs, the divergent integrations become well-defined and can safely be performed to obtain the regularized divergent ILIs $I_2^R$ and $I^R_{2\mu\nu}$ as well as $I_0^R$ and $I^R_{0\mu\nu}$\cite{LORE1,LORE2,LORE3}. We just represent them as follows:
\begin{eqnarray}
I_2^R &=& \frac{-i}{16\pi^2}\{M_c^2-\mu_M^2 - \mu_M^2[\ln\frac{M_c^2}{\mu_M^2}-\gamma_E+ \varepsilon (\frac{\mu_M^2}{M_c^2}) + \varepsilon^{\prime}(\frac{\mu_M^2}{M_c^2}) -1 ]\} , \nn \\
I_0^R &=& \frac{i}{16\pi^2}[\ln\frac{M_c^2}{\mu_M^2}-\gamma_E+  \varepsilon (\frac{\mu_M^2}{M_c^2})],  \nn \\
I^R_{2\mu\nu} & = & \frac{1}{2} g_{\mu\nu}\ I^R_2; \qquad I^R_{0\mu\nu} = \frac{1}{4} g_{\mu\nu} \ I^R_0 ,
\end{eqnarray}
with the definitions,
\be
& & \mu_M^2=\mu_s^2+M^2 , \nn \\
& & M_c^2\equiv \lim_{N,M_R\to \infty } \left( \frac{M_R^2}{\ln N} \right) , \nn \\
& & \varepsilon (x)  = \sum_{n=1}^{\infty} \frac{(-)^{n-1} }{n\ n!} x ^n = \int_0^x d\sigma \frac{1-e^{-\sigma}}{\sigma} .
\ee
Where the primary regulator mass $M_R$ is taken to be infinity so as to recover the original integrals and the regulator number $N$ is set to be infinity so as to make the regularized theory independent of the regularization prescription. $\gamma_E$ is the Euler constant ($\gamma_E=0.577215 \cdots $) and $\varepsilon(x)$ is a special function with $x=\mu_M^2/M_c^2$ and $\varepsilon^{\prime}(x) \equiv \partial_x \varepsilon (x)$. It is noted that $M_c$ and $\mu_s$ are regarded as intrinsic mass scales. In general, the mass scale $M_c$ can always be made to be finite as long as the ratio $M_R^2/\ln N$ is kept as a finite quantity when the primary regulator mass $M_R$ and regulator number $N$ are approaching to infinity. 

It is seen that the LORE method truly brings on an infinity-free and symmetry-preserving regularization method with the presence of two intrinsic mass scales and the satisifaction of symmetry invariance consistency conditions. Specifically, $M_c$ provides an ultraviolet (UV) `cutoff', and $\mu_s $ sets an infrared (IR) `cutoff' for massless case $M^2 =0$.  It is clear that the LORE method does maintain the original divergent structure of integrals when taking $M_c\to \infty$ and $\mu_s\to 0$. In comparison to  the dimensional regularization (DR) scheme,  although the DR scheme can also result in the same symmetry invariance consistency conditions to preserve gauge invariance, while its regularized quadratic divergent ILIs are actually suppressed to be a logarithmic divergence multiplying by the mass scale ${\cal M}^2$ or vanishes ($I^{R}_2=0$) for the massless case ${\cal M}^2 =0$. This is because the dimensional regularization scheme modifies the original theory by making an analytical extension for the space-time dimensions of original theory, which makes the DR scheme to be limited in directly applying to theories which require to keep well-defined spacetime dimensions and maintain quadratic divergent structure, such as chiral type theories and supersymmetric theories as well as theories with dynamically spontaneous symmetry breaking.  

Furthermore, the LORE method becomes simple at one loop level and gets very applicable at high loop level for understanding the overlapping divergence structure of Feynman diagrams. Nevertheless, LORE method is not motivated for working out a much simpler regularization scheme but for figuring out an infinity-free and symmetry-preserving regularization method without modifying the original theory. It has been demonstrated in \cite{HW1,HLW} that the derivation of ILIs from Feynman integrals by adopting UVDP parametrization brings about the so-called Bjorken-Drell's circuit analogy between Feynman diagrams and electric circuits\cite{BD}, which enables us to pick up one-to-one correspondence between the divergences of UVDP parameters and the subdiagrams of Feynman diagrams.

Therefore, the LORE method does overcome the shortages and limitations appearing in some widely-used regularization schemes. In fact, the LORE method has been applied to make numerous interesting applications and carry out various consistent calculations\cite{Cui:2008uv, Cui:2008bk, Ma:2005md, Ma:2006yc, cui:2011, DW,HW, Tang:2008ah, Tang:2010cr,Tang2011}, such as:  how it preserves non-Abelian gauge symmetry~\cite{Cui:2008uv} and supersymmetry~\cite{Cui:2008bk}, how it results in consistent analyses on the chiral anomaly\cite{Ma:2005md} and radiatively induced Lorentz and CPT-violating Chern-Simons term in the extended QED\cite{Ma:2006yc} as well as QED trace anomaly\cite{cui:2011}, how it enables us to derive the gap equation for the dynamically generated spontaneous chiral symmetry breaking at low energy QCD and obtain reliably the dynamic quark masses and mass spectra of light scalar and pseudoscalar mesons in the chiral effective field theory\cite{DW}, and meanwhile understand the chiral symmetry restoration in chiral thermodynamic model\cite{HW}. Especially, the LORE method has been utilized to perform the consistent computations on the quantum gravitational contribution to gauge theories and obtain asymptotic free power-law running\cite{Tang:2008ah,Tang:2010cr,Tang2011}, and also to clarify the important issues raised in\cite{HGG} for the Higgs decay process $H\to \gamma\gamma$\cite{HTW}. It is intriguing to observe that the LORE method has also been adopted to achieve a dynamical spontaneous symmetry breaking mechanism for the electroweak symmetry in the SM and arrive at a sliding electroweak symmetry breaking scale to be around $\mu_{EW} \sim 760$ GeV\cite{BCW}. In general, the LORE method was proved to be applicable to theories in high dimensional spacetime\cite{Chapling1,Chapling2}.

With the above analyses, we should turn to make discussions on the finiteness and renormalizability of GQFT. The crucial point in LORE is the appearance of two intrinsic energy scales $M_c$ and $\mu_s$ which play the role as ultraviolet (UV) cut-off and infrared (IR) cut-off of loop integrals to avoid infinities in QFT without spoiling symmetries of original theory. These two energy scales should get physically meaningful as the characteristic energy scale $M_c$ and sliding energy scale $\mu_s$ indicated from the so-called folk's theorem\cite{SW1,SW2} addressed by Weinberg. Such a folk's theorem states that: any quantum theory that at sufficiently low energy and large distances looks Lorentz invariant and satisfies the cluster decomposition principle will also at sufficiently low energy look like a quantum field theory. It implies that there must exist a characteristic energy scale (CES) $M_c$ which makes the statement of sufficiently low energy to be reliable. It also means that such a CES $M_c$ can be either a fundamental mass scale for a basic theory or a dynamically generated energy scale for an effective theory. In the GQFT, the CES $M_c$ is considered to be a fundamental mass scale, which distinguishes to the QFT in Minkowski spacetime. The reason is that the presence of gravigauge field in GQFT will bring about the scaling gauge invariance of the action, which always allows us to set up a fundamental mass scale $M_{\kappa}$ by imposing a suitable scaling gauge fixing condition. Such a fundamental mass scale $M_{\kappa}$ is naturally identified to the CES $M_c$ for characterizing the ultraviolet behavior of the theory, i.e.:
\be
M_{\kappa} \simeq M_c ,
\ee
which makes any theory in the framework of GQFT to be infinity-free. Namely, there exists in principle no divergences in GQFT and any quantum contribution from high dimensionful interactions in GQFT should be finite and physically meaningful. 

In addition to the above consideration, let us follow along the key concept of biframe spacetime in GQFT to understand the finiteness and renormalizability of GQFT. In general, the biframe spacetime forms the fiber bundle structure with the base spacetime taken to be a globally flat Minkowski spacetime and the fiber bringing on a locally flat gravigauge spacetime. As the base spacetime acts as an inertial reference frame for describing the motion of basic fields, the energy momentum of basic fields in such a base spacetime can approach to be infinite large and small as it is considered to be a continue and differential spacetime to ensure the laws of energy momentum conservation, which becomes the main reason why the path integral in such a globally flat Minkowski spacetime gets divergences as the virtual particles can in principle have very short wavelength/high frequency fluctuations or infinite large energy momentum $k_{\mu} \to \infty$. In contrast, the fiber as a locally flat gravigauge spacetime emerges as a non-commutative geometry characterized by the field strength of gravigauge field and reflects the gravitational interaction, which functions as an interaction frame of basic fields. Therefore, the actual energy momentum of all basic fields in such a locally flat gravigauge spacetime is always associated to the dual gravigauge field and presented explicitly as follows:
\be
& & \mathcal{K}_{a} \equiv \chih_{a}^{\;\; \mu}(x) k_{\mu} \sim \int [dq] e^{-i q_{\mu}x^{\mu} } \hat{k}_{a}(q), \nn \\
& &  \hat{k}_{a} \equiv \chih_{a}^{\;\; \mu}(q) k_{\mu} ,
\ee 
which provides a physically meaningful energy momentum in GQFT. For convenience, we may refer to $\hat{k}_{a}$ (or $ \mathcal{K}_{a}$) as {\it gravigauging enengy momentum}. Unlike the globally flat Minkowski spacetime, the locally flat gravigauge spacetime as a matter field spacetime with noncommutative geometry is in principle no longer to appear as an infinite continue spacetime, so the gravigauging energy momentum $\hat{k}_{a}$ (or $\mathcal{K}_{a}$) should not approach to be an infinite large/small quantity even when the ordinary energy momentum $k_{\mu}$ in globally flat Minkowski spacetime goes to be infinite, i.e.:
\be
\hat{k}_{a} \neq \infty, \qquad k_{\mu} \to \infty ,
\ee
which may be comprehended as follows:  once at a short distance with infinite large energy momentum $k_{\mu} \to \infty$ in the base spacetime of coordinates, the gravigauge field $\chi_{\mu}^{\;\; a}$ will become infinite large $\chi_{\mu}^{\;\; a}\to \infty$ and its dual/inverse gravigauge field $\chih_{a}^{\;\; \mu}$ gets infinitely small $\chih_{a}^{\;\; \mu}\to 0$ , whereas the gravigauging energy momentum $\hat{k}_{a}$ (or $\mathcal{K}_{a}$) as a combination quantity $\hat{k}_{a} \equiv \chih_{a}^{\;\; \mu}(q) k_{\mu}$ (or $\chih_{a}^{\;\; \mu}(x) k_{\mu}$) remains to keep a physically meaningful finite energy momentum. Therefore, it appears in principle no infinite gravigauging energy momentum in GQFT. On the other hand, there should exist a potential horizon $L_{H}$ for the locally flat gravigauge spacetime as a matter field spacetime, which can provide a natural infrared cutoff scale, i.e.:
\be
 \mu_s\to 1/L_{H} , 
\ee
which allows us to remove the infrared divergence in GQFT.

From the above analyses and discussions, we are in the position to conclude that any divergence should in principle disappear in GQFT, which provides a resonable answer to the criticisms raised by Dirac and Feynman about the treatment of infinities and divergences via the usual renormalization schemes in QFT. Actually, the globally flat Minkowski spacetime as an inertial frame of free motion fields is unobservable, only the locally flat gravigauge spacetime as a gravitational frame of dynamic interaction fields becomes observable, which constitutes our observed universe. Moreover, by applying for the concept of renormalization group developed in refs.\cite{KGW,GML}, we should be able to study physical phenomena at any interesting energy scale by integrating out physical effects at higher energy scale, and define a finite renormalized theory at any interesting renormalization scale. In fact, the existence of both the characteristic energy scale (CES) $M_c\simeq M_{\kappa}$ and sliding energy scale(SES) $\mu_s\sim 1/L_{H}$ in GQFT allows us to choose the renormalization scale at any scale of interest between them. In this sense, GQFT becomes more natural to provide a profound theoretical framework for describing all basic forces.


\section{Gravitational relativistic quantum theory on Dirac spinor field}

The above various formalisms of the action enable us to derive the equations of motion for all basic fields based on the least action principle. These equations can be applied to investigate the dynamics of basic fields in the gravitational relativistic quantum theory. Let us first study the equation of motion for the chirality-based Dirac spinor $\Psi_{-}(x)$ in the presence of gravitational interaction and spin gauge interaction as well as electromagnetic interaction, which generalizes the usual relativistic quantum mechanics to gravitational relativistic quantum theory. 


\subsection{ Generalized Dirac equation in gravitational relativistic quantum theory}

Following along the least action principle, we are able to arrive at the following two types of the equations of motion for the chirality-based Dirac spinor field:
\be \label{EMDSF}
& & \hmH^{\mu\nu} \Sigma_{- a} \fA_{\mu}^{\;\; a} i \left(\cD _{\nu} -\fV_{\nu} \right) \Psi_{-}(x) -  \frac{m}{\Mka} \Sigma_{-}^6 \Psi_{-}(x) =   0 , \nn \\
& & \chih^{\mu\nu}\Sigma_{- a} \chi_{\mu}^{\;\; a} i \left(\cD _{\nu} -\mV_{\nu} \right) \vPsi_{-}(x)  - \frac{m}{\Mka}\phi(x) \Sigma_{-}^6 \vPsi_{-}(x) =   0 , 
\ee
which correspond to the scaling gauge invariant dimensionless fields and local scaling charged fields. Where we have introduced the following vector-like field: 
\be
\fV_{\mu} & \equiv & \frac{1}{2} \fkA\, \hfA_{b}^{\;\; \nu}\cD_{\nu}(\hat{\fkA} \fA_{\mu}^{\;\; b}) =  - \frac{1}{2}  \fA_{\mu a} \hfA_{b}^{\;\; \nu} \fA_{\nu}^{ab} ,\nn \\
 \mV_{\mu} & \equiv & \frac{1}{2} \chi \, \chih_{b}^{\;\; \nu}\cD_{\nu}(\chih \chi_{\mu}^{\;\; b}) = -\frac{1}{2}  \chi_{\mu a} \chih_{b}^{\;\; \nu} \mA_{\nu}^{ab} ,
 \ee
which is given by the gravigauge field and spin covariant gauge field. Such a vector-like field preserves the scaling gauge invariance for the equation of motion, which can be verified explicitly from the second equation in Eq.(\ref{EMDSF}) when all fields are represented as local scaling charged fields. For convenience, we may refer to such a vector-like field as a graviscaling induced gauge field, which is distinguished from the ordinary gauge field of Lie group due to the associated imaginary factor in the covariant derivative.   

For simplicity of discussions, let us consider the equation of motion of chirality-based Dirac spinor by  taking the scaling gauge fixing condition to be in Einstein basis, i.e.:
\be
 \chih^{\mu\nu}\Sigma_{- a} \chi_{\mu}^{\;\; a} i \left(\cD _{\nu} - \mV_{\nu} \right) \vPsi_{-}(x) - m \Sigma_{-}^6 \vPsi_{-}(x) =   0 . 
\ee
In terms of the vector-like spinor representation of Dirac spinor field in four-dimensional Hilbert space, the above equation can be simplified into the following form:
\be
\gamma^{a} \chih_{a}^{\;\; \mu} i \left(\cD _{\mu} - \mV_{\mu} \right) \psi 
 - m \psi  =   0 ,
\ee
with the covariant derivative defined as follows:
\be
i(\cD_{\mu} -\mV_{\mu}) \equiv i \p_{\mu} + \frac{1}{2}g_G i  \chi_{\mu a} \chih_{b}^{\;\; \nu} \mA_{\nu}^{ab} + (\mOm_{\mu}^{bc} + g_G  \mA_{\mu}^{bc} ) \frac{1}{2}\Sigma_{bc} + g_E A_{\mu} .
\ee

Let us now investigate the quadratic form of the equation of motion for the Dirac spinor in the presence of gravitational interaction and spin gauge interaction as well as electromagnetic interaction. Its explicit form is found to be:
\be \label{EMDSF2}
& &\chih^{\mu\nu} (\nabla_{\mu} - \mV_{\mu} ) ( \cD _{\nu} - \mV_{\nu} ) \psi + m^2  \psi \nn \\
& & \; \; = \Sigma^{cd}\chih_{c}^{\;\; \mu} \chih_{d}^{\;\; \nu} [ \cF_{\mu\nu}^{ab} \frac{1}{2} \Sigma_{ab} + g_E F_{\mu\nu} - g_G\fF_{\mu\nu}^{a} \chih_{a}^{\;\; \rho} i ( \cD _{\rho} - \mV_{\rho} ) - i \mV_{\mu\nu}\, ] \psi, 
 \ee
with the following definitions:
 \be \label{CD2}
 & & \nabla_{\mu} ( \cD _{\nu}  - \mV_{\nu} ) \equiv \cD _{\mu} ( \cD _{\nu} -\mV_{\nu} ) - \vGa_{\mu\nu}^{\rho} ( \cD _{\rho} -\mV_{\rho} ) ,\nn \\
 & & \vGa_{\mu\nu}^{\rho}  \equiv   \chih_{a}^{\;\; \rho}  \cD_{\mu}\chi_{\nu}^{\;\;a} =  \chih_{a}^{\;\; \rho}(\p_{\mu} \chi_{\nu}^{\;\; a} + \cA_{\mu\, b}^{a}  \chi_{\nu}^{\;\;b}\, ) \equiv
 \Gamma_{\mu\nu}^{\rho} + g_G \mA_{\mu}^{ab}  \chi_{\nu b} \chih_{a}^{\;\; \rho}, \nn \\
 & & \fF_{\mu\nu}^{a} \equiv  \cD_{\mu}\chi_{\nu}^{\;\;a} -  \cD_{\nu}\chi_{\mu}^{\;\;a} \equiv \mA_{\mu}^{ab}  \chi_{\nu b} - \mA_{\nu}^{ab}  \chi_{\mu b} , \nn \\
 & &  \mV_{\mu\nu}  \equiv  \p_{\mu}\mV_{\nu}-  \p_{\nu}\mV_{\mu} .
 \ee
Where $\cF_{\mu\nu}^{ab}$ and $F_{\mu\nu}$ are spin gauge field strength and electromagnetic field strength, respectively. $\Gamma_{\mu\nu}^{\rho}$ denotes the affine connection and $\mV_{\mu\nu}$ represents the graviscaling induced gauge field strength. 

It is intriguing to notice that although the gauge invariant action is hermitian, while the equation of motion of Dirac spinor field gets an emergent dissipative term characterized by the graviscaling induced gauge field $\mV_{\mu}$ and field strength $\mV_{\mu\nu}$ due to the presence of gravitational interaction and spin gauge interaction via the gravigauge field $\chi_{\mu}^{\; a}$ and spin covariant gauge field $\mA_{\mu}^{ab}$. 

To display explicitly the effects of gravitational interaction and spin gauge interaction on the dynamics of Dirac fermion, we represent the above equation of motion into the following form:
\be \label{EMDSF4}
& &\chih^{\mu\nu} ( \cD_{\mu} - \mV_{\mu} ) (\cD _{\nu} -\mV_{\nu} )\psi + \left( \p_{\mu}\chih^{\mu\nu} + 2 \chih^{\mu\nu} \p_{\mu}\ln \chi  + g_G\mA_{\mu}^{ab}  \chih_{a}^{\; \mu} \chih_{b}^{\; \nu} \right) ( \cD _{\nu} - \mV_{\nu} ) \psi \nn \\
& & + ( m^2 + \frac{1}{16} R +  \frac{1}{16} g_G \mF_{\mu\nu}^{ab} \chih_{b}^{\; \mu} \chih_{a}^{\; \nu} -  \frac{1}{8} g_G \epsilon_{abcd} \mF_{\mu\nu}^{ab}  \chih^{c \mu} \chih^{d \nu}  i \gamma_5) \psi 
\nn \\
& & \;  = \varSigma^{cb} [ \chih_{c}^{\;\; \mu} \chih_{b}^{\;\; \nu} ( g_E F_{\mu\nu} - i \mV_{\mu\nu} ) + \chih_{c}^{\; \mu} \left( i g_G\mF_{\mu\nu b}^{a}  - 2 g_G \mA_{\mu b}^{a} i ( \cD _{\nu} - \mV_{\nu} ) \right)  \chih_{a}^{\; \nu} \, ] \psi, 
 \ee
where we have used the algebra relation of $\gamma$-matrices:
\be
\Sigma^{cd}\Sigma^{ab} = \frac{1}{2} i(\Sigma^{cb}\eta^{ad} - \Sigma^{db}\eta^{ac} - \Sigma^{ca}\eta^{db} + \Sigma^{da}\eta^{cb}  ) + \frac{1}{16} ( \eta^{ca}\eta^{db} - \eta^{da}\eta^{cb} ) + \frac{1}{4}\epsilon^{abcd} i \gamma_5 ,
\ee
and the properties of Riemann and Ricci curvature tensors:
\be
R_{\mu\rho\nu\sigma} + R_{\rho\nu\mu\sigma} + R_{\nu\mu\rho\sigma} = 0 , \quad R_{\mu\nu} = R_{\nu\mu}. 
\ee


\subsection{ Gravigauge Dirac equation in locally flat gravigauge spacetime}

From the gauge and scaling invariant action presented in Eqs.(\ref{HSGaction})-(\ref{SGIaction}) and (\ref{EBaction}), it is noticed that the Dirac fermion actually couples to the spin gauge field $\cA_{c}^{ab}\equiv \chih_{c}^{\;\mu} \cA_{\mu}^{ab}$ defined in locally flat gravigauge spacetime. Therefore, it is particularly interesting to derive the equation of motion of Dirac spinor field in locally flat gravigauge spacetime via a hidden coordinate formalism. Taking the scaling gauge fixing condition to be in Einstein basis, we can rewrite the equation of motion for the  Dirac spinor into the following simple form:
\be
\gamma^{c} i \left(\cD _{c} - \mV_{c} \right) \psi  - m \psi  =   0 ,
\ee
which appears formally to be analogous to the Dirac equation in globally flat Minkowski spacetime. Whereas they are essentially distinguished due to the presence of spin gauge field $\cA_{c}^{ab}$ and the emergent non-commutation relation of covariant derivative operator in locally flat gravigauge spacetime shown in Eq.(\ref{NCR}), which can be seen from the following definitions and relations: 
\be
& & i\cD_{c} \equiv i \eth_{c} + g_G \cA_{c}^{ab}  \frac{1}{2}\Sigma_{ab} 
+ g_E A_{a} , 
\nn \\
& & [\eth_{c}\, , \eth_{d} ] =  \mOm_{[cd]}^{a} \1 \eth_{a}, \quad \eth_{c} \equiv \chih_{c}^{\;\mu} \p_{\mu}, \nn \\
& &  \mV_{c} \equiv \frac{1}{2} g_G\mA_{b c}^{b} \equiv \frac{1}{2} g_G \chih_{b}^{\;\; \nu} \mA_{\nu c}^{b} .
\ee
To be more explicit, let us check the covariant quadratic form for the equation of motion in locally flat gravigauge spacetime:
\be \label{EMDSF3}
& &(\nabla_{c} - \mV_{c} ) ( \cD ^{c} - \mV^{c} ) \psi + m^2  \psi \nn \\
& & = \Sigma^{cd}[ \cF_{cd}^{\, ab} \frac{1}{2} \Sigma_{ab} + g_E F_{cd} - g_G\fF_{cd}^{a} i ( \cD _{a} - \mV_{a} ) - i \mV_{cd}\, ] \psi, 
 \ee
with the definitions:
 \be \label{CD3}
 & & \nabla_{c} ( \cD _{d}  - \mV_{d} ) \equiv \cD _{c} ( \cD _{d} -\mV_{d} ) - \cA_{cd}^{a} ( \cD _{a} -\mV_{a} ) ,\nn \\
 & & \fF_{cd}^{a} \equiv  \cA_{[cd]}^a - \mOm_{[cd]}^a  \equiv \mA_{[cd]}^{a}\equiv \mA_{cd}^a - \mA_{dc}^a ,\nn  \\
 & &  \mV_{cd}  \equiv  \eth_{c}\mV_{d}-  \eth_{d}\mV_{c} - \mOm_{[cd]}^a \mV_a .
 \ee
Where all the gauge fields and field strengths are defined in locally flat gravigauge spacetime as shown in Eqs.(\ref{SGFGG}) and (\ref{SCGFS}). It can be further demonstrated that Eq.(\ref{EMDSF3}) can be rewritten into the following explicit form in locally flat gravigauge spacetime:
\be \label{EMDSF4}
& & [ (\cD_{c} - \mV_{c}) (\cD ^{c} - \mV^{c}) + (\p_{\mu}\chih_{c}^{\;\mu} + \eth_{c}\ln \chi +  \mA_{a}^{a c} ) ( \cD ^{c} - \mV^{c} ) ] \psi \nn \\
& & + ( m^2 + \frac{1}{16} R + \frac{1}{16}g_G \mF_{ab}^{\, ba} -  \frac{1}{8} g_G \epsilon^{cdab} \mF_{cd ab}  i \gamma_5) \psi 
\nn \\
& &  = \varSigma^{cb} [ g_E F_{cb}  - 2g_G \mA_{c\, b}^{a} i ( \cD _{a} - \mV_{a} ) + i g_G\mF_{c a \1 b}^{a} - i \mV_{cb} ) \, ] \psi .
 \ee
 
It is interesting to notice that the Ricci curvature scalar $R\equiv R_{cb}\eta^{cb} \equiv R_{\mu\sigma}\chih^{\mu\sigma}$ and the spin covariant gauge field strength scalar $\mF_{cd}^{ab} \eta_{b}^{\; c} \eta_{a}^{\; d} \equiv \mF_{\mu\nu}^{ab} \chih_{b}^{\; \mu} \chih_{a}^{\; \nu} $ emerge as mass-like terms in the quadratic form of equation of motion for the Dirac fermion, which indicates that a non-zero Ricci curvature scalar $R\neq 0$ appears to generate an effective mass for the Dirac fermion when it propagates in locally flat gravigauge spacetime. Meanwhile, the totally antisymmetric tensor $\epsilon^{cdab} \mF_{cd ab}$ arising from the spin covariant gauge field strength behaves as a pseudoscalar.  
 
In conclusion, the equation of motion of Dirac spinor field in the presence of gravitational interaction and spin gauge interaction characterized by the spin gravigauge field (or gravigauge field) and spin gauge field generalizes the relativistic quantum mechanics described by Dirac equation in globally flat Minkowski spacetime to obtain a gravitational relativistic quantum theory in locally flat gravigauge spacetime with the emergence of non-commutative geometry.


\section{Gravidynamics with gauge-type and geometric gravitational equations as extension to Einstein theory of general relativity}

To study the dynamics of gauge fields, we are going to take the gravigauge field $\fA_{\mu}^{\;\; a}(x)\equiv \phi(x) \chi_{\mu}^{\;\; a}(x)$ and spin gauge field $\cA_{\mu}^{ab}(x)$ as independent degrees of freedom based on the inhomogeneous spin gauge symmetry WS(1,3). In particular, we will discuss the potential new effects of gravidynamics beyond the Einstein theory of general relativity.

\subsection{ Gravidynamics with gauge-type gravitational equations of gravigauge field }

To describe the gravidynamics, let us derive the equation of motion for the $\cW_e$-spin invariant-gauge field as gravigauge field. The gauge covariant and scaling invariant equations of motion in correspondence to the scaling gauge invariant gravigauge field $\fA_{\mu}^{\;\; a}$ and scaling charged gravigauge field $\chi_{\mu}^{\;\; a}$ are found to be:
\be  \label{EMGGF}
& & \p_{\nu} \tilde{\mF}^{\mu\nu }_{a}  =  \bs{\cJ}_{a}^{\;\; \mu}  , \nn \\
& & \tilde{d}_{\nu} \tilde{\rF}^{\mu\nu }_{a}  = \cJ_{a}^{\;\; \mu} ,
\ee
where the field strengths $\tilde{\mF}^{\mu\nu }_{a}$ and $\tilde{\rF}^{\mu\nu }_{a}$ are defined as follows:
\be
& & \tilde{\mF}^{\mu\nu }_{a} \equiv \Mka^{-1}\fkA\1\tmH^{[\mu\nu]\mu'\nu'}_{a a'} \mF_{\mu'\nu' }^{a'} , \quad \mF_{\mu\nu}^a  \equiv \p_{\mu} \fA_{\nu}^{\; a} - \p_{\nu} \fA_{\mu}^{\; a} , \nn \\
& & \tilde{\rF}^{\mu\nu }_{a} \equiv \chi \1 \Mka^{-2}\phi^2 \tchi^{[\mu\nu]\mu'\nu'}_{a a'} \rF_{\mu'\nu' }^{a'} , \quad  \rF_{\mu\nu}^a  \equiv d_{\mu} \chi_{\nu}^{\; a} - d_{\nu} \chi_{\mu}^{\; a},  \nn \\
& & \tilde{d}_{\mu} \equiv  \p_{\mu} - S_{\mu} , \quad d_{\mu} =   \p_{\mu} + S_{\mu}, \quad   S_{\mu} \equiv \p_{\mu}\ln \phi ,
\ee
and the currents $\bs{\cJ}_{a}^{\;\; \mu}$ and $\cJ_{a}^{\;\; \mu}$ are given explicitly by:
\be \label{BVCJ}
\bs{\cJ}_{a}^{\;\; \mu} & \equiv &16\pi G_N\1 \hat{\fJ}_{a}^{\;\; \mu} + \tilde{\fJ}_{a}^{\;\; \mu} , \quad  \hat{\fJ}_{a}^{\;\; \mu} \equiv  \fJ_{a}^{\;\; \mu}  +  \widetilde{\fJ}_{a}^{\;\; \mu} , \nn \\
 \tilde{\fJ}_{a}^{\;\; \mu} & = & \hfA_{a}^{\; \rho}  \mF_{\rho\nu}^{c} \tilde{\mF}^{\mu\nu}_{c}  - \frac{1}{4}  \hfA_{a}^{\; \mu} \1 \mF_{\rho\nu}^{c} \tilde{\mF}^{\rho\nu}_{c} , \nn \\
 \fJ_{a}^{\;\; \mu} & = & \Mka \fkA\1 \{ ( \hfA_{a}^{\; \rho} \hfA_{c}^{\; \mu} -  \hfA_{a}^{\; \mu} \hfA_{c}^{\;  \rho}  ) ( \bar{\Psi}_{-} \vSi_{-}^{c} i \cD_{\rho} \Psi_{-} + H.c.) \nn \\
 & + & \hfA_{a}^{\; \mu}  \frac{m}{\Mka} \bar{\Psi}_{-} \Gamma^6 \Psi_{-}  - g_E^{-2} ( \hmH^{\mu\mu'}  \hfA_{a}^{\;  \rho} - \frac{1}{4} \hmH^{\rho\mu'} \hfA_{a}^{\; \mu}  ) \hmH^{\nu\nu'} \1 F_{\rho\nu} F_{\mu'\nu'} \}, \nn \\
 \widetilde{\fJ}_{a}^{\;\; \mu} & = & \Mka \fkA\1 \{- g_G^{-2} ( \hmH^{\mu\mu'} \hfA_{a}^{\; \rho} - \frac{1}{4} \hmH^{\rho\mu'} \hfA_{a}^{\; \mu} ) \hmH^{\nu\nu'}\1 \cF_{\rho\nu}^{bc} \cF_{\mu'\nu' b c}   \nn \\
& + & \frac{m_G^2}{\Mka^2} (  \bmH^{[\mu\nu]\mu'\nu'}_{b a'}  \hfA_{a}^{\;\;\rho} - \frac{1}{4}  \bmH^{[\rho\nu]\mu'\nu'}_{b a'} \hfA_{a}^{\;\;\mu}  ) \cF_{\rho\nu}^{b} \cF_{\mu'\nu'}^{a'} \}  \nn \\
& - & \frac{m_G^2}{\Mka} \cD_{\nu} ( \fkA\, \bmH^{[\mu\nu]\mu'\nu'}_{a a'}   \cF_{\mu'\nu' }^{a'} ) , 
\ee
and
\be
\cJ_{a}^{\;\; \mu} & \equiv &16\pi G_N\1 \hat{\mJ}_{a}^{\;\; \mu} + \tilde{\mJ}_{a}^{\;\; \mu} , \quad \hat{\mJ}_{a}^{\;\; \mu}  \equiv  \mJ_{a}^{\;\; \mu} + \widetilde{\mJ}_{a}^{\;\; \mu}, \nn \\
\tilde{\mJ}_{a}^{\;\; \mu} & = &  \chih_{a}^{\; \rho}  \rF_{\rho\nu}^{c} \tilde{\rF}^{\mu\nu}_{c} -  \frac{1}{4} \chih_{a}^{\; \mu} \1 \rF_{\rho\nu}^{c} \tilde{\rF}^{\rho\nu}_{c} , \nn \\
\mJ_{a}^{\;\; \mu} & = &  \chi \1 \{ ( \chih_{a}^{\; \; \rho} \chih_{c}^{\; \; \mu} -  \chih_{a}^{\; \; \mu} \chih_{c}^{\; \; \rho}  ) ( \bar{\vPsi}_{-} \vSi_{-}^{c} i \cD_{\rho} \vPsi_{-} + H.c.) \nn \\
& + & \chih_{a}^{\;\;\mu}  \frac{m}{\Mka} \phi \bar{\vPsi}_{-} \Gamma^6 \vPsi_{-}   - ( \chih^{\mu\mu'}  \chih_{a}^{\; \; \rho} - \frac{1}{4} \chih^{\rho\mu'} \chih_{a}^{\; \; \mu}  ) \chih^{\nu\nu'} \1 F_{\rho\nu} F_{\mu'\nu'} \}, \nn \\
\widetilde{\mJ}_{a}^{\;\; \mu} & = &  \chi \1 \{- ( \chih^{\mu\mu'} \chih_{a}^{\; \rho} - \frac{1}{4} \chih^{\rho\mu'} \chih_{a}^{\; \mu} ) \chih^{\nu\nu'}  \1 \cF_{\rho\nu}^{bc} \cF_{\mu'\nu' bc } \nn \\
& + & \frac{m_G^2}{\Mka^2} \phi^2 (  \bchi^{[\mu\nu]\mu'\nu'}_{b a'}  \chih_{a}^{\;\;\rho} - \frac{1}{4}  \bchi^{[\rho\nu]\mu'\nu'}_{b a'} \chih_{a}^{\;\;\mu}  ) (\fF_{\rho\nu}^{b}+ S_{\rho\nu}^{b} ) (\fF_{\mu'\nu'}^{a'} + S_{\mu'\nu'}^{a'} )\}  \nn \\
& - & \frac{m_G^2}{\Mka} \cD_{\nu} \left( \chi\, \bchi^{[\mu\nu]\mu'\nu'}_{a a'}  \phi (\fF_{\mu'\nu' }^{a'}  + S_{\mu'\nu'}^{a'} ) \right) .
\ee
In the above equations, we have introduced the following definitions:
\be
& & \cD_{\nu} ( \fkA\, \bmH^{[\mu\nu]\mu'\nu'}_{a a'}   \cF_{\mu'\nu' }^{a'} ) = \p_{\nu}  ( \fkA\, \bmH^{[\mu\nu]\mu'\nu'}_{a a'}   \cF_{\mu'\nu' }^{a'} ) -  \fkA\, g_G\cA_{\nu a}^{b}  \bmH^{[\mu\nu]\mu'\nu'}_{b a'}   \cF_{\mu'\nu' }^{a'} , \nn \\
& & \bmH_{aa'}^{[\mu\nu] \mu'\nu'} \equiv \hfA_{c}^{\;\, \mu}\hfA_{d}^{\;\, \nu} \hfA_{c'}^{\;\, \mu'} \hfA_{d'}^{\;\, \nu'}  \bar{\eta}^{[c d] c' d'}_{a a'} ,  \quad \tmH_{aa'}^{[\mu\nu] \mu'\nu'} \equiv \hfA_{c}^{\;\, \mu}\hfA_{d}^{\;\, \nu} \hfA_{c'}^{\;\, \mu'} \hfA_{d'}^{\;\, \nu'}  \tilde{\eta}^{[c d] c' d'}_{a a'} ,  \nn \\
& & \bchi_{aa'}^{[\mu\nu] \mu'\nu'} \equiv \chih_{c}^{\;\, \mu}\chih_{d}^{\;\, \nu} \chih_{c'}^{\;\, \mu'} \chih_{d'}^{\;\, \nu'}  \bar{\eta}^{[c d] c' d'}_{a a'}  , \quad \tchi_{aa'}^{[\mu\nu] \mu'\nu'} \equiv \chih_{c}^{\;\, \mu}\chih_{d}^{\;\, \nu} \chih_{c'}^{\;\, \mu'} \chih_{d'}^{\;\, \nu'}  \tilde{\eta}^{[c d] c' d'}_{a a'} , \nn \\
& &  \bar{\eta}^{[c d] c' d'}_{a a'} \equiv \frac{1}{2}(  \bar{\eta}^{cd c' d'}_{a a'}  - \bar{\eta}^{ dc c' d'}_{a a'} ) , \quad \tilde{\eta}^{[c d] c' d'}_{a a'} \equiv \frac{1}{2}(  \tilde{\eta}^{cd c' d'}_{a a'}  - \tilde{\eta}^{ dc c' d'}_{a a'} ).
\ee

Taking the scaling gauge fixing condition to be in Einstein basis, we arrive at the following gauge-type gravitational equation: 
\be  \label{GTGE}
 \p_{\nu} \tilde{F}^{\mu\nu }_{a}  = J_{a}^{\; \mu}  , 
\ee
where the field strength $\tilde{F}^{\mu\nu }_{a}$ and current $J_{a}^{\; \mu}$ are defined as follows:
\be
\tilde{F}^{\mu\nu }_{a} & \equiv & \chi \1 \tchi^{[\mu\nu]\mu'\nu'}_{a a'}  F_{\mu'\nu' }^{a'}  , 
\ee
and
\be \label{GTGEC}
J_{a}^{\;\mu} & \equiv & 16\pi G_N \hat{J}_{a}^{\;\; \mu}  + \tilde{J}_{a}^{\;\; \mu} , \quad \hat{J}_{a}^{\;\; \mu} \equiv  \mathrm{J}_{a}^{\;\; \mu} + \widetilde{J}_{a}^{\;\; \mu} , \nn \\
\tilde{J}_{a}^{\;\; \mu} & = & \chih_{a}^{\; \rho}  F_{\rho\nu}^{c} \tilde{F}^{\mu\nu}_{c} - \frac{1}{4} \chih_{a}^{\; \mu} \1 F_{\rho\nu}^{c} \tilde{F}^{\rho\nu}_{c} , \nn \\
\mathrm{J}_{a}^{\;\; \mu} & = &  \chi \1 \{ ( \chih_{a}^{\; \; \rho} \chih_{c}^{\; \; \mu} -  \chih_{a}^{\; \; \mu} \chih_{c}^{\; \; \rho}  ) \frac{1}{2} ( \bar{\psi} \gamma^{c} i \cD_{\rho} \psi + H.c.)  \nn \\
& + & \chih_{a}^{\;\;\mu}  m \bar{\psi} \psi   - ( \chih^{\mu\mu'} \chih_{a}^{\; \; \rho} - \frac{1}{4} \chih^{\rho\mu'} \chih_{a}^{\; \; \mu}  )  \chih^{\nu\nu'}\1 F_{\rho\nu} F_{\mu'\nu'} \}, \nn \\
\widetilde{J}_{a}^{\;\; \mu} & = &  \chi \1 \{ - ( \chih^{\mu\mu'} \chih_{a}^{\; \rho} - \frac{1}{4} \chih^{\rho\mu'} \chih_{a}^{\; \mu} ) \chih^{\nu\nu'}  \1 \cF_{\rho\nu}^{bc} \cF_{\mu'\nu' bc } \nn \\
& + & m_G^2(  \bchi^{[\mu\nu]\mu'\nu'}_{b a'}  \chih_{a}^{\;\;\rho} - \frac{1}{4}  \bchi^{[\rho\nu]\mu'\nu'}_{b a'} \chih_{a}^{\;\;\mu}  ) \fF_{\rho\nu}^{b} \fF_{\mu'\nu'}^{a'} \} \nn \\
& - & m_G^2 \cD_{\nu} ( \chi\, \bchi^{[\mu\nu]\mu'\nu'}_{a a'}  \fF_{\mu'\nu' }^{a'} ) ,
\ee
with the conserved current,
\be
\p_{\mu} J_{a}^{\; \mu} = 0 .
\ee

Such equations of motion for gravigauge field are obtained to describe the gravidynamics and referred to as gauge-type gravitational equations. 


\subsection{ Gauge-type gravidynamics in locally flat gravigauge spacetime}

From the gauge and scaling invariant action built in hidden coordinate formalisms in Eqs.(\ref{HSGaction})-(\ref{SGIaction}) and (\ref{EBaction}), it is seen that all interactions emerge in locally flat gravigauge spacetime spanned by the gravigauge bases. It should be appropriate to derive the gauge-type gravitational equation in locally flat gravigauge spacetime. When taking the scaling gauge fixing condition to be in Einstein basis, we arrive at the following equation of motion:
\be 
 \tilde{D}_{d} \tilde{F}^{cd}_{b}   = J_{b}^{\;\; c} , 
\ee
with the definitions:
\be
& & \tilde{F}^{cd}_{a} \equiv  \tilde{\eta}^{[cd]c'd'}_{a a'}  F_{c'd'}^{a'}  , \quad F_{cd}^{a}  \equiv \chih_{c}^{\;\; \mu} \chih_{d}^{\;\; \nu} F_{\mu\nu}^{a},  \nn \\ 
& & \tilde{D}_{d} \tilde{F}^{cd}_{b} \equiv  \eth_{d} \tilde{F}^{cd}_{b} + \mOm_{d a}^c \tilde{F}^{ad}_{b} + \mOm_{d a}^d \tilde{F}^{ca}_{b}  ,
\ee
for the field strength and covariant derivative, and
\be
J_{b}^{\;\; c} & \equiv & 16\pi G_N \hat{J}_{b}^{\;\; c} + \tilde{J}_{b}^{\; c} ,\quad \hat{J}_{b}^{\; c} \equiv \mathrm{J}_{b}^{\; c} + \widetilde{J}_{b}^{\; c} ,  \nn \\
\tilde{J}_{b}^{\; c} & = &  F_{bd}^a \tilde{F}^{cd}_{a} - \frac{1}{4} \eta_{b}^{\; c} \1 F_{cd}^{a} \tilde{F}^{cd}_{a}, \nn \\
\mathrm{J}_{b}^{\; c} & = &  \frac{1}{2} ( \bar{\psi} \gamma_{c} i \cD_{b} \psi  + H.c.) - \eta_{b}^{\; c} [ \frac{1}{2} ( \bar{\psi} \gamma^{a} i \cD_{a} \psi  + H.c. ) - m \bar{\psi} \psi  ]   \nn \\
& - & (\eta_{b}^{\; a} \eta^{c c'}  - \frac{1}{4}\eta_{b}^{\; c} \eta^{ac'} ) \eta^{dd'}  F_{ad} F_{c'd'}   , \nn \\
\widetilde{J}_{b}^{\; c} & = & - (\eta_{b}^{\; a} \eta^{c c'}  - \frac{1}{4}\eta_{b}^{\; c} \eta^{ac'} ) \eta^{dd'}  F_{ad}^{a'b'} F_{c'd' a' b'} \nn \\
& + & m_G^2 (\eta_{b}^{\; c''} \eta_{b'}^{\; c} \bar{\eta}^{[b' d] c'd'}_{a a'}   - \frac{1}{4} \eta_{b}^{\; c}  \bar{\eta}^{[c''d] c'd'}_{a a'} ) \fF_{c'' d}^{a} \fF_{c'd'}^{a'}  \nn \\
& - & m_G^2  \bar{\cD}_{d} (\bar{\eta}^{[c d]c'd'}_{b a'}  \fF_{c'd' }^{a'} ) ,
\ee
for the currents. The gauge covariant derivative in the last term is defined as follows:
\be
\bar{\cD}_{d} (\bar{\eta}^{[cd]c'd'}_{b a'}  \fF_{c'd' }^{a'} ) & \equiv & \eth_d(\bar{\eta}^{[cd]c'd'}_{b a'} \fF_{c'd' }^{a'} )  + ( \mOm_{da}^{c} \bar{\eta}^{[ad]c'd'}_{b a'} + \mOm_{da}^{d} \bar{\eta}^{[ca]c'd'}_{b a'}- g_G\cA_{db}^{a} \bar{\eta}^{[cd]c'd'}_{a a'} ) \fF_{c'd' }^{a'} .
\ee

Such equations of motion characterize the gauge-type gravidynamics in locally flat gravigauge spacetime.


\subsection{ Geometric gravidynamics and extension to Einstein theory of general relativity}

To obtain geometric gravitational equations, let us take into account the gauge-gravity and gravity-geometry correspondences shown in Eqs.(\ref{GGGR})-(\ref{GGGI}) and (\ref{EBaction2}). When ignoring the total derivative term, the least action principle leads to the following relations: 
\be  \label{EMGGF1}
& & 2\fR_{\rho\sigma} \hfA_{a}^{\; \rho} \hmH^{\mu\sigma} -  \hfA_{a}^{\; \mu} \fR  = -16\pi G_N\1 \hat{\fJ}_{a}^{\;\; \mu}  , \nn \\
& & 2\phi^2 \mR_{\rho\sigma}  \chih_{a}^{\; \rho} \chih^{\mu\sigma} -  \chih_{a}^{\; \mu} \phi^2 \mR  = -16\pi G_N  \Mka^2\1 \hat{\mJ}_{a}^{\;\; \mu} , 
\ee
with 
\be
& & \mR_{\rho\sigma} \equiv R_{\rho\sigma} - 6 \p_{\rho}\ln \phi \p_{\sigma}\ln \phi  , \nn \\
& &  \mR \equiv \chih^{\rho\sigma} \mR_{\rho\sigma} = R - 6\chih^{\rho\sigma} \p_{\rho}\ln\phi \p_{\sigma}\ln\phi . 
\ee
In the Einstein basis, we can rewrite the above relation into the following simple form:
\be \label{BFGE}
& &  R_{a}^{\;\mu} - \frac{1}{2} \chih_{a}^{\; \mu} R   = - 8\pi G_N \1 \hat{J}_{a}^{\;\; \mu} , 
\ee
with
\be
& & R_{a}^{\;\mu}  \equiv  \chih_{a}^{\; \mu'}  \chih_{a'}^{\; \nu'}  \chih_{b'}^{\; \mu} R_{\mu'\nu'}^{a'b'} .
\ee

When projecting the tensor and current in biframe spacetime into the base spacetime of coordinates by applying for the gravigauge field $\chi_{\mu}^{\; a}$ ($\chih_{a}^{\; \mu}$), we arrive at the following two geometric gravitational equations:
\be \label{GGE}
& & R_{\mu\nu} -  \frac{1}{2}\chi_{\mu\nu} R  + 8\pi G_N \1 \widetilde{T}_{\mu\nu}  = - 8\pi G_N \1 T_{\mu\nu} , 
\ee
and
\be \label{GGEAS}
& & \bar{\nabla}_{\rho} \bar{\fF}_{\;\, [\mu\nu]}^{\rho} +  \widetilde{T}_{[\mu\nu]} = - m_G^{-2} T_{[\mu\nu]} .
\ee
The above two equations correspond to the symmetric tensor and antisymmetric tensor in the base spacetime of coordinates.

The geometric gravitational equation presented in Eq.(\ref{GGE}) brings on the extension to Einstein theory of general relativity in the presence of spin gauge field, which is characterized by the symmetric tensor $\widetilde{T}_{\mu\nu}$. The explicit forms of the tensors are given as follows:
\be
& & T_{\mu\nu} + \widetilde{T}_{\mu\nu}  =  T_{\nu\mu} + \widetilde{T}_{\nu\mu} \equiv \frac{1}{2}\chih (  \chi_{\mu\rho} \hat{J}_{a}^{\; \rho} \chi_{\nu}^{\; a} + \chi_{\nu\rho} \hat{J}_{a}^{\; \rho} \chi_{\mu}^{\; a}  ) ,   \nn \\
&& T_{\mu\nu} =   \frac{1}{4} [ \chi_{\mu}^{\; a} \bar{\psi} \gamma_{a} i \cD_{\nu} \psi  + \chi_{\nu}^{\; a} \bar{\psi} \gamma_{a} i \cD_{\mu} \psi   + H.c.]  \nn \\
& & \qquad - \chi_{\mu\nu} [   \frac{1}{2} (\bar{\psi} \gamma^{c} \chih_{c}^{\; \rho} i \cD_{\rho} \psi  + H.c.) - m \bar{\psi} \psi  ]   \nn \\
& & \qquad - (\eta_{\mu}^{\; \rho}\eta_{\nu}^{\; \rho'} - \frac{1}{4}\chi_{\mu\nu} \chih^{\rho\rho'} )   \chih^{\sigma\sigma'}\1 F_{\rho\sigma} F_{\rho'\sigma'} ,  \nn \\
& & \widetilde{T}_{\mu\nu} = - (\eta_{\mu}^{\; \rho}\eta_{\nu}^{\; \rho'} - \frac{1}{4}\chi_{\mu\nu} \chih^{\rho\rho'} )   \chih^{\sigma\sigma'}\1 \cF_{\rho\sigma}^{ab} \cF_{\rho'\sigma' ab}  +  m_G^2  \bar{\nabla}_{\rho}  \bar{\fF}_{\;\,(\mu\nu)}^{\rho}   \nn \\
& & \qquad + m_G^2[ \frac{1}{2}  (\chi_{\mu\lambda}\eta_{\nu}^{\; \rho} +\chi_{\nu\lambda}\eta_{\mu}^{\; \rho}  )  \bchi^{[\lambda\sigma]\rho'\sigma'}_{a a'}   - \frac{1}{4} \chih_{\mu\nu} \bchi^{[\rho\sigma]\rho'\sigma'}_{a a'}   ] \fF_{\rho\sigma}^{a} \fF_{\rho'\sigma'}^{a'} ,
\ee 
with the definitions:
\be
& & \bar{\nabla}_{\rho} \bar{\fF}_{\;\, (\mu\nu)}^{\rho} \equiv \p_{\rho}  \bar{\fF}_{(\mu\nu)}^{\rho} - \Gamma_{\rho \lambda}^{\rho} \bar{\fF}_{(\mu\nu)}^{\lambda} -  \frac{1}{2} ( \Gamma_{\rho\mu}^{\lambda} \bar{\fF}_{\lambda\nu}^{\rho} +\Gamma_{\rho\nu}^{\lambda} \bar{\fF}_{\lambda\mu}^{\rho} + g_G\cA_{\rho\nu}^{\lambda} \bar{\fF}_{\mu\lambda}^{\rho} +  g_G\cA_{\rho\mu}^{\lambda} \bar{\fF}_{\nu\lambda}^{\rho} ) , \nn \\
& & \bar{\fF}_{\mu\nu}^{\rho} \equiv \chi_{\mu\sigma} \chi_{\nu}^{\;\; a} \bchi^{[\rho\sigma]\rho'\sigma'}_{a a'}  \fF_{\rho'\sigma' }^{a'}  \equiv  \bar{\fF}_{(\mu\nu)}^{\rho} + \bar{\fF}_{[\mu\nu]}^{\rho} ,  \nn \\
& &  \bar{\fF}_{(\mu\nu)}^{\rho} \equiv  \frac{1}{2} ( \bar{\fF}_{\mu\nu}^{\rho}  +  \bar{\fF}_{\nu\mu}^{\rho}  )=  \frac{1}{2} ( \chi_{\mu\sigma} \chi_{\nu}^{\;\; a} + \chi_{\nu\sigma} \chi_{\mu}^{\;\; a} ) \bchi^{[\rho\sigma]\rho'\sigma'}_{a a'}  \fF_{\rho'\sigma' }^{a'} .
\ee
Where $T_{\mu\nu}$ is the usual four energy momentum tensor arising from the Dirac fermion and electromagnetic gauge field in the presence of spin gauge field and gravigauge field, and $\widetilde{T}_{\mu\nu}$ is purely attributed to the inhomogeneous spin gauge field governed by the inhomogeneous spin gauge symmetry. 

The geometric gravitational equation given in Eq.(\ref{GGEAS}) is obtained from the antisymmetric tensors, which indicates that the gauge invariant field strength $\tilde{\mF}_{\mu\nu}^{\rho}$ is governed by the relevant antisymmetric tensors $T_{[\mu\nu]}$ and $\widetilde{T}_{[\mu\nu]}$. Their explicit forms are defined as follows:
\be
 & & \bar{\nabla}_{\rho} \bar{\fF}_{\;\, [\mu\nu]}^{\rho}  \equiv  \p_{\rho}  \bar{\fF}_{[\mu\nu]}^{\rho} - \Gamma_{\rho\lambda}^{\rho} \bar{\fF}_{[\mu\nu]}^{\lambda} - \frac{1}{2}(\Gamma_{\rho\mu}^{\lambda} \bar{\fF}_{\lambda\nu}^{\rho} - \Gamma_{\rho\nu}^{\lambda} \bar{\fF}_{\lambda\mu}^{\rho}   + g_G \cA_{\rho\nu}^{\lambda} \bar{\fF}_{\mu\lambda}^{\rho}  - g_G\cA_{\rho\mu}^{\lambda} \bar{\fF}_{\nu\lambda}^{\rho} ), \nn \\
 & & \bar{\fF}_{[\mu\nu]}^{\rho} \equiv \frac{1}{2} ( \bar{\fF}_{\mu\nu}^{\rho}  -  \bar{\fF}_{\nu\mu}^{\rho}  )=  \frac{1}{2} ( \chi_{\mu\sigma} \chi_{\nu}^{\;\; a} - \chi_{\nu\sigma} \chi_{\mu}^{\;\; a}) \bchi^{[\rho\sigma]\rho'\sigma'}_{a a'}  \fF_{\rho'\sigma' }^{a'} , \nn \\
& &  \widetilde{T}_{[\mu\nu]}  \equiv  \frac{1}{2} ( \chi_{\mu\lambda}\eta_{\nu}^{\; \rho}  -  \chi_{\nu\lambda}\eta_{\mu}^{\; \rho}  ) \bchi^{[\lambda\sigma]\rho'\sigma'}_{a a'}   \fF_{\rho\sigma}^{a} \fF_{\rho'\sigma'}^{a'} , \nn \\
& & T_{[\mu\nu]}  \equiv   \frac{1}{4} \left( \chi_{\mu}^{\; a} \bar{\psi} \gamma_{a} i \cD_{\nu} \psi  - \chi_{\nu}^{\; a} \bar{\psi} \gamma_{a} i \cD_{\mu} \psi  + H.c \right) ,
\ee
which reflect the basic feature of gravigauge field $\chi_{\mu}^{\; a}$ as the fundamental gravitational field instead of the symmetric gravimetric field $\chi_{\mu\nu}$. 


\subsection{ Geometric gravidynamics in locally flat gravigauge spacetime}

Let us now project all tensors and currents into the ones in locally flat gravigauge spacetime by adopting the gravigauge field $\chi_{\mu}^{\; a}$ ($\chih_{a}^{\; \mu}$) as Goldstone-like boson. The following gravitational equations are obtained:
\be \label{GEGST}
& & R_{cb} -  \frac{1}{2}\eta_{cb}R +  8\pi G_N \1 \widetilde{T}_{cb}  = - 8\pi G_N \1 T_{cb} , 
\ee
and
\be \label{GEGSTAS}
& & \bar{\cD}_{d} \bar{\mF}_{[cb]}^{d} +  \widetilde{T}_{[cb]} = - m_G^{-2} T_{[cb]},
\ee
where the tensors are given by the field strength of spin gravigauge field $\mOm_{c}^{ab}$:
\be
& &  R_{cb} =  \eta^{da}R_{cdab}, \quad R= \eta^{cb} R_{cb} , \nn \\
& & R_{cdab} = \eth_{c}\mOm_{d ab} - \eth_{d}\mOm_{c ab} + \mOm_{ca e}\mOm_{db}^{e} - \mOm_{da e}\mOm_{cb}^{e} - \mOm_{[cd]}^{e}\mOm_{e ab} .
\ee
The symmetric tensors $T_{cb}$ and $\widetilde{T}_{cb}$ are provided by the following explicit forms:
\be
T_{cb} & = &  \frac{1}{4} [ \bar{\psi} \gamma_{c} i \cD_{b} \psi  + \bar{\psi} \gamma_{b} i \cD_{c} \psi + H.c.] \nn \\
 & - & \eta_{cb} [ \frac{1}{2} ( \bar{\psi} \gamma^{a} i \cD_{a} \psi  + H.c. ) - m \bar{\psi} \psi  ]   \nn \\
& - & (\eta_{c}^{\; a} \eta_{b}^{\; c'}  - \frac{1}{4}\eta_{cb} \eta^{ac'} ) \eta^{dd'}  F_{ad} F_{c'd'}  , \nn \\
\widetilde{T}_{cb} & = &  - (\eta_{c}^{\; c''} \eta_{b}^{\; c'}  - \frac{1}{4}\eta_{cb} \eta^{c''c'} ) \eta^{dd'}  \cF_{c''d}^{ab'} \cF_{c'd' a b'} + m_G^2 \bar{\cD}_{d} \bar{\fF}_{\; (cb)}^{d} \nn \\
& + & m_G^2[ \frac{1}{2}  (\eta_{c b'}\eta_{b}^{\; c''} + \eta_{b b'}\eta_{c}^{\; c''} )  \bar{\eta}^{[b' d] c'd'}_{a a'}   - \frac{1}{4} \eta_{cb}  \bar{\eta}^{[c''d] c'd'}_{a a'} ] \fF_{c'' d}^{a} \fF_{c'd'}^{a'} , 
\ee
with the definitions:
\be
& & \bar{\cD}_{d} \bar{\fF}_{\; (cb)}^{d} \equiv \eth_{d}  \bar{\fF}_{(cb)}^{d} + \mOm_{d a}^{d}  \bar{\fF}_{(cb)}^{a} - \frac{1}{2} (\mOm_{d c}^{a}  \bar{\fF}_{ab}^{d} + \mOm_{d b}^{a}  \bar{\fF}_{ac}^{d}  + g_G\cA_{d b}^{a}  \bar{\fF}_{ca}^{d}  + g_G\cA_{d c}^{a}  \bar{\fF}_{ba}^{d} ) , \nn \\
& &  \bar{\fF}_{cb}^{d}  \equiv  \eta_{c b'} \eta_{b}^{\; a} \bar{\eta}^{[d b']c'd'}_{a a'}  \fF_{c'd' }^{a'} \equiv  \bar{\fF}_{(cb)}^{d}  +  \bar{\fF}_{[cb]}^{d} , \nn \\
& & \bar{\fF}_{(cb)}^{d} \equiv \frac{1}{2} ( \bar{\fF}_{cb}^{d} +  \bar{\fF}_{bc}^{d} )  = \frac{1}{2} ( \eta_{c b'} \eta_{b}^{\; a} + \eta_{b b'}\bar{\eta}_{c}^{\; a} ) \bar{\eta}^{[d b']c'd'}_{a a'}  \fF_{c'd' }^{a'} , 
\ee
The antisymmetric tensors are defined as follows:
\be
& & \bar{\cD}_{d} \bar{\fF}_{\; [cb]}^{d} \equiv \eth_{d}  \bar{\fF}_{[cb]}^{d} + \mOm_{d a}^{d}  \bar{\fF}_{[cb]}^{a} - \frac{1}{2} (\mOm_{d c}^{a}  \bar{\fF}_{ab}^{d} -  \mOm_{d b}^{a}  \bar{\fF}_{ac}^{d}  + g_G\cA_{d b}^{a}  \bar{\fF}_{ca}^{d}  -  g_G\cA_{d c}^{a}  \bar{\fF}_{ba}^{d}  ) ,\nn \\
& & \bar{\fF}_{[cb]}^{d} \equiv  \frac{1}{2} ( \bar{\fF}_{cb}^{d} -  \bar{\fF}_{bc}^{d} ) = \frac{1}{2}( \eta_{cb'} \eta_{b}^{\; a} - \eta_{bb'}\eta_{c}^{\; a} ) \bar{\eta}^{[db']c'd'}_{a a'}  \fF_{c'd' }^{a'} , 
\ee
and
\be
\widetilde{T}_{[cb]} & =  &  \frac{1}{2}  (\eta_{c b'}\eta_{b}^{\; c''} - \eta_{b b'}\eta_{c}^{\; c''} )  \bar{\eta}^{[b' d] c'd'}_{a a'}  \mF_{c'' d}^{a} \mF_{c'd'}^{a'} , \nn \\
T_{[cb]} & = &   \frac{1}{4} [ \bar{\psi} \gamma_{c} i \cD_{b} \psi  - \bar{\psi} \gamma_{b} i \cD_{c} \psi + H.c.] .
 \ee

The gravitational equations in Eqs.(\ref{GEGST}) and (\ref{GEGSTAS}) provide the equivalent formalisms to characterize the gravidynamics in locally flat gravigauge spacetime spanned by the gravigauge bases. 


\subsection{New effects of gravidynamics beyond Einstein theory of general relativity}

The gravidynamics described by the geometric gravitational equation presented in Eq.(\ref{GGE}) or Eq.(\ref{GEGST}) brings about the extension to Einstein theory of general relativity, which is explicitly shown from the additional energy momentum tensor $\widetilde{T}_{\mu\nu}$ in coordinate spacetime or $\widetilde{T}_{bc}$ in gravigauge spacetime. Such an additional energy momentum tensor arises solely from the inhomogeneous spin gauge field which is governed by the inhomogeneous spin gauge symmetry. 

To explore potential new effects caused from such a gravidynamics beyond the Einstein theory of general relativity, let us turn to investigate the property of spin gauge field. It is interesting to notice from the action shown in Eq.(\ref{EBaction}) that the term given by the spin gauge invariant gravigauge field strength $\fF_{cd}^{a}$ can be rewritten into the following form:
\be \label{MSG}
 \frac{1}{4} \bar{\eta}^{c d c' d'}_{a a'}  m_G^2 \fF_{cd}^{a}\fF_{c'd'}^{a'} & = & \frac{3}{4}(\alpha_G+\frac{3}{2}\alpha_W) m_G^2 g_G^2 ( \cA_{[cab]}-\mOm_{[cab]}/g_G)  ( \cA^{[cab]}-\mOm^{[cab]}/g_G) \nn \\
 & + &  \frac{1}{4}(\alpha_G-\frac{1}{2}\alpha_W) m_G^2 g_G^2( \cA_{cab}-\mOm_{cab}/g_G)  ( \cA^{cab}-\mOm^{cab}/g_G) \nn \\
 & \equiv & \frac{1}{2} m_T^{2} ( \cA_{[cab]}-\mOm_{[cab]}/g_G)  ( \cA^{[cab]}-\mOm^{[cab]}/g_G) \nn \\
 & + &  \frac{1}{2}m_S^2 ( \cA_{(cab]}-\mOm_{(cab]}/g_G)  ( \cA^{(cab]}-\mOm^{(cab]}/g_G) ,
\ee
with 
\be
& & m_T^{2} \equiv 2(\alpha_G+\alpha_W) m_G^2 g_G^2, \nn \\
& & m_S^2 \equiv  \frac{1}{2}(\alpha_G-\frac{1}{2}\alpha_W) m_G^2 g_G^2.
\ee
Where we have introduced the following definitions:
\be
\cA_{cab} \equiv \cA_{[cab]} + \cA_{(cab]} , \quad \mOm_{cab} \equiv \mOm_{[cab]} + \mOm_{(cab]} ,
\ee   
with
\be
& & \cA_{[cab]} \equiv \frac{1}{3} ( \cA_{cab} + \cA_{abc} + \cA_{bac} ), \nn \\
& & \cA_{(cab]} \equiv \frac{1}{3} ( 2\cA_{cab} - \cA_{abc} -\cA_{bac} ) .
\ee
A similar definition holds for $\mOm_{[cab]}$ and $\mOm_{(cab]}$. Where $\cA_{[cab]}$ defines a totally antisymmetric spin gauge field which couples directly to spinor field, and  $\cA_{(cab]}$ represents a mixing symmetric spin gauge field. 

It indicates that the action of spin gauge invariant gravigauge field strength $\fF_{cd}^{a}$ brings a gauge invariant mass term for the spin gauge field. In general, the totally antisymmetric spin gauge field $\cA_{[cab]}$ as a torsion like field and the mixing symmetric spin gauge field $\cA_{(cab]}$ have different masses. Only when the coupling constants satisfy a special relation, they arrive at the same mass, i.e.:
\be \label{SMR}
& & \alpha_W = -\frac{2}{3} \alpha_G, \nn \\
& &  m_T^{2} = m_S^2 = \frac{2}{3}\alpha_G m_G^2 g_G^2 \equiv m_A^2,
\ee 
which leads to the mass $m_A$ for whole spin gauge field $\cA_{\mu}^{ab}$.

As the spin gauge field couples to all spinor fields, i,e., leptons and quarks in the standard model, so that it can in general decay into leptons and quarks when its mass becomes much larger than the masses of those particles. On the other hand, if the spin gauge field is not heavy enough, it should be produced from the colliders of the electron and hadron. As there is no any signal indicated from the experiments on such a spin gauge field, the large hadron collider LHC as the highest energy collider at present provides the most stringent constraint on the mass of spin gauge field up to several TeV for the spin gauge coupling constant being compatible to the electroweak coupling constant. In fact, various tests on the general relativity at short distance arrive at the same order of magnitude though its consistency has well been established in the macroscopic world. 

Let us first consider the gravidynamics at the distance $l_{G}$ which is larger than the one determined by the mass scale of spin gauge field, $l_{G} > l_{A} = 1/m_{A}$. In such a case, the spin gauge field goes away due to its decay into other light particles, so the gravidynamics is characterized by the following simplified equation:
\be \label{GGESE}
& & R_{\mu\nu} -  \frac{1}{2}\chi_{\mu\nu} R +  \gamma_G \widetilde{R}_{\mu\nu}  = - 8\pi G_N \1 T_{\mu\nu} , 
\ee
with the definitions:
\be
& & \gamma_G \equiv 8\pi G_N m_G^2 , \nn \\
& & \widetilde{R}_{\mu\nu} =  \bar{\nabla}_{\rho}  \bar{F}_{\;\,(\mu\nu)}^{\rho} + [ \frac{1}{2}  (\chi_{\mu\lambda}\eta_{\nu}^{\; \rho} +\chi_{\nu\lambda}\eta_{\mu}^{\; \rho}  )  \bchi^{[\lambda\sigma]\rho'\sigma'}_{a a'}   - \frac{1}{4} \chih_{\mu\nu} \bchi^{[\rho\sigma]\rho'\sigma'}_{a a'}   ] F_{\rho\sigma}^{a} F_{\rho'\sigma'}^{a'} , \nn \\
& & \bar{\nabla}_{\rho} \bar{F}_{\;\, (\mu\nu)}^{\rho} \equiv \p_{\rho}  \bar{F}_{(\mu\nu)}^{\rho} - \Gamma_{\rho \lambda}^{\rho} \bar{F}_{(\mu\nu)}^{\lambda} - \frac{1}{2} ( \Gamma_{\rho\mu}^{\lambda} \bar{F}_{\lambda\nu}^{\rho} + \Gamma_{\rho\nu}^{\lambda} \bar{F}_{\lambda\mu}^{\rho} )  ,\nn \\
& & \bar{F}_{\mu\nu}^{\rho} \equiv \chi_{\mu\sigma} \chi_{\nu}^{\;\; a} \bchi^{[\rho\sigma]\rho'\sigma'}_{a a'} F_{\rho'\sigma' }^{a'} , \quad F_{\rho\sigma }^{a}=\p_{\rho}\chi_{\sigma}^{\;\; a} - \p_{\sigma}\chi_{\rho}^{\;\; a} ,
\ee
and
\be
&& T_{\mu\nu} =   \frac{1}{4} [ \chi_{\mu}^{\; a} \bar{\psi} \gamma_{a} i \cD_{\nu} \psi  + \chi_{\nu}^{\; a} \bar{\psi} \gamma_{a} i \cD_{\mu} \psi   + H.c.]  \nn \\
& & \qquad - \chi_{\mu\nu} [   \frac{1}{2} (\bar{\psi} \gamma^{c} \chih_{c}^{\; \rho} i \cD_{\rho} \psi  + H.c.) - m \bar{\psi} \psi  ]   \nn \\
& & \qquad - (\eta_{\mu}^{\; \rho}\eta_{\nu}^{\; \rho'} - \frac{1}{4}\chi_{\mu\nu} \chih^{\rho\rho'} )   \chih^{\sigma\sigma'}\1 F_{\rho\sigma} F_{\rho'\sigma'} , \nn \\
& & i\cD_{\mu} = i\p_{\mu} + g_E A_{\mu} , 
\ee
where $T_{\mu\nu} $ is the usual energy momentum tensor arising from the Dirac fermion and electromagnetic gauge field. 

It is seen from the above equation that the new effect beyond Einstein theory of general relativity will arise from the tensor $\widetilde{R}_{\mu\nu}$ which is characterized by the gravigauge field $\chi_{\mu}^{\;\;a}$. As the spin gauge symmetry always allows us to make the gravigauge field to be symmetric $\chi_{\mu a}=\chi_{a\mu}$ by taking the physical unitary gauge, so that its degrees of freedom are the same as those of gravimetric field $\chi_{\mu\nu}$ with the relation $\chi_{\mu\nu} = \chi_{\mu a}\eta^{ab}\chi_{b\nu} \equiv (\chi)^2_{\mu\nu}$. As the tensor $\widetilde{R}_{\mu\nu}$ involves a quadratic derivative over the gravigauge field, it brings on a similar effect as that of Ricci tensor $R_{\mu\nu}$. To be more explicit, let us write down its gauge-type gravitational equation based on Eqs.(\ref{GTGE})-(\ref{GTGEC}) as follows:
\be 
 \p_{\nu} \tilde{F}^{\mu\nu }_{a}  =  \tilde{J}_{a}^{\;\; \mu} + \gamma_G \check{J}_{a}^{\;\; \mu} + 16\pi G_N \mathrm{J}_{a}^{\;\; \mu}  
\ee
with
\be
\tilde{F}^{\mu\nu }_{a} & \equiv & \chi \1 \tchi^{[\mu\nu]\mu'\nu'}_{a a'}  F_{\mu'\nu' }^{a'}  ,\nn \\
\tilde{J}_{a}^{\;\; \mu} & = & \chih_{a}^{\; \rho}  F_{\rho\nu}^{c} \tilde{F}^{\mu\nu}_{c} - \frac{1}{4} \chih_{a}^{\; \mu} \1 F_{\rho\nu}^{c} \tilde{F}^{\rho\nu}_{c} , \nn \\
\check{J}_{a}^{\;\; \mu} & = & \chi\, ( \bchi^{[\mu\nu]\mu'\nu'}_{b a'}  \chih_{a}^{\;\;\rho} - \frac{1}{4}  \bchi^{[\rho\nu]\mu'\nu'}_{b a'} \chih_{a}^{\;\;\mu}  ) F_{\rho\nu}^{b} F_{\mu'\nu'}^{a'} 
-  \p_{\nu} ( \chi\, \bchi^{[\mu\nu]\mu'\nu'}_{a a'}  F_{\mu'\nu' }^{a'} ) , \nn \\
\mathrm{J}_{a}^{\;\; \mu} & = &  \chi \1 \{ ( \chih_{a}^{\; \; \rho} \chih_{c}^{\; \; \mu} -  \chih_{a}^{\; \; \mu} \chih_{c}^{\; \; \rho}  ) \frac{1}{2} ( \bar{\psi} \gamma^{c} i \cD_{\rho} \psi + H.c.)  \nn \\
& + & \chih_{a}^{\;\;\mu}  m \bar{\psi} \psi   - ( \chih^{\mu\mu'} \chih_{a}^{\; \; \rho} - \frac{1}{4} \chih^{\rho\mu'} \chih_{a}^{\; \; \mu}  )  \chih^{\nu\nu'}\1 F_{\rho\nu} F_{\mu'\nu'} \}, 
\ee
where $\check{J}_{a}^{\;\; \mu}$ reflects the new effect of gravidynamics beyond Einstein theory of general relativity. It is seen that the usual gravitational current $\tilde{J}_{a}^{\;\; \mu}$ and the extra current $\check{J}_{a}^{\;\; \mu}$ are all described by the gravigauge field strength $F_{\mu\nu}^a$. 

It is manifest that the magnitude of new effect caused from the gravidynamics based on the inhomogeneous spin gauge symmetry at a distance larger than $1/m_G$ is characterized by the mass scale $m_G$ via the mass ratio $\gamma_G$:
\be
\gamma_{G} = 8\pi G_N m_G^2 = \frac{8\pi m_G^2}{M_P^2}  \equiv \frac{m_G^2}{\bar{M}_P^2} ,
\ee
with $M_P$ the Planck mass. Such a ratio indicates that when the mass of spin gauge field is much smaller than the Planck mass, i.e., $m_G^2 \ll \bar{M}_P^2$ and $\gamma_G\ll 1$, the resulting new effect only brings a tiny modification to the general relativity. It is expected that the future experiments could provide a reliable constraint on the mass of spin gauge field. 

Let us now discuss an alternative case that the considering energy scale $\mu_G$ is higher than the mass scale of spin gauge field, i.e., $\mu_G > m_A$. For such a short distance case, it is useful to take into account the gravitational origin of spin gauge symmetry, which brings the spin gauge field decomposed into two parts as shown in Eq.(\ref{SGFDC1}), i.e., $\cA_{\mu}^{ab} \equiv \mOm_{\mu}^{ab} + \mA_{\mu}^{ab}$.  By applying for such a decomposition of spin gauge field with the replacement $\mA_{\mu}^{ab}\to g_G \mA_{\mu}^{ab}$,  we can rewrite the gravidynamic equation given in Eq.(\ref{GGE}) into the following form:
\be \label{GGESE}
& & R_{\mu\nu} -  \frac{1}{2}\chi_{\mu\nu} R   + 8\pi G_N  \widetilde{G}_{\mu\nu}  = - 8\pi G_N \1 (T_{\mu\nu} + \check{T}_{\mu\nu} ), 
\ee
with the definitions:
\be
& & \widetilde{G}_{\mu\nu} = (\eta_{\mu}^{\; \rho}\eta_{\nu}^{\; \rho'} - \frac{1}{4}\chi_{\mu\nu} \chih^{\rho\rho'} )   \chih^{\sigma\sigma'}\1 R_{\rho\sigma\lambda'}^{\lambda} R_{\rho'\sigma' \lambda}^{\lambda'} =  R_{\mu\rho}^{\; \sigma\lambda} R_{\; \nu\sigma\lambda}^{\rho} -  \frac{1}{4}\chi_{\mu\nu}  R_{\rho\sigma}^{\; \rho'\sigma'} R_{\; \rho' \sigma' }^{\sigma\rho} , \nn \\
& & \check{T}_{\mu\nu} = - (\eta_{\mu}^{\; \rho}\eta_{\nu}^{\; \rho'} - \frac{1}{4}\chi_{\mu\nu} \chih^{\rho\rho'} )   \chih^{\sigma\sigma'} ( 2g_G^{-1}\mF_{\rho\sigma}^{ab} R_{\rho'\sigma' ab}  +  \mF_{\rho\sigma}^{ab}\mF_{\rho'\sigma' ab} ) + m_G^2 g_G \bar{\nabla}_{\rho}  \bar{\mA}_{\;\,(\mu\nu)}^{\rho}   \nn \\
& & \qquad + m_G^2 g_G^2[ \frac{1}{2}  (\chi_{\mu\lambda}\eta_{\nu}^{\; \rho} +\chi_{\nu\lambda}\eta_{\mu}^{\; \rho}  )  \bchi^{[\lambda\sigma]\rho'\sigma'}_{a a'}   - \frac{1}{4} \chih_{\mu\nu} \bchi^{[\rho\sigma]\rho'\sigma'}_{a a'}   ] \mA_{[\rho\sigma]}^{a} \mA_{[\rho'\sigma']}^{a'} ,
\ee 
and
\be
& & \bar{\nabla}_{\rho} \bar{\mA}_{\;\, (\mu\nu)}^{\rho} \equiv \p_{\rho}  \bar{\mA}_{(\mu\nu)}^{\rho} - \Gamma_{\rho \lambda}^{\rho} \bar{\mA}_{(\mu\nu)}^{\lambda} -\Gamma_{\rho\mu}^{\lambda} \bar{\mA}_{\lambda\nu}^{\rho} -\Gamma_{\rho\nu}^{\lambda} \bar{\mA}_{\lambda\mu}^{\rho} -  \frac{1}{2} g_G (  \mA_{\rho\nu}^{\lambda} \bar{\mA}_{\mu\lambda}^{\rho} +  \mA_{\rho\mu}^{\lambda} \bar{\mA}_{\nu\lambda}^{\rho} ) , \nn \\
& & \bar{\mA}_{\mu\nu}^{\rho} \equiv \chi_{\mu\sigma} \chi_{\nu}^{\;\; a} \bchi^{[\rho\sigma]\rho'\sigma'}_{a a'}  \mA_{[\rho'\sigma'] }^{a'}   ,  \quad \bar{\mA}_{(\mu\nu)}^{\rho} \equiv  \frac{1}{2} ( \bar{\mA}_{\mu\nu}^{\rho}  +  \bar{\mA}_{\nu\mu}^{\rho}  ) , \quad \mA_{[\rho\sigma] }^{a} \equiv \mA_{\rho b}^{a} \chi_{\sigma}^{\; b} - \mA_{\sigma b}^{a} \chi_{\rho}^{\; b} ,
\ee
with $\mF_{\rho\sigma}^{ab}$ and $R_{\rho\sigma}^{ab}$ defined in Eq.(\ref{DCFS3}).

It is seen that the tensor $\widetilde{G}_{\mu\nu}$ is solely described by the Riemann tensor $R_{\mu\nu\sigma}^{\rho}$, while the tensor $\check{T}_{\mu\nu}$ is proportionally given by the spin covariant gauge field $\mA_{\mu}^{ab}$. Actually, the tensor $\widetilde{G}_{\mu\nu}$ provides the radiative energy momentum of gravitational field with the following traceless property:
\be
\chih^{\mu\nu} \widetilde{G}_{\mu\nu} = \chih^{\mu\nu} (\eta_{\mu}^{\; \rho}\eta_{\nu}^{\; \rho'} - \frac{1}{4}\chi_{\mu\nu} \chih^{\rho\rho'} )   \chih^{\sigma\sigma'}\1 R_{\rho\sigma\lambda'}^{\lambda} R_{\rho'\sigma' \lambda}^{\lambda'} = 0 .
\ee
As the tensor term $\widetilde{G}_{\mu\nu}$ concerns the quadratic form of Riemann tensor, it involves high derivatives over the gravimetric field. Therefore, when the energy scale $\mu_G$ is considered to be much below the Planck scale but above the mass scale $m_A$ of spin gauge field, the effect of gravidynamics is small, i.e.:
\be
8\pi G_N  \widetilde{G}_{\mu\nu} \ll R_{\mu\nu} , \quad m_A < \mu_G \ll M_P ,
\ee
Nevertheless, once the energy scale $\mu_G$ is compatible to the Planck mass scale,  such an effect becomes significant:
\be
8\pi G_N  \widetilde{G}_{\mu\nu} \sim R_{\mu\nu} , \quad m_A < \mu_G \sim M_P .
\ee

In conclusion, at a very short distance close to or even smaller than the Planck length $l_P = 1/M_P$, the new effect caused from the gravidynamics based on the inhomogeneous spin gauge symmetry will get significantly important, it may even change the statement on the singularity of spacetime based on the Einstein theory of general relativity. It is interesting to make a quantitive study and general proof, while it is beyond the scope of present paper and will be investigated elsewhere. Actually, our starting point in this paper is mainly based on the consideration of effective theory. When approaching to the Planck scale, we should begin with exploring a more fundamental theory which is expected to involve more contents of new physics.


\section{Spinodynamics via equations of motion of spin gauge field in biframe spacetime }

Following along the least action principle and taking the spin gauge field and gravigauge field as independent degrees of freedom, we are able to derive the following equations of motion to describe the spinodynamics:
\be \label{EMSCGF}
& & \cD_{\nu} \bs{\hcF}_{ab}^{\mu\nu} - \frac{m_G^2}{\Mka^2}  \bar{\cF}_{a}^{\mu\nu} \fA_{\nu b} = \hat{\fJ}_{ab}^{\mu} , \nn \\
& & \cD_{\nu} \hcF_{ab}^{\mu\nu} - m_G^2g_G \bar{\fF}_{a}^{\mu\nu} \chi_{\nu b}= \hat{\mJ}_{ab}^{\mu} ,
\ee
where the second equation is resulted by making the scaling gauge fixing condition to be in Einstein basis. The field strength and gauge covariant derivative are defined as follows:
\be
& & \bs{\hcF}_{ab}^{\mu\nu}  \equiv  \fkA \hmH^{\mu\mu'}\hmH^{\nu\nu'}g_G^{-2} \cF_{\mu'\nu' ab} , \nn \\
& & \cD_{\nu} \bs{\hcF}_{ab}^{\mu\nu} \equiv \p_{\nu} \bs{\hcF}_{ab}^{\mu\nu} - \cA_{\nu a}^{c} \bs{\hcF}_{cb}^{\mu\nu},\nn \\
& &  \bcF_{a}^{\mu\nu} \equiv  \fkA \bar{\mH}_{aa'}^{[\mu\nu] \mu'\nu'} \cF_{\mu'\nu'}^{a'} , 
\ee
and
\be
& & \hcF_{ab}^{\mu\nu}  \equiv  \chi \chih^{\mu\mu'}\chih^{\nu\nu'} \cF_{\mu'\nu' ab} , \nn \\
& & \cD_{\nu} \hcF_{ab}^{\mu\nu} \equiv \p_{\nu} \hcF_{ab}^{\mu\nu} - g_G \cA_{\nu a}^{c} \hcF_{cb}^{\mu\nu},\nn \\
& &  \bfF_{a}^{\mu\nu} \equiv  \chi \bar{\chi}_{aa'}^{[\mu\nu] \mu'\nu'} \fF_{\mu'\nu'}^{a'} , 
\ee
with
\be
& & \bmH_{aa'}^{[\mu\nu]\mu'\nu'} \equiv \frac{1}{2} ( \bmH_{a a'}^{\mu\nu \mu'\nu'} 
- \bmH_{a a'}^{\nu\mu\mu'\nu'} ), \nn \\
& &  \bchi_{a a'}^{[\mu\nu]\mu'\nu'} \equiv \frac{1}{2} ( \bchi_{a a'}^{\mu\nu \mu'\nu'} 
- \bchi_{a a'}^{\nu\mu\mu'\nu'} ) .
\ee
Where $\bmH^{[\mu\nu]\mu'\nu'}_{a a'}$ and $\bchi^{[\mu\nu]\mu'\nu'}_{a a'}$ are defined in Eqs.(\ref{CDTensors}) and (\ref{CDTensors2}), respectively. The field strengths $\cF_{\mu\nu}^{ab}$, $\cF_{\mu\nu}^{a}$ and $\fF_{\mu\nu}^{a}$ are defined correspondingly in Eqs.(\ref{GFS}), (\ref{WESFS}) and (\ref{GIGGFS}). The corresponding currents are given by the following explicit forms:
\be
& & \hat{\fJ}_{ab}^{\mu} \equiv  \fkA \hfA^{c\mu} \bar{\Psi}_{-} \Sigma_{cab}\Psi_{-} , \nn \\
& & \hat{\mJ}_{ab}^{\mu} \equiv  \chi \chih^{c\mu} g_G\bar{\psi}\Sigma_{cab}\psi  .
\ee
The above equations of motion characterize the spinodynamics of spin gauge field.  

The inhomogeneous spin gauge symmetry brings the spin gauge field $\cA_{\mu}^{ab}$ as a vector field in Minkowski spacetime to couple with Dirac fermion via a vector gauge field $\bs{\cA}_{c}^{ab} \equiv \hfA_{c}^{\; \mu} \cA_{\mu}^{ab}$ or $\cA_{c}^{ab} \equiv \chih_{c}^{\; \mu} \cA_{\mu}^{ab}$ due to the presence of gravigauge field, which is regarded as a vector spin gauge field defined in locally flat gravigauge spacetime, which motivates us to derive the equation of motion of spin gauge field in locally flat gravigauge spacetime. The equation of motion for such a spin gauge field can directly be derived from the actions presented in Eqs.(\ref{SGIaction}) and (\ref{EBaction}) based on the least action principle. In the Einstein basis, the equation of motion is obtained as follows:
\be
\cD_{d} \cF_{ab}^{\; cd} - m_G^2g_G \bar{\fF}_{a}^{cd} \eta_{d b} = \mJ_{ab}^{\; c} ,
\ee
with the definitions:
\be
& & \cF_{ab}^{cd}  \equiv  \eta^{cc'}\eta^{dd'} \cF_{c'd' ab} , \quad  \bfF_{a}^{cd} \equiv  \bar{\eta}_{aa'}^{[cd] c'd'} \fF_{c' d'}^{a'} , \nn \\
& & \cD_{d} \cF_{ab}^{\; cd} \equiv \eth_{d} \cF_{ab}^{\; cd} + \mOm_{dc'}^{c} \cF_{ab}^{\; c'd} + \mOm_{dd'}^{d} \cF_{ab}^{\; cd'} - g_G\cA_{d a}^{a'} \cF_{a'b}^{\; cd}, 
\ee
for the field strength and gauge covariant derivative, and
\be
& & \mJ_{ab}^{\; c} \equiv  g_G\eta^{cd} \bar{\psi}\Sigma_{dab}\psi ,
\ee
for the corresponding current. 

By applying the relations shown in Eqs.(\ref{MSG})-(\ref{SMR}), we are able to rewrite the above equation of motion into the following forms:
\be
& & \cD_{d} \cF_{\; \; [cab]}^{d} +  m_T^2  \cA_{[cab]}  = m_T^2g_G^{-1} \mOm_{[cab]} - \mJ_{cab}, \nn \\
& & \cD_{d} \cF_{\; \; (cab]}^{d} +  m_S^2 \cA_{(cab]}  = m_S^2g_G^{-1} \mOm_{(cab]},
\ee
with
\be
& & \cF_{\;\; cab}^{d} \equiv \eta^{dd'} \cF_{d'c ab} \equiv \cF_{\;\; [cab]}^{d} + \cF_{\; \;(cab]}^{d} , \nn \\
& & \cF_{\; \; [cab]}^{d} \equiv \frac{1}{3} ( \cF_{\;\; cab}^{d} + \cF_{\;\; abc}^{d}  + \cF_{\;\; bca}^{d}  ) , \nn \\
& &  \cF_{\; \; (cab]}^{d} \equiv \frac{1}{3}  (2 \cF_{\;\; cab}^{d} - \cF_{\;\; abc}^{d}  - \cF_{\;\; bca}^{d}  ) , \nn \\
& & \mJ_{cab} \equiv  g_G \bar{\psi}\Sigma_{cab}\psi , 
\ee
where the totally antisymmetric spin gauge field $\cA_{[cab]}$ with mass $m_T$ reflects the so-called torsion like field. 

The above equations of motion for the spin gauge field describe the spinodynamics in locally flat gravigauge spacetime, which can also be derived from Eq.(\ref{EMSCGF}) by projecting it into locally flat gravigauge spacetime via the gravigauge field $\chi_{\mu}^{\; a}$.

\section{Electrodynamics in gravitational quantum field theory and generalized Maxwell equations in biframe spacetime }

Based on the maximal joint symmetry for the action of chirality-based Dirac spinor, namely the inhomogeneous spin gauge symmetry and inhomogeneous Lorentz symmetry (or Poincar\'e symmetry), we have provided a detailed analysis and investigation on the gravitational relativistic quantum theory of Dirac spinor as well as the gravidynamics and spinodynamics, which brings about the theoretical framework of gravitational quantum field theory. In this section, we are going to investigate systematically the electrodynamics in the presence of gravitational interaction and to derive the generalized Maxwell equations in Minkowski spacetime of coordinates and also in locally flat gravigauge spacetime spanned by the gravigauge basis.

\subsection{Gravigeometry-medium electrodynamics and generalized Maxwell equations  with  gravigeometry-medium electromagnetic field}

Following along the least action principle, we obtain the following equations of motion for the electromagnetic gauge field:
\be \label{GMED}
& & \p_{\nu} \widehat{\fF}^{\mu\nu} = - \wh{\fJ}^{\mu} , \nn \\
& & \p_{\nu} \widehat{F}^{\mu\nu} = -\wh{J}^{\mu} ,
\ee
with the field strengths and currents defined by
\be \label{GFS0}
& & \widehat{\fF}^{\mu\nu} \equiv \fkA \hat{\mH}^{[\mu\nu]\mu'\nu'} F_{\mu'\nu'} , \nn \\
& &  \widehat{F}^{\mu\nu} \equiv \chi \hat{\chi}^{[\mu\nu]\mu'\nu'}  F_{\mu'\nu'} , \nn \\
& & \hat{\mH}^{[\mu\nu]\mu'\nu'} \equiv \frac{1}{2} ( \hmH^{\mu\mu'}\hmH^{\nu\nu'} - \hmH^{\nu\mu'}\hmH^{\mu\nu'} ) , \nn \\
& &  \hat{\chi}^{[\mu\nu]\mu'\nu'} \equiv \frac{1}{2} ( \chih^{\mu\mu'}\chih^{\nu\nu'} - \chih^{\nu\mu'}\chih^{\mu\nu'} ) ,
\ee
and 
\be \label{GFS1}
& & \wh{\fJ}^{\mu} \equiv \fkA \hfA_{a}^{\; \mu} \fJ^{a} , \quad \fJ^{a}  = - g_E \bar{\Psi}_{-} \Gamma^{a} \Psi_{-} , \nn \\
& & \wh{J}^{\mu}  \equiv \chi \chih_{a}^{\; \mu} J^{a} , \quad  J^{a} = - g_E \bar{\psi} \gamma^{a} \psi .
\ee 
The second equation in Eq.(\ref{GGE}) holds in the Einstein basis for the gauge fixing condition of scaling gauge symmetry. Where $\widehat{\fF}^{\mu\nu}$ ($\widehat{F}^{\mu\nu}$) and $\wh{\fJ}^{\mu}$ ($\wh{J}^{\mu}$) define the electromagnetic gauge field strength and the corresponding current density in the presence of gravimetric field, and $\fJ^{a}$ ($J^{a}$) is regarded as a free current density in the absence of gravitational interaction.

Meanwhile, the gauge field strength $F_{\mu\nu}$ satisfies the Bianchi identity,
\be
\varepsilon^{\mu\nu\rho\sigma} \p_{\nu} F_{\rho\sigma} = 0 .
\ee

The above equations of motion in the Einstein basis can be rewritten into the following four sets of equations:
\be \label{EME}
& & \nabla \cdot \widehat{\fE} = \wh{J}^0 , \nn \\
& & \nabla \times \widehat{\fB} = \boldsymbol{\wh{J}} + \frac{\p \widehat{\fE}}{\p x^0} , \nn \\
& & \nabla \cdot \fB = 0 , \nn \\
& & \nabla \times \fE = - \frac{\p \fB}{\p x^0} , 
\ee
where we have introduced the definitions:
\be 
& & \widehat{\fE} \equiv (\widehat{E}^{1}, \widehat{E}^{2}, \widehat{E}^{3} ) = ( -\widehat{F}^{0 1},   -\widehat{F}^{0 2},  -\widehat{F}^{0 3} ) , \nn \\
& &  \widehat{\fB} \equiv (\widehat{B}^{1}, \widehat{B}^{2}, \widehat{B}^{3} )  = ( -\widehat{F}^{23},  -\widehat{F}^{31},  -\widehat{F}^{12} )  , \nn \\
& & \fE \equiv (E^{1}, E^{2}, E^{3} ) = ( -F^{0 1},   - F^{0 2},  - F^{0 3} ) , \nn \\
& &  \fB \equiv (B^{1}, B^{2}, B^{3} ) = ( - F^{23},  - F^{31},  - F^{12} )  . 
\ee

The vector fields $\widehat{\fE}$ and $\widehat{\fB}$ define the electric field and magnetic field in the presence of gravimetric field which is introduced to characterize the Riemann geometry of curved spacetime. For convenience, we may refer to $\widehat{\fE}$ and $\widehat{\fB}$ as {\it gravigeometry-medium electric field} and {\it gravigeometry-medium magnetic field}, respectively. The generalized Maxwell equations presented in Eq.(\ref{EME}) may be mentioned as {\it gravigeometry-medium Maxwell equations}, and the corresponding electrodynamics is referred to as {\it gravigeometry-medium electrodynamics}.

In the ordinary Maxwell equations yielded in globally flat Minkowski spacetime, the corresponding field strength with $\fE\equiv \{ E^{i} = -F^{0i} \}$ (i=1,2,3) and $\fB\equiv \{ B^{i} = -\frac{1}{2}\epsilon^{i}_{\;\; jk}F^{jk} \}$ (i,j,k=1,2,3) defines a free-motion electromagnetic field in the absence of gravitational interaction. In the gravigeometry-medium Maxwell equations shown in Eq.(\ref{EME}) due to the presence of gravimetric field, the field strength defined via $\widehat{\fE}\equiv \{ \widehat{E}^{i} = -\widehat{F}^{0i} \}$ and $\widehat{\fB}\equiv \{ \widehat{B}^{i} = -\frac{1}{2}\epsilon^{i}_{\;\; jk}\widehat{F}^{jk} \}$ are regarded as gravigeometry-medium electromagnetic field. The relations between the gravigeometry-medium electromagnetic field and the free motion electromagnetic field can directly be written down from the definitions in Eq.(\ref{GFS}) as follows:
\be
& & \widehat{F}^{0i} = \chi (\chih^{00}\chih^{ij} - \chih^{0i}\chih^{0j}) F_{0j} +  \chi \chih^{ik}\chih^{0j} F_{jk} \equiv -\widehat{E}^i , \nn \\
& &  \widehat{F}^{ij} = \chi (\chih^{i0}\chih^{jk}- \chih^{j0} \chih^{ik} ) F_{0k} +  \chi \chih^{ik}\chih^{jl} F_{kl} \equiv \epsilon^{ij}_{\;\;\; k} \widehat{B}^{k} ,
\ee
which can be expressed into the following simple forms:
\be
& & \widehat{E}^{i} = \hat{\varepsilon}^{i}_{\; j} E^{j} + \hat{\alpha}^{i}_{\; j} B^{j} , \nn \\
& & \widehat{B}^{i} = \hat{\mu}^{i}_{\; j} B^{j} - \hat{\alpha}^{\; \, i}_{j} E^{j} , 
\ee
with the coefficient functions:
\be
& & \hat{\varepsilon}^{i}_{\; j}(x) \equiv \chi( \chih^{0}_{\;0}\chih^{i}_{\;j} - \chih^{i}_{\;0}\chih^{0}_{\;j} ) ,  \nn \\
& & \hat{\mu}^{i}_{\; j}(x) \equiv -  \frac{1}{2} \chi \epsilon^{i}_{\;\; kl}\chih^{kk'}\chih^{l l'} \epsilon_{k'l'j} , \nn \\
& &  \hat{\alpha}^{i}_{\; j}(x) \equiv   \chi \chih^{ik}\chih^{l0} \epsilon_{kl j}  .
\ee
Note that all the coefficients are functions of coordinate spacetime and they are in general the vectors and tensors in spatial dimensions. 

It is useful to rewrite the above relations into the following forms in the vector representation:
\be \label{GGMFS}
& & \widehat{\fE} \equiv \widehat{\bs{\varepsilon}}\cdot \fE + \widehat{\bs{\alpha}}\cdot \fB , \nn \\
& & \widehat{\fB} \equiv \widehat{\bs{\mu}}\cdot \fB - \widehat{\bs{\alpha}}^{T} \cdot \fE ,
\ee
with the bi-vector coefficient functions:
\be
& & \widehat{\bs{\varepsilon}}(x) \equiv \{ \hat{\varepsilon}^{i}_{\; j}(x) \}, 
\quad \widehat{\bs{\mu}}(x) \equiv \{ \hat{\mu}^{i}_{\; j}(x) \} , \nn \\
& & \widehat{\bs{\alpha}}(x) \equiv \{ \hat{\alpha}^{i}_{\; j}(x) \} , 
\quad \widehat{\bs{\alpha}}^{T}(x) \equiv \{ \hat{\alpha}_{j}^{\; i}(x) \} .
\ee
These coefficient functions are in principle governed by the gravidynamics.


\subsection{ Gravigauge-mediated electrodynamics and generalized Maxwell equations  in locally flat gravigauge spacetime }

In the absence of gravitational interaction, the electromagnetic gauge field $A_{\mu}$ is regarded as a vector field in globally flat Minkowski spacetime. The inhomogeneous spin gauge symmetry with the introduction of gravigauge field $\chi_{\mu}^{\; a}$ ($\chih_{a}^{\; \mu}$) brings $A_{\mu}$ to couple with Dirac fermion via a vector gauge field $A_{a} \equiv \chih_{a}^{\; \mu} A_{\mu}$ defined in locally flat gravigauge spacetime. The least action principle leads the equation of motion of gauge field $A_{a}$ to have the following form in locally flat gravigauge spacetime:
\be \label{EMEM}
D_{b} F^{ab} = - J^{a} , 
\ee
with the field strength and covariant derivative defined as follows:
\be
D_{b} F^{ab} & \equiv & \eth_{b} F^{ab} + \mOm_{bc}^{a} F^{cb} + \mOm_{bc}^{b} F^{ac} , \nn \\
F^{ab} & \equiv & \eta^{ac}\eta^{bd} F_{cd} \equiv \chih^{ a \mu} \chih^{b\nu} F_{\mu\nu}, \quad F_{\mu\nu} = \p_{\mu}A_{\nu} - \p_{\nu}A_{\mu} , \nn \\
F_{cd} & \equiv & D_{c}A_{d} - D_{d}A_{c} = \eth_{c}A_{d} - \eth_{d}A_{c} + \mOm_{cd}^{b} A_{b} -  \mOm_{dc}^{b} A_{b} , 
\ee
and the current density is simply given by
\be
J^{a} = - g_E \bar{\psi} \gamma^a \psi ,
\ee
which is considered as the current density for freely moving Dirac fermion in locally flat gravigauge spacetime.

The field strength $F_{ab}$ satisfies the following identity:
\be \label{EMEM1}
\epsilon^{cabd}D_{c}F_{ab} = 0 ,
\ee 
which results from the Bianchi identity of Riemann curvature tensor $\epsilon^{cabd} R_{cab}^{b'}= 0$ in locally flat gravigauge spacetime. It can directly be verified as follows:
\be
\epsilon^{cabd}D_{c}F_{ab} = 2\epsilon^{cabd}D_{c}(D_{a}A_{b} ) 
= - \epsilon^{cabd} R_{cab}^{b'} A_{b'} = 0 .
\ee 

It can be proved from Eq.(\ref{EMEM}) that the current density $J^{a}$ is conserved in a gauge covariant form, i.e.:
\be
D_{a}J^{a} & = & -D_{a}D_{b}F^{ab} = \frac{1}{2}( R_{abc}^{a}F^{cb} + R_{abc}^{b}F^{ac} ) \nn \\
& = & R_{cb} F^{cb} = \frac{1}{2} (R_{cb} - R_{bc} ) F^{cb} = 0 ,
\ee
where the last equality is due to the symmetric feature of Ricci curvature tensor $R_{cb}=R_{bc}$ and the antisymmetric property of field strength $F^{cb}= - F^{bc}$. It is straightforward to write down the quadratic form for the equation of motion of gauge field $A_{a}$ shown in Eq.(\ref{EMEM}) as follows:
\be
D_{c}D^{c} A_{a} +  R_{ac} A^{c} - D_{a} (D_{c} A^{c}) = J_{a} .
\ee
When taking the following covariant gauge fixing condition:
\be
D_{c} A^{c} = \eth_{c}A^{c} + \mOm_{cd}^{c} A^{d} = 0,
\ee
and utilizing the equation of motion of gravidynamics presented in Eq.(\ref{GEGST}), we arrive at the following equation of motion for the gauge field $A_{a}$ in locally flat gravigauge spacetime:
\be
D_{c}D^{c} A_{a} +  \frac{1}{2}R\1 A_{a} - 8\pi G_N (T_{ac} + \widetilde{T}_{ac}) A^{c} = J_{a} ,
\ee
which indicates that a non-zero Ricci curvature scalar $R\neq 0$ brings on an effective mass to the electromagnetic gauge field $A_{a}$ when it propagates in locally flat gravigauge spacetime. 


It is useful to rewrite the equations of motion in Eqs.(\ref{EMEM}) and (\ref{EMEM1}) into the following forms:
\be \label{EMEM2}
& & \eth_{b}F^{ba} - \bvA_{b} F^{ba} - \frac{1}{2} F_{bc}^{a} F^{bc} = J^{a} , \nn \\
& & \epsilon^{abcd} ( \eth_{b}F_{cd} + F_{bc}^{c'} F_{c' d} ) = 0 , 
\ee
with the definitions:
\be
& & F_{bc}^{a} \equiv - \mOm_{[bc]}^{a} = \mOm_{cb}^{a} - \mOm_{bc}^{a} \equiv \chih_{b}^{\;\;\mu} \chih_{c}^{\;\; \nu} F_{\mu\nu}^{a} \equiv \hat{F}_{cb}^{\mu}\chi_{\mu}^{\;\; a} , \nn \\
& & \hat{F}_{cb}^{\mu} \equiv \eth_{c}\chih_{b}^{\;\; \mu} - \eth_{b}\chih_{c}^{\;\; \mu} , \quad \eth_{b} \equiv \chih_{b}^{\;\; \nu}\p_{\nu} , \nn \\
& &  \bvA_{b} \equiv F_{cb}^{c} \equiv  - \mOm_{cb}^{c}  = -\chih \p_{\nu}(\chi \chih_{b}^{\;\; \nu} ) , 
\ee
where $F_{bc}^{a}$ is the gravigauge field strength in locally flat gravigauge spacetime. The equations of motion in locally flat gravigauge spacetime with $a, b =(0, \alpha)$ ($\alpha=1,2,3$) can be expressed as follows:
\be\label{GME0}
& & \eth_{\alpha}\wt{E}^{\alpha} - \bvA_{\alpha}\wt{E}^{\alpha} - \bvE_{\alpha} \wt{E}^{\alpha}  = J^0 - \bvB_{\alpha} \wt{B}^{\alpha}, \nn \\
& &  \epsilon^{\beta\alpha\gamma} ( \eth_{\beta} \wt{B}_{\gamma} -  \bvA_{\beta} \wt{B}_{\gamma} )  + \bvB_{\beta}^{\;\;\alpha}\wt{B}^{\beta} = J^{\alpha}  + \eth_{0}\wt{E}^{\alpha} - \bvA_{0}\wt{E}^{\alpha} +\bvE_{\beta}^{\;\;\alpha} \wt{E}^{\beta}, \nn \\
& & \eth_{\alpha}\wt{B}^{\alpha}  + 2\bvA_{\alpha}\wt{B}^{\alpha} + 2\bv{E}_{\alpha}\wt{B}^{\alpha} =  \bv{B}_{\alpha}\wt{E}^{\alpha} , \nn \\
& & \eth_{0}\wt{B}^{\alpha} + 2\bvA_{0}\wt{B}^{\alpha} - 2\bvE_{\beta}^{\;\; \alpha} \wt{B}^{\beta} = -  \epsilon^{\beta\alpha\gamma} ( \eth_{\beta} \wt{E}_{\gamma} +  \bv{E}_{\beta} \wt{E}_{\gamma} )  - \bvB^{\;\; \alpha}_{\beta} \wt{E}^{\beta} , 
\ee
with the field strengths defined to be:
\be \label{GGFS}
& & \wt{E}^{\alpha} = - F^{0\alpha} , \quad \wt{B}^{\alpha} = -\frac{1}{2}\epsilon^{\alpha}_{\;\; \beta\gamma} F^{\beta\gamma} , \nn \\
& & \bv{E}_{\alpha} = - F_{0\alpha}^{0} , \quad \bv{B}_{\alpha} = -\frac{1}{2}\epsilon_{\alpha}^{\;\; \gamma\delta} F_{\gamma\delta}^{0} , \nn \\
& &  \bv{E}_{\alpha}^{\;\; \beta} = - F_{0\alpha}^{\beta} , \quad \bv{B}_{\alpha}^{\;\; \beta} = -\frac{1}{2}\epsilon_{\alpha}^{\;\; \gamma\delta} F_{\gamma\delta}^{\beta} .
\ee

The field strengths $ \wt{E}^{\alpha}$ and $\wt{B}^{\alpha}$ are defined as electromagnetic field in locally flat gravigauge spacetime, which may be called as {\it gravigauge-dressed electric field} and {\it gravigauge-dressed magnetic field}, respectively, in order to distinguish with the free-motion electromagnetic field $\fE$ and $\fB$ defined in Minkowski spacetime. $\bv{E}_{\alpha}$ and $\bv{B}_{\alpha}$ as well as $\bv{E}_{\alpha}^{\;\; \beta}$ and $\bv{B}_{\alpha}^{\;\; \beta}$ are defined as {\it electromagnetic-like gravigauge fields} to characterize the gravitational interactions. For convenience, $\bv{E}_{\alpha}$ and $\bv{B}_{\alpha}$ are referred to as {\it electric-like gravigauge field} and {\it magnetic-like gravigauge field}, respectively, while $\bv{E}_{\alpha}^{\;\; \beta}$ and $\bv{B}_{\alpha}^{\;\; \beta}$ are mentioned as {\it electric-like gravigauge tensor field} and {\it magnetic-like gravigauge tensor field}, respectively. $\bvA_{0}$ and $\bvA_{\alpha}$ are related to the electromagnetic-like gravigauge field as follows:
\be
& &  \bvA_{0} = \eta_{\alpha}^{\;\; \beta} \bvE_{\beta}^{\;\; \alpha}, \quad \bvA_{\alpha} = - \bvE_{\alpha} - \epsilon_{\alpha\beta}^{\;\;\;\;\, \gamma}\bvB_{\gamma}^{\;\;\beta} .
\ee

The equations of motion given in Eqs.(\ref{EMEM}) and (\ref{EMEM1}) or Eqs.(\ref{EMEM2})-(\ref{GGFS}) characterize the electrodynamics in locally flat gravigauge spacetime, which may be referred to as {\it gravigauge-mediated electrodynamics}. 

In light of vector representation, the equations of motion for the gravigauge-dressed electromagnetic fields $\wt{\fE}$ and $\wt{\fB}$ as shown in Eq.(\ref{GME0}) can be rewritten into the following forms in locally flat gravigauge spacetime:
\be\label{GME}
& & \bvnabla \cdot \wt{\fE} - \bvfA\cdot \wt{\fE} -\bvfE \cdot \wt{\fE}   = J^0 + \bvfB \cdot \wt{\fB},   \nn \\
& & \bvnabla \times \wt{\fB} - \bv{\fA}\times \wt{\fB} + \bv{\mathbb{B}}\cdot \wt{\fB} = \fJ + \eth_0 \wt{\fE} - \bv{A}_0 \wt{\fE} + \bv{\mathbb{E}}\cdot \wt{\fE}, \nn \\
& & \bvnabla \times \wt{\fE} + \bv{\fE}\times \wt{\fE} - \bv{\mathbb{B}}\cdot \wt{\fE} = - \eth_0 \wt{\fB} - 2\bv{A}_0 \wt{\fB} + \bv{\mathbb{E}}\cdot \wt{\fB}, \nn \\
& & \bvnabla \cdot \wt{\fB} +2 \bvfA\cdot \wt{\fB} + 2\bvfE \cdot \wt{\fB}   =  \bvfB \cdot \wt{\fE},   
\ee  
where we have introduced the following vector-like notations:
\be
& & \wt{\fE}\equiv (\wt{E}^{1}, \wt{E}^{2}, \wt{E}^{3}), \quad \wt{\fB}\equiv  (\wt{B}^{1}, \wt{B}^{2}, \wt{B}^{3}), \nn \\
& & \bv{\fE}\equiv (\bv{E}_{1}, \bv{E}_{2}, \bv{E}_{3}), \quad \bv{\fB}\equiv  (\bv{B}_{1}, \bv{B}_{2}, \bv{B}_{3}), \nn \\
& & \bv{\fE}^{\alpha} \equiv (\bv{E}_{1}^{\;\;\alpha}, \bv{E}_{2}^{\;\;\alpha}, \bv{E}_{3}^{\;\;\alpha}), \quad \bv{\fB}^{\alpha} \equiv (\bv{B}_{1}^{\;\;\alpha}, \bv{B}_{2}^{\;\;\alpha}, \bv{B}_{3}^{\;\;\alpha}), \nn \\
& & \bv{\fA} \equiv  (\bv{A}_{1}, \bv{A}_{2}, \bv{A}_{3}) ,
\ee
and 
\be
& & \bv{\mathbb{E}}\equiv \{ \bv{\fE}^{\alpha}\} \equiv \{\bv{\fE}^{1}, \bv{\fE}^{2}, \bv{\fE}^{3}\} \nn \\
& & \quad \equiv \{ (\bv{E}_{1}^{\;\; 1}, \bv{E}_{2}^{\;\; 1}, \bv{E}_{3}^{\;\; 1} ) , (\bv{E}_{1}^{\;\; 2}, \bv{E}_{2}^{\;\; 2}, \bv{E}_{3}^{\;\; 2} ), (\bv{E}_{1}^{\;\; 3}, \bv{E}_{2}^{\;\; 3}, \bv{E}_{3}^{\;\; 3}) \} , \nn \\
& & \bv{\mathbb{B}} \equiv \{ \bv{\fB}^{\alpha}\} \equiv \{\bv{\fB}^{1}, \bv{\fB}^{2}, \bv{\fB}^{3}\} \nn \\
& & \quad \equiv \{ (\bv{B}_{1}^{\;\; 1}, \bv{B}_{2}^{\;\; 1}, \bv{B}_{3}^{\;\; 1} ) , (\bv{B}_{1}^{\;\; 2}, \bv{B}_{2}^{\;\; 2}, \bv{B}_{3}^{\;\; 2} ), (\bv{B}_{1}^{\;\; 3}, \bv{B}_{2}^{\;\; 3}, \bv{B}_{3}^{\;\; 3}) \} , \nn \\
& & \bv{\nabla} \equiv (\eth_{1}, \eth_{2}, \eth_{3}) \equiv \{ \chih_{\alpha}^{\;\; 0} \p_{0} + \chih_{\alpha}^{\;\; i} \p_{i}  \} , \quad \eth_0  \equiv \chih_{0}^{\;\; \mu}\p_{\mu} \equiv \chih_{0}^{\;\; 0}\p_{0} +  \chih_{0}^{\; \; i} \p_{i}  ,
\ee
with $\bv{\mathbb{E}}$ and $\bv{\mathbb{B}}$ denoting electromagnetic-like gravigauge bi-vector fields. Where $\bv{\nabla}$ and $\eth_{0}$ represent the gravi-coordinate derivatives, which are related to the ordinary coordinate derivatives as follows:
\be
& & \bv{\nabla} \equiv (\eth_{1}, \eth_{2}, \eth_{3}) \equiv \{ \chih_{\alpha}^{\;\; 0} \p_{0} + \chih_{\alpha}^{\;\; i} \p_{i}  \}  \equiv \bv{\bs{\kappa}} \frac{\p}{\p x^{0}} + \mathring{\bs{\kappa}}\cdot \nabla, \nn \\
& & \eth_0  \equiv \chih_{0}^{\; \; \mu} \p_{\mu} = \chih_{0}^{\;\; 0}\p_{0} +  \chih_{0}^{\; \; i} \p_{i}  \equiv \bv{\kappa} \frac{\p}{\p x^{0}} + \hat{\bs{\kappa}}\cdot \nabla , 
\ee
with the definitions:
\be \label{GGkappa}
& & \bv{\bs{\kappa}}(x) \equiv \{\chih_{\alpha}^{\;\; 0}\} \equiv (\chih_{1}^{\;\; 0}, \chih_{2}^{\;\; 0}, \chih_{3}^{\;\; 0} ) , \quad \hat{\bs{\kappa}}_{\alpha}(x) \equiv \{ \chih_{\alpha}^{\;\; i} \} \equiv (\chih_{\alpha}^{\;\; 1}, \chih_{\alpha}^{\;\; 2}, \chih_{\alpha}^{\;\; 3} ) , \nn \\
& & \mathring{\bs{\kappa}}(x) \equiv \{ \hat{\bs{\kappa}}_{1}, \hat{\bs{\kappa}}_{2}, \hat{\bs{\kappa}}_{3} \} \equiv \{ ( \chih_{1}^{\;\; 1}, \chih_{1}^{\;\; 2}, \chih_{1}^{\;\; 3} ),  ( \chih_{2}^{\;\; 1}, \chih_{2}^{\;\; 2}, \chih_{2}^{\;\; 3} ),  ( \chih_{3}^{\;\; 1}, \chih_{3}^{\;\; 2}, \chih_{3}^{\;\; 3} ) \} , \nn \\
& & \bv{\kappa}(x) \equiv \chih_{0}^{\;\; 0} , \quad \hat{\bs{\kappa}}(x) \equiv \{ \chih_{0}^{\; \; i}\}\equiv (\chih_{0}^{\;\; 1}, \chih_{0}^{\;\; 2}, \chih_{0}^{\;\; 3} ) .
\ee

In terms of the above definitions and notations for the gravi-coordinate derivatives, the gravigauge Maxwell equations in Eq.(\ref{GME}) can be expressed as follows:
\be\label{GME1}
& &  \bv{\bs{\kappa}}  \cdot \frac{\p \wt{\fE}}{\p x^{0}} + (\mathring{\bs{\kappa}}\cdot \nabla) \cdot \wt{\fE} - \bvfA\cdot \wt{\fE} -\bvfE \cdot \wt{\fE}   = J^0 + \bvfB \cdot \wt{\fB},   \nn \\
& &   \bv{\bs{\kappa}} \times \frac{ \p \wt{\fB}}{\p x^{0}} + (\mathring{\bs{\kappa}}\cdot \nabla) \times \wt{\fB}  - \bv{\fA}\times \wt{\fB} + \bv{\mathbb{B}}\cdot \wt{\fB} = \fJ + \bv{\kappa} \frac{\p \wt{\fE}}{\p x^{0}} + (\hat{\bs{\kappa}}\cdot \nabla) \wt{\fE} - \bv{A}_0 \wt{\fE} + \bv{\mathbb{E}}\cdot \wt{\fE}, \nn \\
& & \bv{\bs{\kappa}} \times \frac{ \p \wt{\fE}}{\p x^{0}} + (\mathring{\bs{\kappa}}\cdot \nabla) \times \wt{\fE}  + \bv{\fE}\times \wt{\fE} - \bv{\mathbb{B}}\cdot \wt{\fE} = - \bv{\kappa} \frac{\p \wt{\fB}}{\p x^{0}} - (\hat{\bs{\kappa}}\cdot \nabla) \wt{\fB} - \bv{A}_0 \wt{\fE} - 2\bv{A}_0 \wt{\fB} + \bv{\mathbb{E}}\cdot \wt{\fB}, \nn \\
& & \bv{\bs{\kappa}}  \cdot \frac{\p \wt{\fB}}{\p x^{0}} + (\mathring{\bs{\kappa}}\cdot \nabla) \cdot \wt{\fB} +2 \bvfA\cdot \wt{\fB} + 2\bvfE \cdot \wt{\fB}   =  \bvfB \cdot \wt{\fE}.   
\ee  

It is noticed that the equations in Eq.(\ref{GME}) and Eq.(\ref{GME1}) (or Eq.(\ref{GME0})) bring on the generalized Maxwell equations  in locally flat gravigauge spacetime. To distinguish with the gravigeometry-medium Maxwell equations presented in Eq.(\ref{GMED}) and Eq.(\ref{EME})-(\ref{GGMFS}) in coordinate spacetime, they may be referred to as {\it gravigauge-mediated Maxwell equations} which are derived to describe the {\it gravigauge-mediated electrodynamics} in locally flat gravigauge spacetime.

The gravigauge-mediated Maxwell equations show explicitly how the gravigauge-dressed electromagnetic fields $\wt{\fE}$ and $\wt{\fB}$ interact directly with the electromagnetic-like gravigauge fields $\bv{\fE}$ and $\bv{\fB}$ as well as electromagnetic-like gravigauge vector fields $\bv{\mathbb{E}}$ and $\bv{\mathbb{B}}$ in locally flat gravigauge spacetime, which reveals how the gravitational effect emerges to bring the gravigauge-mediated electrodynamics in locally flat gravigauge spacetime.


\subsection{Gravimetric-gauge-mediated electrodynamics and generalized Maxwell equations in dynamic Riemannian spacetime}

 The equations of motion for the gauge field $A_{\mu}$ can also be expressed as the following form in dynamic Riemannian spacetime:
\be \label{GMED}
& &  \nabla^{\nu} F_{\mu\nu} = -J_{\mu} ,  \nn \\
& &  \varepsilon^{\mu\nu\rho\sigma} \p_{\rho} F_{\mu\nu}  = 0 ,
\ee
with the covariant derivative defined as follows:
\be
& & \nabla^{\nu} F_{\mu\nu} \equiv  \chih^{\nu\rho} \nabla_{\rho} F_{\mu\nu} =  \chih^{\nu\rho} ( \p_{\rho} F_{\mu\nu}  - \Gamma_{\rho\mu}^{\lambda} F_{\lambda\nu}  - \Gamma_{\rho\nu}^{\lambda} F_{\mu\lambda} ) .\ee
The second equation in Eq.(\ref{GMED}) is the Bianchi identity which can also be rewritten in terms of the covariant derivative:
\be
\varepsilon^{\mu\nu\rho\sigma} \p_{\rho} F_{\mu\nu} \equiv \varepsilon^{\mu\nu\rho\sigma} \nabla_{\rho} F_{\mu\nu} \equiv \varepsilon^{\mu\nu\rho\sigma} ( \p_{\rho} F_{\mu\nu} - \Gamma_{\rho\mu}^{\lambda} F_{\lambda\nu} - \Gamma_{\rho\nu}^{\lambda} F_{\mu\lambda} )   = 0 ,
\ee
where we have used the symmetric property of the affine connection (Christoffel symbol) $\Gamma_{\rho\mu}^{\lambda} = \Gamma_{\mu\rho}^{\lambda}$.
  
The current density in Eq.(\ref{GMED}) is defined by,
\be
J_{\mu} \equiv \chi_{\mu}^{\;\; a} J_{a} = - g_E \chi_{\mu}^{\;\; a}\bar{\psi} \gamma_a \psi .
\ee
which is conserved via the covariant derivative:
\be
\nabla^{\mu}J_{\mu} & = & -\nabla^{\mu}\nabla^{\nu}F_{\mu\nu} = \frac{1}{2}( R^{\mu\nu\rho}_{\mu}F_{\rho\nu} + R^{\mu\nu\rho}_{\nu} F_{\mu\rho} ) = R^{\mu\rho} F_{\mu\rho} =  0 ,
\ee
where the last equality is due to the symmetric feature of Ricci curvature tensor $R^{\mu\rho}=R^{\rho\mu}$ and the antisymmetric property of electromagnetic field strength $F_{\mu\rho}= - F_{\rho\mu}$.

In can be checked that the equation of motion of gauge field $A_{\mu}$ gets the following form:
\be
\nabla^{\nu}\nabla_{\nu} A_{\mu}  + R_{\mu\nu} A^{\nu} - \nabla_{\mu} (\nabla^{\nu} A_{\nu}) = J_{\mu} .
\ee
By imposing the covariant gauge fixing condition $\nabla^{\nu} A_{\nu} = 0$, we come to the following equation of motion for the gauge field $A_{\mu}$:
\be
& & \nabla^{\nu}\nabla_{\nu} A_{\mu}  + R_{\mu\nu} A^{\nu} = J_{\mu} . 
\ee
When applying for the geometric gravitational equation shown in Eq.(\ref{GGE}), we have,
\be
& &  ( \nabla^{\nu}\nabla_{\nu} + \frac{1}{2}R )  A_{\mu}  - 8\pi G_N (T_{\mu\nu} + \widetilde{T}_{\mu\nu}) A^{\nu}  = J_{\mu} ,
\ee
which indicates that a non-zero Ricci curvature scalar $R\neq 0$ leads to an effective mass-like term for the electromagnetic gauge field $A_{\mu}$ when it is propagating in a curved Riemannian spacetime. 

To reflect the effect of geometric gravidynamics on the electrodynamics, it is useful to express the equations of motion in Eq.(\ref{GMED}) into the following form:
\be 
(\hat{\p}^{\nu}  + \ck{A}^{\nu} ) F_{\nu\mu}  + \frac{1}{2} \chih^{\rho\nu}\chih^{\sigma\nu'}F_{\rho\sigma}^{\lambda} \eta_{\lambda\mu} F_{\nu\nu'}   = J_{\mu}, 
\ee
where we have introduced the definitions:
\be
& & F_{\rho\sigma}^{\lambda} \equiv \p_{\rho}\chi_{\sigma}^{\; \lambda} - \p_{\sigma}\chi_{\rho}^{\; \lambda} , \quad \chi_{\rho}^{\; \lambda} \equiv  \chi_{\rho \mu} \eta^{\mu\lambda}, \nn \\
& &  \ck{A}^{\nu}  \equiv \frac{1}{2} \chih^{\nu\rho} \chih^{\mu\sigma} F_{\rho\sigma}^{\lambda}\eta_{\mu\lambda} \equiv \hat{\p}^{\nu}\ln \chi + \p_{\rho} \chih^{\rho\nu}, \nn \\
& &  \hat{\p}^{\nu} \equiv \chih^{\nu\rho} \p_{\rho} .
\ee
Such a defined $F_{\rho\sigma}^{\lambda}$ is considered to be a gauge field strength when viewing the gravimetric field $\chi_{\rho}^{\; \lambda}$ to be a gauge-type field (which is actually a combined field of gauvigauge field $\chi_{\rho}^{\;\lambda} \equiv \chi_{\rho}^{\; a} \chi_{\; a}^{\lambda}$). For convenience of mention,  $\chi_{\rho}^{\; \lambda}$ is referred to as {\it gravimetric-gauge field} and $F_{\rho\sigma}^{\lambda}$ is called as {\it electromagnetic-like gravimetric-gauge field strength}.

The equations of electromagnetic field in dynamic Riemannian spacetime are obtained as follows:
\be \label{GMGME}
& & (\hat{\p}^{i}  + \ck{A}^{i} ) E_i + \chih^{\rho i}\chih^{\sigma 0}F_{\rho\sigma}^{0} E_i +  \frac{1}{2} \chih^{\rho j}\chih^{\sigma k} F_{\rho\sigma}^{0} \epsilon_{jk}^{\;\;\;\; i} B_i = J_0 , \nn \\
& & (\hat{\p}^{j}  + \ck{A}^{j} )\epsilon_{jk}^{\;\;\;\;\; i} B_i  +  \frac{1}{2} \chih^{\rho k}\chih^{\sigma l} F_{\rho\sigma i} \epsilon_{kl}^{\;\;\;\; j} B_j = J_{i} + (\hat{\p}^{0}  + \ck{A}^{0} ) E_i + \chih^{\rho 0}\chih^{\sigma j}F_{\rho\sigma i} E_j  ,
\ee
which can simply be expressed as follows:
\be \label{GMGME1}
& & (\hat{\p}^{i}  + \ck{A}^{i} ) E_i + \ck{\cE}^{i} E_i + \ck{\cB}^{i} B_i = J_0 , \nn \\
& & (\hat{\p}^{j}  + \ck{A}^{j} )\epsilon_{jk}^{\;\;\;\;\; i} B_i  +  B_j \ck{\cB}^{j}_{\;\; i} = J_{i} + (\hat{\p}^{0} + \ck{A}^{0} ) E_i -  E_j \ck{\cE}^{j}_{\;\; i} .
\ee
Where we have introduced the definitions:
\be
& & \ck{\cE}^{i} \equiv \varepsilon^{ij} \ck{E}_j  + \alpha^{ij} \ck{B}_{j} , \nn \\
& &  \ck{\cB}^{i} \equiv \mu^{ij} \ck{B}_j - \alpha^{ji} \ck{E}_j , \nn \\
& & \ck{\cE}^{j}_{\;\; i} \equiv \varepsilon^{jk} \ck{E}_{k\, i} + \alpha^{jk}\ck{B}_{k\, i} , \nn \\
& &   \ck{\cB}^{j}_{\;\; i} \equiv \mu^{jk} \ck{B}_{k\, i} + \alpha^{k j} \ck{E}_{k\, i} ,
\ee
with the coefficient functions and field strengths defined as follows:
\be
& & \varepsilon^{ij}(x) \equiv \chih^{ij} \chih^{00}- \chih^{i0}\chih^{j0}  ,  \nn \\
& & \mu^{ij}(x) \equiv \frac{1}{2} \epsilon^{i}_{\;\; kl}\chih^{kk'}\chih^{l l'} \epsilon_{k'l'}^{\;\;\; \; \; j} , \nn \\
& &  \alpha^{ij}(x) \equiv \chih^{ik}\chih^{0l}  \epsilon^{j}_{\;\; kl}  ,
\ee
and
\be \label{GGFS}
& & \ck{E}_{i} = - F_{0i}^{0} , \quad \ck{B}_{i} = -\frac{1}{2}\epsilon_{i}^{\;\; jl} F_{jl}^{0} , \nn \\
& &  \ck{E}_{k}^{\; \; i} = - F_{0 k}^{i} , \quad \ck{B}_{k}^{\; \; i} = -\frac{1}{2}\epsilon_{k}^{\;\; jl} F_{j l}^{i} .
\ee

It is noticed that $\ck{E}_{i}$ and $\ck{B}_{i}$ as well as $\ck{E}_{k}^{\;\; i}$ and $\ck{B}_{k}^{\;\; i}$ are introduced as {\it electromagnetic-like gravimetric-gauge fields} to characterize the gravitational interactions in dynamic Riemannian spacetime. In analogous to electromagnetic field, $\ck{E}_{i}$ and $\ck{B}_{i}$ are referred to as {\it electric-like gravimetric-gauge field} and {\it magnetic-like gravimetric-gauge field}, respectively, while $\ck{E}_{k}^{\;\; i}$ and $\ck{B}_{k}^{\;\; i}$ are called as {\it electric-like gravimetric-gauge tensor field} and {\it magnetic-like gravimetric-gauge tensor field}, respectively. 

In terms of vector representation, the equations presented in Eq.(\ref{GMGME1}) and Eq.(\ref{GMED}) can be rewritten into the following forms:
\be \label{GMGME2}
& & \hat{\nabla} \cdot \fE  + \ck{\fA}\cdot \fE + \ck{\bs{\cE}}\cdot \fE + \ck{\bs{\cB}}\cdot \fB = J_0 , \nn \\
& & \hat{\nabla} \times \fB  + \ck{\fA} \times \fB  +  \wh{\bs{\cB}} \cdot \fB = \fJ + (\hat{\p}^{0} + \ck{A}^{0} ) \fE -  \wh{\bs{\cE}}\cdot \fE , \nn \\
& & \nabla \cdot \fB = 0 , \nn \\
& & \nabla \times \fE = - \frac{\p \fB}{\p x^0} , 
\ee
which bring on the generalized Maxwell equations in dynamic Riemannian spacetime. Where we have used the vector notations $\ck{\bs{\cE}} \equiv \{ \ck{\cE}_{i} \}$ , $\ck{\bs{\cB}} \equiv \{ \ck{\cB}_{i} \}$, $\ck{\fA} \equiv \{ \ck{\fA}_{i} \}$, $\hat{\nabla}\equiv \{ \hat{\p}_{i} \}$ and bi-vector notations $\wh{\bs{\cE}} \equiv \{ \ck{\cE}_{j}^{\;\; i} \} $, $\wh{\bs{\cB}} \equiv \{ \ck{\cB}_{j}^{\;\; i} \} $.

To distinguish with the gravigauge-mediated Maxwell equations in locally flat gravigauge spacetime and gravigeometry-medium Maxwell equations in coordinate spacetime, we may refer to the above generalized Maxwell equations as {\it gravimetric-gauge-mediated Maxwell equations}, which are provided to describe the {\it gravimetric-gauge-mediated electrodynamics} in dynamic Riemannian spacetime.


\section{General covariance of dynamic equations and electrodynamics in any motional and spinning reference frames}

It is shown that the hermitian action constructed in locally flat gravigauge spacetime based on the gauge invariance principle becomes independent of the choice of coordinates. Namely, such an action holds in any coordinate system and gets invariant under the general coordinate transformation. In other word, such a gauge invariant action possesses automatically a hidden general linear group symmetry GL(1,3,R) which is known to lay the foundation of Einstein's general theory of relativity. Such a consequence indicates that the gauge invariance principle brings naturally on the emergence of general linear group symmetry GL(1,3,R) as a local symmetry in coordinate spacetime. 


\subsection{General covariance of dynamic equations}

To be concrete, the general coordinate transformation is defined as an arbitrary reparametrization of coordinate systems, i.e.:
\be \label{GCT}
x^{\mu} \to x^{' \mu} \equiv x^{' \mu}(x), 
\ee
which provides a local transformation in curved Riemannian spacetime and leads the displacement and derivative of coordinate systems to satisfy the following transformation laws:
\be \label{GCT2}
& & dx^{\mu} \to   dx^{'\mu}= T^{\mu}_{\;\; \; \nu}\, dx^{\nu}, \quad  T^{\mu}_{\;\; \; \nu}  \equiv \frac{\p x^{'\mu}}{\p x^{\nu}} , \nn \\
& &  \p_{\mu} \equiv \frac{\p}{\p x^{\mu}}  \to  \p'_{\mu} \equiv \frac{\p}{\p x^{'\mu}} = T_{\mu}^{\;\;\nu}\, \p_{\nu} , \quad  T_{\mu}^{\;\; \nu} \equiv \frac{\p x^{\nu}}{\p x^{'\mu}} ,
\ee
which preserves the inner product to be invariant under the general coordinate transformation,
\be \label{IPCS}
\langle \p'_{\mu}, dx^{'\nu} \rangle = T_{\mu}^{\;\, \rho} T^{\nu}_{\;\; \; \sigma}  
\langle \p_{\rho}, dx^{\sigma} \rangle = T_{\mu}^{\;\, \rho} T^{\nu}_{\;\; \; \sigma} \eta_{\rho}^{\;\; \sigma} = \eta_{\mu}^{\;\; \nu} \equiv \langle \p_{\mu}, dx^{\nu} \rangle .
\ee

From the above transformation law, $\p_{\mu}$ is usually regarded as covariant vector and $dx^{\mu}$ as contravariant vector in coordinate spacetime. In general, any covariant vector field $\mV_{\mu}(x)$ with a lower index and contravariant vector field $\hat{\mV}^{\mu}(x)$ with an upper index transform as follows: 
\be \label{TTL}
\mV'_{\mu}(x') = T_{\mu}^{\;\; \nu}\1 \mV_{\nu}(x), \quad  \hat{\mV}^{'\mu}(x') = T^{\mu}_{\;\; \; \nu}\1 \hat{\mV}^{\nu}(x),
\ee
which keeps the scalar product between the covariant vector field and contravariant vector field to be invariant:
\be
 \mV'_{\mu}(x') \hat{\mV}^{'\mu}(x') = T_{\mu}^{\;\; \nu} T^{\mu}_{\;\; \; \nu'} \mV_{\nu}(x) \hat{\mV}^{\nu'}(x) = \eta_{\; \nu'}^{\nu} \mV_{\nu}(x) \hat{\mV}^{\nu'}(x) \equiv \mV_{\mu}(x) \hat{\mV}^{\mu}(x).
\ee

The gravigauge field $\fA_{\mu}^{\; a}(x)$($\chi_{\mu}^{\; a}(x)$) and its dual field $\hfA_{a}^{\; \mu}(x)$($\chih_{a}^{\; \mu}(x)$) are considered as bi-covariant vector fields in biframe spacetime. Where $\fA_{\mu}^{\; a}(x)$($\chi_{\mu}^{\; a}(x)$) is regarded as a covariant vector field in coordinate spacetime and a contravariant vector field in locally flat gravigauge spacetime, while $\hfA_{a}^{\; \mu}(x)$($\chih_{a}^{\; \mu}(x)$) is viewed as a contravariant vector field in coordinate spacetime and a covariant vector field in locally flat gravigauge spacetime. Namely, they transform as follows:
\be
& & \fA_{\mu}^{'\; a}(x') = T_{\mu}^{\;\; \nu}\1 \fA_{\nu}^{\; a}(x), \quad  \chi_{\mu}^{'\; a}(x') = T_{\mu}^{\;\; \nu}\1 \chi_{\nu}^{\; a}(x),  \nn \\
& & \hfA_{a}^{'\;\mu}(x') = T^{\mu}_{\;\; \; \nu}\1 \hfA_{a}^{\; \nu}(x), \quad  \chih_{a}^{'\;\mu}(x') = T^{\mu}_{\;\; \; \nu}\1 \chih_{a}^{\; \nu}(x) ,
\ee
under the general coordinate transformation of general linear group symmetry GL(1,3; R), and 
\be
& & \fA_{\mu}^{'\; a}(x) = \Lambda^{a}_{\; \; b}(x) \1 \fA_{\mu}^{\; b}(x), \quad  \chi_{\mu}^{'\; a}(x) = \Lambda^{a}_{\; \; b}(x) \1 \chi_{\mu}^{\; b}(x)  \nn \\
& & \hfA_{a}^{'\;\mu}(x) = \Lambda_{a}^{\;\; b}(x) \1 \hfA_{b}^{\; \mu}(x), \quad  \chih_{a}^{'\;\mu}(x) = \Lambda_{a}^{\;\; b}(x) \1 \chih_{b}^{\; \mu}(x),
\ee
under the gauge transformation of spin gauge group symmetry SP(1,3). So the gravicoordinate displacement $\delta \vka^{a}$ ($\delta \chi^{a}$) and derivative $\heth_{a}$ ($\eth_a$) defined in locally flat gravigauge spacetime become invariant under the general coordinate transformation. Explicitly, we have,
\be
& & \delta \vka^{a} \equiv \fA_{\mu}^{\; a}(x) dx^{\mu} = \fA_{\mu}^{'\; a}(x') dx^{' \mu}, \quad \delta \chi^{a} \equiv \chi_{\mu}^{\; a}(x) dx^{\mu} = \chi_{\mu}^{'\; a}(x') dx^{' \mu}, \nn \\
& & \heth_{a} \equiv \hfA_{a}^{\; \mu}(x)\p_{\mu}= \hfA_{a}^{'\; \mu}(x') \p'_{\mu} , \quad \eth_{a} \equiv \chih_{a}^{\; \mu}(x)\p_{\mu}= \chih_{a}^{'\; \mu}(x') \p'_{\mu} , 
\ee
which indicates that the gravicoordinate displacement $\delta \vka^{a}$ ($\delta \chi^{a}$) and derivative $\heth_{a}$ ($\eth_a$) introduced in Eqs.(\ref{GCDO}) and (\ref{DGCDO}) are the scalar product between the covariant vector and contravariant vector in curved Riemannian spacetime of coordinates. Therefore, the gravicoordinate displacement $\delta \vka^{a}$ ($\delta \chi^{a}$) and derivative $\heth_{a}$ ($\eth_a$) naturally possess the general linear group symmetry GL(1,3,R). It can be checked that the inner product of $\delta \vka^{a}$ ($\delta \chi^{a}$) and derivative $\heth_{a}$ ($\eth_a$) becomes invariant under the spin gauge transformation,
\be
\langle \heth'_{a}, \delta \vka^{' b} \rangle = \Lambda_{a}^{\;\, c} \Lambda^{b}_{\;\; \; d}  
\1 \langle \heth_{c}, \delta \vka^{d} \rangle = \Lambda_{a}^{\;\, c} \Lambda^{b}_{\;\; \; d} \1 \eta_{c}^{\;\; d} = \eta_{a}^{\;\; b} \1 \equiv \langle \heth_{a}, \delta \vka^{b} \rangle .
\ee

On the other hand, the gravimetric field $\mH_{\mu\nu}(x)$ ($\chi_{\mu\nu}(x)$) and dual gravimetric field $\hmH^{\mu\nu}(x)$ ($\chih^{\mu\nu}(x)$) defined in Eqs.(\ref{DTF}) and (\ref{DTF1}) as the scalar products of bi-covariant vector field $\fA_{\mu}^{\; a}(x)$($\chi_{\mu}^{\; a}(x)$) and $\hfA_{a}^{\; \mu}(x)$($\chih_{a}^{\; \mu}(x)$) in locally flat gravigauge spacetime correspond to the covariant tensor and contravariant tensor fields in curved Riemannian spacetime of coordinates. They obey the following transformation laws:
\be
\mH'_{\mu\nu}(x')  = \frac{\p x^{\sigma}}{\p x^{'\mu}} \frac{\p x^{\sigma}}{\p x^{'\nu}} \1\mH_{\rho\sigma}(x) =  T_{\mu}^{\;\; \rho}  T_{\nu}^{\;\; \sigma}\, \mH_{\rho\sigma}(x), \quad \chi'_{\mu\nu}(x')  =  T_{\mu}^{\;\; \rho}  T_{\nu}^{\;\; \sigma}\, \chi_{\rho\sigma}(x),
 \nn \\
 \hmH^{'\mu\nu}(x')  = \frac{\p x^{'\mu}}{\p x^{\rho}} \frac{\p x^{'\nu}}{\p x^{\sigma}} \1\hmH^{\rho\sigma}(x) =  T^{\mu}_{\;\; \rho}  T^{\nu}_{\;\; \sigma}\, \hmH^{\rho\sigma}(x), \quad \chih^{'\mu\nu}(x')  =  T^{\mu}_{\;\; \rho}  T^{\nu}_{\;\; \sigma}\, \chih^{\rho\sigma}(x).
\ee
So the Christoffel symbols given in Eq.(\ref{HSGMF}) transform as follows:
\be \label{GLGT}
\fGa_{\mu\nu}^{'\rho} (x') & = &  ( \frac{\p x^{\mu'}}{\p x^{'\mu}}\frac{\p x^{\nu'}}{\p x^{'\nu}} ) \left(\, \frac{\p x^{'\rho}}{\p x^{\rho'}} \fGa_{\mu'\nu'}^{\rho'} (x) - \frac{\p^2 x^{'\rho}}{\p x^{\mu'}\p x^{\nu'}} \, \right)\nn \\
& \equiv & T_{\mu}^{\;\; \mu'}  T_{\nu}^{\;\; \nu'} T^{\rho}_{\;\; \; \rho'}  \left(\, \fGa_{\mu'\nu'}^{\rho'} (x) -  T^{\rho'}_{\;\; \, \sigma} \p_{\mu'} T^{\sigma}_{\;\; \, \nu'}  \right) ,
\ee
which provides the local transformation property of Christoffel symbols with appearance of inhomogeneous term under general linear group symmetry GL(1,3,R), which is analogous to the gauge-type transformation of spin gauge group symmetry SP(1,3) in locally flat gravigauge spacetime. Such a property of transformations keeps the covariant derivative on the covariant vector field $\mV_{\mu}(x)$ and contravariant vector field $\hat{\mV}^{\mu}(x)$ to be general covariance under the general coordinate transformations, i.e.:
\be  \label{GC}
& & \nabla'_{\mu}\mV'_{\nu}(x') =  T_{\mu}^{\;\; \rho}  T_{\nu}^{\;\; \sigma} \nabla_{\rho}\mV_{\sigma}(x) , \quad \nabla_{\mu}\mV_{\nu}(x) = \p_{\mu} \mV_{\nu}(x) - \fGa_{\mu\nu}^{\rho}(x) \mV_{\rho}(x)  ,  \nn \\
& & \nabla'_{\mu}\hat{\mV}^{'\nu}(x') =  T_{\mu}^{\;\; \rho}  T^{\nu}_{\;\; \sigma}  \nabla_{\rho}\hat{\mV}^{\sigma}(x) , \quad \nabla_{\mu}\hat{\mV}^{\nu}(x) = \p_{\mu} \hat{\mV}^{\nu}(x) + \fGa_{\mu\rho}^{\nu}(x) \hat{\mV}^{\rho}(x)  .
\ee 

With the above transformation laws, it can be verified explicitly that all actions constructed in the previous sections are invariant under the general coordinate transformation via an arbitrary reparametrization of coordinates shown in Eqs.(\ref{GCT})-(\ref{IPCS}), which displays the emergent general linear group symmetry GL(1,3,R) of the actions. As a consequence, we come to the statement that the gauge invariance principle naturally leads the action formulated in locally flat gravigauge spacetime to possess a maximal joint symmetry: 
\be \label{EGLS}
G_S = \mbox{GL(}1,3,\mbox{R)} \Join \mbox{WS(}1, 3) , \nn
\ee
which implies that the Poincar\'e group symmetry PO(1,3) in globally flat Minkowski spacetime will automatically be extended to the local general linear group symmetry GL(1,3,R), which brings on the appearance of Riemann geometry in curved Riemannian spacetime.

Such a joint symmetry of the action indicates that the dynamic equations of motion for the basic fields derived based on the least action principle all become covariance in biframe spacetime. Namely, the gravidynamics and spinodynamics as well as electrodynamics discussed in the previous sections should obey the general coordinate covariance in curved Riemannian spacetime and spin gauge covariance in locally flat gravigauge spacetime. 


\subsection{Gravigeometry-medium electrodynamics in any motional reference frame}

Let us examine explicitly the general coordinate covariance and spin gauge covariance for the electrodynamics. The general coordinate covariance indicates that the dynamic equations of electromagnetic gauge field presented in Eqs. (\ref{GGE})-(\ref{GFS1}) hold in any reference frame in the Einstein basis for the scaling gauge symmetry, i.e.:
\be
& & \p'_{\nu} \wh{F}^{'\mu\nu}(x') = -\wh{J}^{'\mu}(x') ,
\ee
with
\be
& &  \widehat{F}^{'\mu\nu}(x') \equiv \chi'(x') \tilde{\chi}^{'[\mu\nu]\rho\sigma}(x')  F'_{\rho\sigma}(x') , \nn \\
& & F'_{\rho\sigma}(x') \equiv \p'_{\rho}A'_{\sigma}(x') - \p'_{\sigma}A'_{\rho}(x') , \nn \\
& &  \tilde{\chi}^{'[\mu\nu]\rho\sigma}(x') \equiv \frac{1}{2} [ \chih^{' \mu\rho}(x')\chih^{'\nu\sigma}(x') - \chih^{'\nu\rho}(x')\chih^{'\mu\sigma}(x') ] , \nn \\
& & \wh{J}^{'\mu}(x') = \chi'(x') \chih_{a}^{'\; \mu}(x') J^{a}(x'), \quad J^{a}(x') \equiv - g_E \bar{\psi}(x') \gamma^{a} \psi(x') ,
\ee
where the relevant fields and vectors in arbitrary new reference frame obey the following transformation properties:
\be
& & \p_{\mu} \equiv \frac{\p}{\p x^{\mu} } \to \p'_{\mu} \equiv \frac{\p}{\p x^{'\mu}} = T_{\mu}^{\; \; \nu}  \p_{\nu},  \nn \\
& & A_{\mu}(x) \to A'_{\mu}(x') = T_{\mu}^{\; \; \nu} A_{\nu}(x), \nn \\
& & F_{\mu\nu}(x) \to F'_{\mu\nu}(x') = T_{\mu}^{\; \; \rho} T_{\nu}^{\; \; \sigma} F_{\rho\sigma}(x) , \nn \\
& & \chih_{a}^{\;\; \mu}(x) \to \chih_{a}^{'\; \mu}(x') = T^{\mu}_{\; \; \nu}\1 \chih_{a}^{\;\; \nu}(x) , \nn \\
& & \chih^{\mu\nu}(x) \equiv \eta^{ab}\chih_{a}^{\;\; \mu}(x) \chih_{b}^{\;\; \nu}(x) \to \chih^{'\mu\nu}(x') = T^{\mu}_{\; \; \rho} T^{\nu}_{\; \; \sigma} \chih^{\rho\sigma}(x) .
\ee
It is noticed that the above dynamic equations possess a hidden gauge invariance. 

To demonstrate the general coordinate covariance, it is useful to rewrite the equation of motion in terms of the covariant derivative. Explicitly, we have the following identity: 
\be
& & \p_{\nu} \widehat{F}^{\mu\nu}(x) \equiv \chi(x) \nabla_{\nu} \hat{F}^{\mu\nu}(x) \equiv \chi(x) [ \p_{\nu}\hat{F}^{\mu\nu}(x) + \Gamma_{\nu\rho}^{\mu}  \hat{F}^{\rho\nu}(x) + \Gamma_{\nu\rho}^{\nu} \hat{F}^{\mu\rho}(x) ] , 
\ee
where we have used the following definition and property:
\be
& & \hat{F}^{\mu\nu}(x) \equiv \chih^{\mu \rho}(x) \chih^{\nu\sigma}(x) F_{\rho\sigma}(x) \equiv \tilde{\chi}^{[\mu\nu]\rho\sigma}(x) F_{\rho\sigma}(x) = - \hat{F}^{\nu\mu}(x) , \nn \\
& & \Gamma_{\nu\rho}^{\mu} \equiv \Gamma_{\rho\nu}^{\mu}, \quad \Gamma_{\nu\rho}^{\nu} \equiv \p_{\rho}\ln \chi(x) , \nn \\
& & \Gamma_{\nu\rho}^{\mu} \wh{F}^{\rho\nu}(x)  \equiv \frac{1}{2} ( \Gamma_{\nu\rho}^{\mu} -  \Gamma_{\rho\nu}^{\mu} ) \wh{F}^{\rho\nu}(x) = 0, \nn \\
& & \chi(x)\Gamma_{\nu\rho}^{\nu} \wh{F}^{\mu\rho}(x) \equiv (\p_{\rho} \chi(x) ) \wh{F}^{\mu\rho}(x) . 
\ee
So the equations of motion in terms of the explicit covariant form can be expressed as follows:
\be
\nabla_{\nu} \hat{F}^{\mu\nu}(x) = - \hat{J}^{\mu}(x) , 
\ee
with the current density,
\be
 \hat{J}^{\mu}(x) \equiv  \chih_{a}^{\; \mu}(x) J^{a}(x) = - g_E \chih_{a}^{\; \mu}(x) \bar{\psi}(x)  \gamma^{a} \psi(x) . 
\ee


Therefore, the gravigeometry-medium Maxwell equations hold in any motional reference frame and can be expressed into the following forms:
\be \label{GME2}
& & \nabla' \cdot \widehat{\fE}' = \wh{J}^{'0} , \nn \\
& & \nabla' \times \widehat{\fB}' = \wh{\fJ}' + \frac{\p \widehat{\fE}'}{\p x^{'0}} , \nn \\
& & \nabla' \cdot \fB' = 0 , \nn \\
& & \nabla' \times \fE' = - \frac{\p \fB'}{\p x^{'0}} , 
\ee
with
\be
& & \widehat{\fE}'(x') \equiv \widehat{\boldsymbol{\varepsilon}}' \cdot \fE' + \widehat{\boldsymbol{\alpha}}'\cdot \fB' , \quad \mbox{or} \quad \widehat{E}^{' i}(x') = \hat{\varepsilon}^{' i}_{\; j} E^{' j} + \hat{\alpha}^{' i}_{\;\;  j} B^{' j} ,  \nn \\
& & \widehat{\fB}'(x') \equiv \widehat{\boldsymbol{\mu}}' \cdot \fB' - \widehat{\boldsymbol{\alpha}}^{' T} \cdot \fE', \quad \mbox{or} \quad \widehat{B}^{' i}(x') = \hat{\mu}^{' i}_{\; j} B^{' j} - \hat{\alpha}^{'\; i}_{j} E^{' j} , \nn \\
& & \hat{J}^{'0}(x') \equiv \bv{\kappa}' J^{'0} +  \bv{\bs{\kappa}}' \cdot \fJ' , \quad \hat{\fJ}'(x') \equiv  \mathring{\bs{\kappa}} \cdot \fJ' + \hat{\bs{\kappa}}' J^{'0} ,
\ee
and 
\be \label{GME3}
& & \widehat{\boldsymbol{\varepsilon}}'(x') \equiv \{ \hat{\varepsilon}^{' i}_{\; j}(x') \}, \quad \mbox{or} \quad \hat{\varepsilon}^{' i}_{\; j}(x') \equiv \chi'(x') ( \chih^{' 0}_{\;0}(x')\chih^{' i}_{\;j}(x') - \chih^{' i}_{\;0}(x')\chih^{' 0}_{\;j}(x') ) ,  \nn \\
& & \widehat{\boldsymbol{\mu}}'(x') \equiv \{ \hat{\mu}^{' i}_{\; j}(x') \} , \quad \mbox{or} \quad  \hat{\mu}^{' i}_{\; j}(x') \equiv -  \frac{1}{2} \chi'(x') \epsilon^{i}_{\;\; kl}(x')\chih^{kk'}(x')\chih^{' l l'}(x') \epsilon_{k'l'j} ,  \nn \\
& & \widehat{\boldsymbol{\alpha}}'(x') \equiv \{ \hat{\alpha}^{' i}_{\; j}(x') \} , \quad  \mbox{or} \quad \hat{\alpha}^{' i}_{\; j}(x') \equiv   \chi'(x') \chih^{' ik}(x')\chih^{' l0}(x') \epsilon_{kl j} , \nn \\
& & \bv{\kappa}'(x') \equiv \chih_{0}^{'\; 0}(x') , \quad  \bv{\bs{\kappa}}'(x') \equiv \{\chih_{\alpha}^{'\; 0}(x')\}, \nn \\
& &  \mathring{\bs{\kappa}}'(x')\equiv  \{ \chih_{\alpha}^{'\; i}(x') \},\quad \hat{\bs{\kappa}}'(x') \equiv \{ \chih_{0}^{'\; i}(x')\}, 
\ee
where the coefficient functions are in general governed by the gravidynamics in the new reference frame.


\subsection{Gravigauge-mediated electrodynamics in any spinning reference frame}

We now turn to discuss the electrodynamics in locally flat gravigauge spacetime. The dynamic equation presented in Eq.(\ref{EMEM}) is spin gauge covariance, which holds in any spinning reference frame in locally flat garvigauge spacetime, i.e.:
\be \label{EMEM3}
D'_{b} F^{'ab}(x) = - J^{'a}(x) , 
\ee
with 
\be
& & D'_{b} F^{'ab}(x) \equiv \eth'_{b} F^{'ab}(x) + \mOm_{bc}^{'a}(x) F^{'cb}(x) + \mOm_{bc}^{'b}(x) F^{'ac}(x) , \nn \\
& & F^{'ab}(x) \equiv  \eta^{ac}\eta^{bd} F'_{cd}(x) \equiv \chih^{' a \mu}(x) \chih^{' b\nu}(x) F_{\mu\nu}(x), \nn \\
& & F'_{cd}(x) \equiv  D'_{c}A'_{d}(x) - D'_{d}A'_{c}(x) = \eth'_{c}A'_{d}(x) - \eth'_{d}A'_{c}(x) + \mOm_{cd}^{'b}(x) A'_{b}(x) -  \mOm_{dc}^{'b}(x) A'_{b}(x)  , \nn \\
& & J^{' a}(x) = - g_E \bar{\psi}'(x) \gamma^a \psi'(x)  ,
\ee
where all fields in arbitrary spinning reference frame obey the following transformation properties  under the gauge transformation of spin gauge symmetry SP(1,3):
\be
& & F^{ab}(x) \to F^{'ab}(x) = \Lambda^{a}_{\;\; c}(x)\Lambda^{b}_{\;\; d}(x) F^{cd}(x) , \quad A_{c}(x) \to A'_{c}(x) =  \Lambda_{c}^{\;\; d}(x) A_{d}(x), \nn \\
& & J^{a}(x) \to J^{'a}(x) =  \Lambda^{a}_{\;\; c}(x) J^{c}(x) , \quad \psi(x) \to \psi'(x) = S(\Lambda) \psi(x), \quad  S(\Lambda) = e^{i \varpi_{ab}(x) \Sigma^{ab} } ,\nn \\
& & \mOm_{cb}^{a}(x) \to \mOm_{cb}^{'a}(x) = \Lambda_{c}^{\;\; c'}(x) [ \Lambda^{a}_{\;\; a'}(x)  \Lambda_{b}^{\;\; b'}(x) \mOm_{c'b'}^{a'}(x) + \frac{1}{2}\Lambda^{a}_{\;\; d}(x)\eth_{c'} \Lambda_{b}^{\;\; d}(x) -  \Lambda_{b}^{\;\; d}(x)\eth_{c'} \Lambda^{a}_{\;\; d}(x) ], \nn \\
& &  \eth_{c} \equiv \chih_{c}^{\;\; \mu}(x)\p_{\mu} \to \eth'_{c} =  \Lambda_{c}^{\;\; d}(x) \eth_{d}, \quad \chih_{c}^{\;\;\mu}(x) \to \chih_{c}^{'\;\, \mu}(x) = \Lambda_{c}^{\;\; d}(x)\chih_{d}^{\;\;\mu}(x) . 
\ee  
So the gravigauge-mediated Maxwell equations in Eq.(\ref{GME}) are valid in any spinning reference frame, which can be expressed as follows:
\be \label{GGME3}
& & \bvnabla' \cdot \wt{\fE}' - \bvfA' \cdot \wt{\fE}' -\bvfE' \cdot \wt{\fE}'   = J^{'0} + \bvfB' \cdot \wt{\fB}',   \nn \\
& & \bvnabla' \times \wt{\fB}' - \bv{\fA}' \times \wt{\fB}' + \bv{\mathbb{B}}' \cdot \wt{\fB}' = \fJ' + \eth'_0 \wt{\fE}' - \bv{A}'_0 \wt{\fE}' + \bv{\mathbb{E}}' \cdot \wt{\fE}', \nn \\
& & \bvnabla' \times \wt{\fE}' + \bv{\fE}' \times \wt{\fE}' - \bv{\mathbb{B}}' \cdot \wt{\fE}' = - \eth'_0 \wt{\fB}' - 2\bv{A}'_0 \wt{\fB}' + \bv{\mathbb{E}}' \cdot \wt{\fB}', \nn \\
& & \bvnabla' \cdot \wt{\fB}' +2 \bvfA' \cdot \wt{\fB}' + 2\bvfE' \cdot \wt{\fB}'   =  \bvfB' \cdot \wt{\fE}',   
\ee
with
\be
& & \wt{\fE}' \equiv \{ \wt{E}^{' \alpha} = - F^{' 0\alpha} \} , \quad \wt{\fB}' \equiv \{ \wt{B}^{' \alpha} = -\frac{1}{2}\epsilon^{\alpha}_{\;\; \beta\gamma} F^{' \beta\gamma} \} , \nn \\
& &  \bv{\fE}' \equiv \{ \bv{E}'_{\alpha} = - F_{0\alpha}^{' 0} \}, \quad \bv{\fB}' \equiv \{ \bv{B}'_{\alpha} = -\frac{1}{2}\epsilon_{\alpha}^{\;\; \gamma\delta} F_{\gamma\delta}^{' 0} \} , \nn \\
& & \bv{\mathbb{E}}' \equiv \{  \bv{E}_{\alpha}^{'\;\, \beta} = - F_{0\alpha}^{' \beta} \}, \quad  \bv{\mathbb{B}}' \equiv \{ \bv{B}_{\alpha}^{'\;\, \beta} = -\frac{1}{2}\epsilon_{\alpha}^{\;\; \gamma\delta} F_{\gamma\delta}^{' \beta}  \} , \nn \\
& &  \bvA'_{0} = \eta_{\alpha}^{\;\; \beta} \bvE_{\beta}^{'\;\, \alpha}, \quad \bv{\fA}' \equiv \{ \bvA'_{\alpha} = - \bvE'_{\alpha} - \epsilon_{\alpha\beta}^{\;\;\;\;\, \gamma}\bvB_{\gamma}^{' \;\, \beta} \} , \nn \\
& & \bv{\nabla}' \equiv \{ \eth'_{\alpha} \} , \quad \eth'_0  \equiv \chih_{0}^{'\;\, \mu}\p_{\mu} .
\ee


\subsection{Gravigeometry-medium Maxwell equations in special background medium and general coordinate covariance of equations in motional reference frame}

In general, either the gravigeometry-medium electric field $\widehat{\fE}$ or gravigeometry-medium magnetic field $\widehat{\fB}$ depends on both free-motion electric field $\fE$ and magnetic field $\fB$ defined in globally flat Minkowski spacetime. Only when the symmetric gravimetric field $\chih^{0i}(x)= \chih^{i0}(x)$ goes to be vanishing and brings the coefficient function $\hat{\alpha}^{i}_{\; j}(x)$ to be zero, so that the gravigeometry-medium electric field $\widehat{\fE}$ or gravigeometry-medium magnetic field $\widehat{\fB}$ becomes to be proportional only to the electric field $\fE$ or magnetic field $\fB$. Explicitly, such a case can be expressed into the following relations:
\be
& & \chih^{0i}(x) = \chih^{i0}(x) \to 0, \quad \hat{\alpha}^{i}_{\; j}(x) \to 0, \nn \\
& &  \widehat{E}^{i} = \hat{\varepsilon}^{i}_{\; j}(x) E^{j}, \quad \widehat{B}^{i} = \hat{\mu}^{i}_{\; j}(x) B^{j} .
\ee

As a simple case, let us examine a special background medium in which the gravimetric field is supposed to have the following simple geometric distributions:
\be \label{SBM}
\chih^{00} = \alpha_0^{2}\eta^{00}, \quad \chih^{ij} = \alpha_s^{2}\eta^{ij}, \quad \chih^{0i} =0, 
\ee
with $\alpha_0$ and $\alpha_s$ being constants. Such a special background medium is homogeneous and isotropic in three spatial dimensions. The global scaling invariance of the action enables us to make the following scale transformations for the gravimetric field and coordinates:
\be
& & \chih_{a}^{\;\; \mu} \to \alpha_{s}  \chih_{a}^{\;\; \mu}, \quad \chih^{\mu\nu} \to \alpha_{s}^{2} \chih^{\mu\nu} , \nn \\
& &  \chi_{\mu}^{\;\; a} \to \alpha_{s}^{-1}  \chi_{\mu}^{\;\; a}, \quad \chi_{\mu\nu} \to \alpha_{s}^{-2} \chi_{\mu\nu} , \nn \\
& & x^{\mu} \to \alpha_{s}^{-1} x^{\mu} ,
\ee
which simplifies the gravimetric distribution into the following form:
\be \label{SBM1}
\chih^{00} = c_s^{-2} \eta^{00}, \quad \chih^{ij} = \eta^{ij}, \quad \chih^{0i} =0, 
\ee
with
\be
c_s \equiv \frac{\alpha_s}{\alpha_0} .
\ee
which indicates that the basic feature of such a special background medium is solely charactrized by the constant ratio $c_s$. So the corresponding gravigeometry-medium electric field $\widehat{\fE}$ and gravigeometry-medium magnetic field $\widehat{\fB}$ get the following simple relations:
\be
& & \widehat{\fE} = \varepsilon_s \fE \equiv \fD, \nn \\
& &  \widehat{\fB} = \frac{\fB}{\mu_s} \equiv \fH, 
\ee
with the constant coefficients:
\be
& & \varepsilon_s \equiv c_s^{-1} , \quad  \mu_s \equiv c_s^{-1}, \quad \frac{1}{\sqrt{\varepsilon_s \mu_s}} = c_s.
\ee
The current density in such a special background medium has a simple form:
\be
& & \wh{J}^0 =  J^{0} , \quad J^{0} = -g_E \bar{\psi}(x)  \gamma^{0} \psi(x) ,\nn \\
& & \wh{J}^i = c_s J^i, \quad J^i = - g_E \bar{\psi}(x)  \gamma^{i} \psi(x) ,
\ee
with $J^0$ and $J^{i}$ being the free charge density and current density, respectively. 
From the above analyses and the general gravigeometry-medium Maxwell equations, we arrive at the following Maxwell equations in such a special background medium:
\be \label{IME}
& & \nabla \cdot \fD = J^0 , \nn \\
& & \nabla \times \fH =   \frac{1}{\sqrt{\varepsilon_s \mu_s}}  \fJ +  \frac{\p \fD }{\p x^0} , \nn \\
& & \nabla \cdot \fB = 0 , \nn \\
& & \nabla \times \fE = - \frac{\p \fB}{\p x^0} , 
\ee
which can be expressed as the following forms: 
\be \label{IME}
& & \nabla \cdot \fE = c_s J^0 , \nn \\
& & \nabla \times \fB =  \fJ + \frac{1}{c_s^2} \frac{\p \fE }{\p x^0} , \nn \\
& & \nabla \cdot \fB = 0 , \nn \\
& & \nabla \times \fE = - \frac{\p \fB}{\p x^0} , 
\ee
where the electromagnetic fields $\fE$ and $\fB$ as well as the gravigeometry-medium electromagnetic fields $\widehat{\fE} \equiv \fD $ and $\widehat{\fB} \equiv \fH$ in the special background medium all have the same unit since the ratio $c_s$ is a dimensionless constant in the present definition. 

Let us now examine the Lorentz covariance of gravigeometry-medium Maxwell equations under the global Lorentz transformation of group symmetry SO(1,3) in globally flat Minkowski spacetime, i.e.:
\be
& & x^{\mu} \to x^{'\mu} = L^{\mu}_{\;\; \nu} x^{\nu}, \quad A_{\mu}(x) \to A'_{\mu}(x') = L_{\mu}^{\;\; \nu} A_{\nu}(x) , \nn \\
& & \chih^{\mu\nu}(x) \to \chih^{'\mu\nu}(x') = L^{\mu}_{\;\; \rho}  L^{\nu}_{\;\; \sigma} \chih^{\rho\sigma}(x).
\ee  
For simplicity, we should consider the gravigeometry-medium Maxwell equations in special background medium under the following global Lorentz boost transformation $L^{\mu}_{\;\; \nu}$:
\be
& &L^{\mu}_{\;\; \nu} \equiv \eta^{\mu}_{\;\; \nu} +  \tL^{\mu}_{\;\; \nu} ,  \nn \\
& & (\tL)^{00} = \gamma - 1, \quad (\tL)^{i0} = - \gamma \beta^{i}  = (\tL)^{0 i} , \nn \\
& &  (\tL)^{ij} = (\gamma-1)\frac{\beta^{i}\beta^{j}}{\beta^2} = (\tL)^{ji} , \quad  \gamma  \equiv \frac{1}{\sqrt{1 - \beta^2}} , \nn \\
& &  \boldsymbol{\beta} \equiv ( \beta^1, \beta^2, \beta^3) , \quad \beta^2 \equiv \bs{\beta}\cdot \bs{\beta} \equiv \beta_1^2 + \beta_2^2 + \beta_3^2  ,
\ee
where $\boldsymbol{\beta}$ is regarded as a boost vector.

To obtain the gravigeometry-medium electromagnetic field in the boosted new coordinate reference frame, it is useful to present the gravimetric of the special background medium in the boosted new coordinate reference frame as follows:
\be
& &  \chih^{'00} = \gamma^2c_s^{-2}( 1- \beta^2c_s^2) , \quad  \chih^{'0i} = \chih^{'i0} = - \gamma^2 c_s^{-2}( 1- c_s^2) \beta^{i} , \nn \\
& & \chih^{'ij} = \chih^{' ji} =  \eta^{ij} + \gamma^2 c_s^{-2}( 1- c_s^2) \beta^{i} \beta^{j} , 
\ee
which enables us to figure out from Eqs.(\ref{GME2}) and (\ref{GME3}) the following gravigeometry-medium electromagnetic field in the boosted new coordinate reference frame:
\be
\widehat{\fE}'(x') & = &  \gamma^2( 1- \beta^2c_s^2) c_s^{-1} \fE'(x')  - \gamma^2( c_s^{-2} - 1)  c_s\boldsymbol{\beta}\times \fB'(x') \nn \\
&-& \gamma^2 ( 1- c_s^2) c_s^{-1}\boldsymbol{\beta}\cdot \fE'(x') \boldsymbol{\beta} , \nn \\
\widehat{\fB}'(x') & = &  [1- 2\gamma^2 \beta^2(c_s^{-2} -1) ] c_s \fB'(x')  - \gamma^2 c_s( c_s^{-2} - 1) \boldsymbol{\beta}\times \fE'(x') \nn \\
& + & 2 \gamma^2 (c_s^{-2} - 1)  c_s \boldsymbol{\beta}\cdot \fB'(x') \boldsymbol{\beta} , 
\ee  
which may be rewritten into the following forms:
\be
\fD'(x') \equiv \widehat{\fE}'(x') & = &  \gamma^2( 1- \frac{\beta^2}{\varepsilon_s\mu_s} ) \varepsilon_s \fE'(x')  - \gamma^2( \varepsilon_s\mu_s - 1) \frac{1}{\mu_s} \boldsymbol{\beta}\times \fB'(x') \nn \\
&-& \gamma^2 ( 1- \frac{1}{\varepsilon_s\mu_s} ) \varepsilon_s \boldsymbol{\beta}\cdot \fE'(x') \boldsymbol{\beta} , \nn \\
\fH'(x') \equiv \widehat{\fB}'(x') & = &  [1- 2\gamma^2 \beta^2(\varepsilon_s\mu_s -1) ] \frac{1}{\mu_s} \fB'(x')  - \gamma^2 ( 1- \frac{1}{\varepsilon_s\mu_s} ) \varepsilon_s \boldsymbol{\beta}\times \fE'(x') \nn \\
& + & 2 \gamma^2 (1- \frac{1}{\varepsilon_s\mu_s})  \frac{1}{\mu_s} \boldsymbol{\beta}\cdot \fB'(x') \boldsymbol{\beta} . 
\ee 
Correspondingly, the current density in the boosted new coordinate reference frame is found to be,
\be
& & \wh{J}^{' 0}(x') = \gamma [ J^{0}(x') - c_s \boldsymbol{\beta} \cdot \fJ(x') ] \nn \\
& & \quad \quad \;\; \; = \gamma [ J^{0}(x') - \frac{1}{\sqrt{\varepsilon_s \mu_s}}  \boldsymbol{\beta} \cdot \fJ(x') ] , \nn \\ 
& & \wh{\fJ}'(x') = c_s \fJ(x') - \gamma J^{0}(x') \boldsymbol{\beta} + (\gamma -1) c_s \boldsymbol{\beta} \cdot \fJ(x') \boldsymbol{\beta} \nn \\
& & \quad \quad \;\; \;  = \frac{1}{\sqrt{\varepsilon_s \mu_s}}  \fJ(x') - \gamma J^{0}(x') \boldsymbol{\beta} + (\gamma -1) \frac{1}{\sqrt{\varepsilon_s \mu_s}}  \boldsymbol{\beta} \cdot \fJ(x') \boldsymbol{\beta} .
\ee

For the case $\beta^2 \ll 1$, when keeping only the leading term of $\beta_i$ with $\gamma \simeq 1$, we arrive at the following simplified relations:
\be
\fD'(x') \equiv \widehat{\fE}'(x') & \simeq & \varepsilon_s \fE'(x')  - ( \varepsilon_s\mu_s - 1) \frac{1}{\mu_s} \boldsymbol{\beta}\times \fB'(x') \nn \\
\fH'(x') \equiv \widehat{\fB}'(x') & \simeq &  \frac{1}{\mu_s} \fB'(x')  - ( 1- \frac{1}{\varepsilon_s\mu_s} ) \varepsilon_s \boldsymbol{\beta}\times \fE'(x') ,
\ee 
and
\be
& & \wh{J}^{' 0}(x') \simeq J^{0}(x') - \frac{1}{\sqrt{\varepsilon_s \mu_s}}  \boldsymbol{\beta} \cdot \fJ(x')  , \nn \\ 
& & \wh{\fJ}'(x') \simeq \frac{1}{\sqrt{\varepsilon_s \mu_s}}  \fJ(x') - J^{0}(x') \boldsymbol{\beta} ,
\ee
where $J^{0}(x')$ and $\fJ(x')$ are free current density. It can be verified explicitly that the gravigeometry-medium electromagnetic field in the boosted new coordinate reference frame does satisfy the gravigeometry-medium Maxwell equations obeying the general coordinate covariance, i.e.: 
\be \label{IME2}
& & \nabla' \cdot \fD' = \wh{J}^{'0} , \nn \\
& & \nabla' \times \fH' =   \wh{\fJ}' +  \frac{\p \fD' }{\p x^{'0}} , \nn \\
& & \nabla' \cdot \fB' = 0 , \nn \\
& & \nabla' \times \fE' = - \frac{\p \fB'}{\p x^{'0}} . 
\ee

In conclusion, we have demonstrated how the gravigeometry-medium Maxwell equations in the special background medium obey the general coordinate covariance by explicitly making the global Lorentz boost transformation into a new coordinate system with a constant boost vector $\boldsymbol{\beta}$. Actually, when the boost vector $\boldsymbol{\beta}$ becomes local $\boldsymbol{\beta} =\boldsymbol{\beta}(x')$, the gravigeometry-medium Maxwell equations in such a boosted new reference frame remain valid. This is because the above derivation does not depend on whether the Lorentz boost transformation is global or local. In general, the gravigeometry-medium Maxwell equations for the gravigeometry-medium electromagnetic field hold in any motional coordinate reference frame due to the general coordinate covariance of electrodynamics. 


\subsection{Gravigauge-mediated electrodynamics in special background medium and spin gauge covariance of generalized Maxwell equations in any spinning reference frame}

To show the spin effect arising from the interaction of spin gravigauge field $\mOm_{\mu}^{ab}(x)$, it is useful to examine the gravigauge-mediated Maxwell equations in the special background medium and demonstrate explicitly the spin gauge covariance under the spin boost transformation of spin gauge symmetry SP(1,3) in locally flat gravigauge spacetime. 

From the simple gravimetric form of special background medium shown in Eq.(\ref{SBM1}), the corresponding dual gravigauge field $\chih_{a}^{\;\;\mu}$ is simply given by,
\be \label{SBM3}
\chih_{0}^{\;\; 0} = c_s^{-1} \eta_{0}^{\; 0}, \quad \chih_{0}^{\;\; i} = 0 = \chih_{\alpha}^{\; \;0} , \quad \chih_{\alpha}^{\; \;i}  = \eta_{\alpha}^{\; i} ,
\ee 
with $a=( 0, \alpha)$ and $\mu= ( 0, i)$ ($\alpha, i = 1,2,3$). In this simple case, the background gravigauge field strength vanishes, and the gravigauge-dressed electromagnetic field gets the following  simple forms:
\be
\wt{\fE} = c_s^{-1} \fE, \quad \wt{\fB} =  \fB .
\ee
The gravigauge Maxwell equations presented in Eq.(\ref{GME}) are simplified to be:
\be
&& \bvnabla \cdot \wt{\fE} = J^0 , \nn \\
& & \bvnabla \times \wt{\fB}  = \fJ + \eth_0 \wt{\fE} , \nn \\
& & \bvnabla \times \wt{\fE} = - \eth_0 \wt{\fB} , \nn \\
& & \bvnabla \cdot \wt{\fB} = 0 ,   
\ee
where the gravicoordinate derivatives in such a special background medium are given by,
\be
\bvnabla = \nabla, \quad \eth_0 = c_s^{-1} \p_{0} \equiv c_{s}^{-1} \frac{\p}{\p x^0}  .
\ee

Let us now consider the following spin gauge transformation in locally flat gravigauge spacetime:
\be
& & \chih_{a}^{\;\;\mu}(x) \to \chih_{a}^{'\;\, \mu}(x) = \Lambda_{a}^{\;\; b}(x)\chih_{b}^{\;\;\mu}(x) ,\quad \eth_{a}  \to \eth'_{a} =  \Lambda_{a}^{\;\; b}(x) \eth_{b}, \nn \\
& & F^{ab}(x) \to F^{'ab}(x) = \Lambda^{a}_{\;\; c}(x)\Lambda^{b}_{\;\; d} F^{cd}(x) ,\nn \\
& & J^{a}(x) \to J^{'a}(x) =  \Lambda^{a}_{\;\; c}(x) J^{c}(x) , 
\ee
with the spin boost transformation given explicitly as follows:
\be
& &\Lambda_{a}^{\;\; b}(x) \equiv \eta_{a}^{\;\; b} +  \tLam_{a}^{\;\; b}(x) ,  \nn \\
& & (\tLam)_{00} = \cosh\varpi - 1, \quad (\tLam)_{\alpha 0} = -  \varsigma_{\alpha} \sinh\varpi   = (\tLam)_{0 \alpha} , \nn \\
& &  (\tLam)_{\alpha\beta} = (\cosh\varpi -1)\varsigma_{\alpha}\varsigma_{\beta} = (\tLam)_{\beta\alpha},  \nn \\
&& \boldsymbol   {\varsigma} \equiv ( \varsigma_1, \varsigma_2, \varsigma_3 ) , \quad   \varsigma^2 \equiv \varsigma_1^2 + \varsigma_2^2 + \varsigma_3^2 = 1  , \nn \\
& &  \varpi \equiv \varpi(x), \quad \varsigma_{\alpha} \equiv \varsigma_{\alpha}(x), \quad \boldsymbol{\varsigma} \equiv \boldsymbol{\varsigma}(x) .
\ee
The gravigauge-dressed electromagnetic field in the spin boosted new reference frame has the following relations:
\be
& & \wt{\fE}'(x) = \cosh \varpi \, \wt{\fE} + \sinh \varpi \,  \boldsymbol{\varsigma} \times \wt{\fB}  - ( \cosh\varpi - 1) \boldsymbol{\varsigma} \cdot \wt{\fE}\, \boldsymbol{\varsigma} , \nn \\
& & \wt{\fB}'(x) = ( 1- 2(\cosh \varpi -1) ) \1 \wt{\fB} - \sinh \varpi \,  \boldsymbol{\varsigma} \times \wt{\fE}  - 2( \cosh\varpi - 1) \boldsymbol{\varsigma} \cdot \wt{\fB}\, \boldsymbol{\varsigma} .
\ee
The corresponding relations for the current density and gravicoordinate derivative are found to be:
\be
& & J^{'0}(x) = \cosh \varpi \, J^{0} - \sinh \varpi \,  \bs{\varsigma} \cdot \fJ , \nn \\
& & \fJ'(x) = \fJ + [ -J^{0}\sinh \varpi + (\cosh \varpi -1) \bs{\varsigma} \cdot \fJ ] \, \bs{\varsigma} ,
\ee
and
\be
& & \bvnabla' = \bvnabla - \bs{\varsigma} [ \sinh\varpi \,  \eth_0 - (\cosh\varpi - 1) \bs{\varsigma}\cdot \bvnabla ]  = \nabla - \bs{\varsigma} c_s^{-1}\sinh\varpi \, ( \frac{\p}{\p x^0} - c_s \tanh\frac{\varpi}{2} \1 \bs{\varsigma}\cdot \nabla ) , \nn \\
& & \eth'_0 = \cosh\varpi \, \eth_0 - \sinh\varpi \, \boldsymbol{\varsigma}\cdot \bvnabla =  c_s^{-1} \cosh\varpi \, (\frac{\p}{\p x^0} - c_s \tanh \varpi \, \bs{\varsigma}\cdot \nabla ) .
\ee

It can be verified that in the spin boosted new reference frame, the gravigauge field stength in the special background medium becomes no vanishing and gets the following form:
\be
F_{cb}^{' a}(x) =  -  \Lambda^{a}_{\;\; d} (\eth'_{c} \Lambda_{b}^{\;\; d}  - \eth'_{b}\Lambda_{c}^{\;\; d} ) \equiv - \mOm_{[cb]}^{'a}(x),
\ee
which results from a pure spin gauge field $\mOm_{c}^{ab}(x)$ after the spin gauge transformation $\Lambda^{a}_{\;\; b}(x)$. To be more explicit, the electromagnetic-like gravigauge fields are given as follows:
\be
& & \bv{E}'_{\alpha} = - [ \eth'_0\varpi + \sinh\varpi\, (\eth'_0 \bs{\varsigma})\cdot \bs{\varsigma} ] \1 \varsigma_{\alpha} - \sinh \varpi [ \eth'_0 \varsigma_{\alpha} + \sinh \varpi\1 (\eth_{\alpha}' \bs{\varsigma}) \cdot \bs{\varsigma} ] , \nn \\
& & \bv{B}'_{\alpha} = - \epsilon_{\alpha}^{\;\; \beta\gamma} [ \sinh \varpi \, \eth'_{\beta} \varsigma_{\gamma} + \eth'_{\beta}\varpi  \varsigma_{\gamma}  - \sinh \varpi\1 (\cosh \varpi - 1)  \varsigma_{\beta} (\eth'_{\gamma} \bs{\varsigma}) \cdot \bs{\varsigma} ] , \nn \\
& & \bv{E}_{\alpha}^{'\beta} = - [ \eth'_{\alpha}\varpi -2 \varsigma_{\alpha} \sinh\varpi\, (\eth'_0 \varpi)] \1 \varsigma^{\beta} - \sinh \varpi \, \eth'_{\alpha} \varsigma^{\beta} - (\cosh \varpi -1) [ \varsigma_{\alpha} \eth_{0}' \varsigma^{\beta} -  (\eth_{0}' \varsigma_{\alpha}) \varsigma^{\beta} ]  , \nn \\
& & \quad \quad \quad + (\cosh \varpi -1) [ (\cosh \varpi -1) \varsigma_{\alpha} (\eth_{0}' \bs{\varsigma})\cdot  \bs{\varsigma}) + \sinh \varpi \, (\eth_{\alpha}' \bs{\varsigma})\cdot \bs{\varsigma} ]  \varsigma^{\beta} , \nn \\
& & \bv{B}_{\alpha}^{'\; \beta}  = \epsilon_{\alpha}^{\;\; \gamma\delta} [ 2\sinh \varpi \, \eth'_{\gamma} \varsigma_{\delta} + 3  (\cosh \varpi -1) \eth'_{\gamma} \varsigma_{\delta}  - (\cosh \varpi - 1)^2  \varsigma_{\gamma} (\eth'_{\delta} \bs{\varsigma}) \cdot \bs{\varsigma} ] \varsigma^{\beta} , \nn \\
& & \quad \quad \quad - (\cosh \varpi -1) \epsilon_{\alpha}^{\;\; \gamma\delta} \varsigma_{\gamma} \eth'_{\delta}\varsigma^{\beta} ,
\ee  
which can be represented by the following vector representations:
\be
& & \bv{\fE}'(x) = - [ \eth'_0\varpi + \sinh\varpi\, (\eth'_0 \bs{\varsigma})\cdot \bs{\varsigma} ] \1 \bs{\varsigma} - \sinh \varpi [ \eth'_0 \bs{\varsigma} + \sinh \varpi\1 (\bvnabla' \bs{\varsigma}) \cdot \bs{\varsigma} ] , \nn \\
& & \bv{\fB}'(x) =  \sinh \varpi \, \bvnabla' \times \bs{\varsigma} + \bvnabla'\varpi \times \bs{\varsigma}  - \sinh \varpi\1 (\cosh \varpi - 1)  \bs{\varsigma}\times \left( (\bvnabla'\bs{\varsigma}) \cdot \bs{\varsigma} \right)  , \nn \\
& & \bv{\mathbb{E}}'(x) = - [ \bvnabla'\varpi -2 \bs{\varsigma} \sinh\varpi\, (\eth'_0 \varpi)] \1 \bs{\varsigma} - \sinh \varpi \, \bvnabla' \bs{\varsigma} - (\cosh \varpi -1) [ \bs{\varsigma} \eth_{0}' \bs{\varsigma} -  (\eth_{0}' \bs{\varsigma}) \bs{\varsigma} ]  , \nn \\
& & \quad \quad \quad + (\cosh \varpi -1) [ (\cosh \varpi -1) \bs{\varsigma} (\eth_{0}' \bs{\varsigma})\cdot  \bs{\varsigma}) + \sinh \varpi \, (\bvnabla' \bs{\varsigma})\cdot \bs{\varsigma} ]  \bs{\varsigma} , \nn \\
& & \bv{\mathbb{B}}'(x)  = -[ 2\sinh \varpi \, \bvnabla' \times \bs{\varsigma} + 3  (\cosh \varpi -1) \bvnabla' \times \bs{\varsigma}  - (\cosh \varpi - 1)^2  \bs{\varsigma} \times (\bvnabla' \bs{\varsigma}) \cdot \bs{\varsigma} ] \bs{\varsigma} , \nn \\
& & \quad \quad \quad + (\cosh \varpi -1) (\bs{\varsigma}\times \bvnabla' ) \bs{\varsigma} .
\ee  

The gauge covariance of gravigauge-mediated Maxwell equations in the special background medium can be expressed into the following forms in the spin boosted new reference frame:
\be \label{GGME4}
& & \nabla \cdot \wt{\fE}' - c_s^{-1} \sinh\varpi \, \bs{\varsigma} \cdot ( \frac{\p}{\p x^{0}}    - c_s\tanh\frac{\varpi}{2}  \bs{\varsigma}\cdot \nabla ) \wt{\fE}'  - \bvfA' \cdot \wt{\fE}' -\bvfE' \cdot \wt{\fE}'  = J^{'0} + \bvfB' \cdot \wt{\fB}',   \nn \\
& & \nabla \times \wt{\fB}' - c_s^{-1} \sinh\varpi \, \bs{\varsigma} \times ( \frac{\p}{\p x^{0}}  - c_s \tanh\frac{\varpi}{2} \bs{\varsigma}\cdot \nabla) \wt{\fB}'  - \bv{\fA}' \times \wt{\fB}' + \bv{\mathbb{B}}' \cdot \wt{\fB}'    \nn \\
& & \qquad \quad  =  \fJ'  + c_s^{-1} \cosh\varpi \, (\frac{\p }{\p x^0}   - c_s \tanh\varpi \, \bs{\varsigma}\cdot \nabla )\wt{\fE}'  - \bv{A}'_0 \wt{\fE}' + \bv{\mathbb{E}}' \cdot \wt{\fE}', \nn \\
& & \nabla\cdot \wt{\fB}' -  c_s^{-1}\sinh\varpi \, \bs{\varsigma} \cdot ( \frac{\p}{\p x^{0}}    - c_s\tanh\frac{\varpi}{2} \bs{\varsigma}\cdot \nabla ) \wt{\fB}'   +2 \bvfA' \cdot \wt{\fB}' + 2\bvfE' \cdot \wt{\fB}'   =  \bvfB' \cdot \wt{\fE}',  \nn \\
& & \nabla \times \wt{\fE}' -  c_s^{-1}\sinh\varpi \,  \bs{\varsigma} \times  ( \frac{\p }{\p x^{0}}  -c_s \tanh\frac{\varpi}{2} \bs{\varsigma}\cdot \nabla ) \wt{\fE}'   + \bv{\fE}' \times \wt{\fE}' - \bv{\mathbb{B}}' \cdot \wt{\fE}' \nn \\
& & \qquad \quad = - c_s^{-1} \cosh\varpi \, ( \frac{\p}{\p x^0}   - c_s \tanh\varpi \, \bs{\varsigma}\cdot \nabla )\wt{\fB}'   - 2\bv{A}'_0 \wt{\fB}' + \bv{\mathbb{E}}' \cdot \wt{\fB}' ,  
\ee
where both the gravigauge-dressed electromagnetic fields and electromagnetic-like gravigauge fields in the gravigauge-mediated Maxwell equations are located at the same point $x^{\mu}$ in the coordinate spacetime as the spin gauge transformation operates on the locally flat gravigauge spacetime and Dirac spinor field in Hilbert space.

For a special case that the spin transformation is taken to be a global transformation, namely the boost parameters $\varpi$ and $\varsigma_{\alpha}$ are all constant parameters, the  electromagnetic-like gravigauge fields get to be vanishing. As a consequence, the above gravigauge-mediated Maxwell equations in the globally spin boosted new reference frame are simplified to be:
\be \label{GGME5}
& & \nabla \cdot \wt{\fE}' - c_s^{-1} \sinh\varpi  ( \frac{\p}{\p x^{0}}    - c_s\tanh\frac{\varpi}{2} \bs{\varsigma}\cdot \nabla ) \bs{\varsigma} \cdot \wt{\fE}'  = J^{'0}  ,   \nn \\
& & \nabla \times \wt{\fB}' - c_s^{-1} \sinh\varpi   ( \frac{\p}{\p x^{0}}  - c_s \tanh\frac{\varpi}{2} \bs{\varsigma}\cdot \nabla) \bs{\varsigma} \times \wt{\fB}'  \nn \\
& & \qquad \quad =  \fJ'  + c_s^{-1} \cosh\varpi \,(\frac{\p }{\p x^0}   - c_s \tanh\varpi \, \bs{\varsigma}\cdot \nabla )  \wt{\fE}'  , \nn \\
& & \nabla\cdot \wt{\fB}' -  c_s^{-1}\sinh\varpi  ( \frac{\p }{\p x^{0}}    - c_s\tanh\frac{\varpi}{2} \bs{\varsigma}\cdot \nabla )  \bs{\varsigma} \cdot \wt{\fB}'   =0,  \nn \\
& & \nabla \times \wt{\fE}' -  c_s^{-1}\sinh\varpi  ( \frac{\p }{\p x^{0}}  - c_s \tanh\frac{\varpi}{2} \bs{\varsigma}\cdot \nabla ) \bs{\varsigma} \times \wt{\fE}'  \nn \\
& & \quad \qquad  = - c_s^{-1} \cosh\varpi \, (\frac{\p }{\p x^0}   - c_s \tanh\varpi \, \bs{\varsigma}\cdot \nabla ) \wt{\fB}'  .  
\ee

Therefore, we come to the observation that the gravigauge-mediated Maxwell equations for gravigauge-dressed electromagnetic fields in locally flat gravigauge spacetime maintain the spin gauge covariance with hidden general coordinate invariance, and the gravigeometry-medium Maxwell equations for gravigeometry-medium electromagnetic fields in coordinate spacetime keep the general covariance of coordinates with hidden spin gauge invariance.


\section{Conclusions and discussions}

In this paper, we have begun from the fact that the electron and charged leptons and quarks as basic constituents of matter in SM are actually Weyl fermions although they appear as Dirac fermions in electromagnetic interaction. This is because their weak interactions in SM produce the maximal parity violation. Based on such an observation, we have represented such a Dirac fermion in four-dimensional Hilbert space as the superposition of left-handed and right-handed Weyl fermions and treated it as a chirality-based Dirac spinor in the chiral spinor representation of eight-dimensional Hilbert space. We have demonstrated that such a chirality-based Dirac spinor possesses the inhomogeneous spin symmetry WS(1,3) in eight-dimensional Hilbert space instead of the usual spin symmetry SP(1,3) in four-dimensional Hilbert space. Such an inhomogeneous spin symmetry group WS(1,3) is a semi-direct product group WS(1,3) = SP(1,3)$\rtimes$W$^{1,3}$ with SP(1,3) being the spin group and W$^{1,3}$ referred to as $\cW_e$-spin group, which realizes an internal Poincar\'e-type group (or internal inhomogeneous Lorentz-type group) in eight-dimensional Hilbert space. It has been shown that the $\cW_e$-spin symmetry group emerges as a translation-like chiral-type spin symmetry group and the electric charge symmetry group U(1) appears as an internal Abelian group symmetry for the chirality-based Dirac spinor formulated in eight dimensional Hilbert space. 

Following along the gauge invariance principle which states that the laws of nature should be independent of the choice of local field configurations, we have gauged the inhomogeneous spin symmetry WS(1,3) and electric charge symmetry U(1) for the chirality-based Dirac spinor in eight dimensional Hilbert space to be local symmetries. Meanwhile, the inhomogeneous Lorentz group symmetry (or Poincar\'e group symmetry) PO(1,3) remains to be global symmetry in four-dimensional Minkowski spacetime of coordinates. The inhomogeneous spin gauge symmetry WS(1,3) as well as electric charge gauge symmetry U(1) lead to the introduction of spin gauge field and $\cW_e$-spin gauge field as well as electromagnetic gauge field. It has been demonstrated that the translation-like $\cW_e$-spin invariant-gauge field provides a gravigauge field as basic gravitational field, which reveals the nature of gravity and brings on the gravitational origin of spin gauge symmetry. So such an inhomogeneous spin gauge symmetry WS(1,3) naturally leads to an internal Poincar\'e-type gauge symmetry. Furthermore, based on the displacement correspondence between the gravivector field in Hilbert space and coordinate vector in Minkowaski spacetime through the gravigauge field, a locally flat gravigauge spacetime spanned by the gravigauge basis emerges to form a fiber bundle structure of biframe spacetime, where the globally flat Minkowski spacetime appears as base spacetime and the locally flat gravigauge spacetime behaves as a fiber. By noticing the intriguing property that the gravigauge field plays a role as a Goldstone-like boson, we are able to define all gauge fields and field strengths in locally flat gravigauge spacetime, where the gravitational effects are shown to be characterized by the emergent non-commutative geometry of locally flat gravigauge spacetime.

Based on the action principle of path integral formulation, we have presented various formalisms of gauge and scaling invariant actions for the chirality-based Dirac spinor in eight dimensional Hilbert space, which has been shown to possess the maximal joint symmetry PO(1,3)$\Join$WS(1,3)$\times$U(1) and generate a hidden general linear group symmetry GL(1,3,R). It has been verified that when the basic symmetry of chirality-based Dirac spinor in eight dimensional Hilbert space is gauged to be local symmetry via the gauge invariance principle, the symmetry of coordinate system automatically turns out to possess a hidden general linear group symmetry GL(1,3,R), which indicates that the laws of nature should be independent of the choice of coordinate systems and meanwhile brings on the genesis of Riemann geometry in curved Riemannian spacetime. Such a feature enables us to demonstrate the gauge-gravity and gravity-geometry correspondences and obtain the gauge-geometry duality relation within the framework of gravitational quantum field theory. 

From the least action principle, we have derived the generalized Dirac equation to characterize the gravitational relativistic quantum theory. In particular, the quadratic gravigauge Dirac equation is derived in locally flat gravigauge spacetime to show the gravitational effect. The gauge-type gravitational equations of gravigauge field have been obtained to describe the gravidynamics in biframe spacetime and also in locally flat gravigauge spacetime. Meanwhile, we have investigated in detail the geometric gravidynamics based on the inhomogeneous spin gauge symmetry and discussed its new effects beyond the Einstein theory of general relativity. The equation of motion for the spin gauge field has also been derived to describe the spinodynamics. 

In particular, we have made a detailed analysis on the various formalisms of electrodynamics via deriving different generalized Maxwell equations in the presence of gravitational interactions, such as: the gravigeometry-medium electrodynamics by redarding the gravimetric field as the medium, the gravigauge-mediated electrodynamics in locally flat gravigauge spacetime and the gravimetric-gauge-mediated electrodynamics in dynamic Riemannian spacetime. In analyzing the gravimetric-gauge-mediated Maxwell equations, we have shown that it is useful to introduce the electric-like and magnetic-like gravimetric-gauge fields. In general, we have explicitly demonstrated that the dynamics of all basic fields should maintain the general coordinate covariance in curved Riemannian spacetime and the spin gauge covariance in locally flat gravigauge spacetime, so the gravigeometry-medium Maxwell equations hold in any motional reference frame and the gravigauge-mediated Maxwell equations validate in any spinning reference frame. We have also presented a full discussion on the generalized Maxwell equations in special background medium and verified explicitly the general coordinate covariance in motional reference frame and the spin gauge covariance in spinning reference frame. 

It is straightforward to apply the present analyses to extend the standard model by including the gravitational interaction, which enables us to provide a unified description on four basic forces, i.e., electromagnetic, weak, strong and gravitational interactions, within the framework of gravitational quantum field theory. Meanwhile, such an extended standard model with inhomogeneous spin gauge symmetry is expected to result from a more fundamental theory, such as the hyperunified field theory \cite{HUFT1,HUFT2,HUFTTK,FHUFT1,FHUFT2} which is found to be governed by inhomogeneous hyperspin gauge symmetry WS(1,18) and inhomogeneous Lorentz-type symmetry (or Pincar\'e-type symmetry) PO(1,18) in 19-dimensional hyper-spacetime. In particular, the foundation of the hyperunified field theory\cite{FHUFT1,FHUFT2,FHUFT} has turned out to be laid by two simple guiding principles, i.e.: maximum locally entangled-qubits motion principle and gauge invariance principle.
\\
\\

\centerline{{\bf Acknowledgement}}

I would like to thank the reviewers for their useful comments and suggestions. This work was supported in part by the National Key Research and Development Program of China under Grant No.2020YFC2201501, the National Science Foundation of China (NSFC) under Grants No.~12147103 (special fund to the center for quanta-to-cosmos theoretical physics), No.~11821505, and the Strategic Priority Research Program of the Chinese Academy of Sciences under Grant No. XDB23030100.

\end{document}